\documentclass[prb,twocolumn,showpacs,amsmath,amssymb,eqsecnum]{revtex4-1}

\usepackage{graphicx}
\usepackage{amsmath,amssymb,bm}
\usepackage{color}

\begin{document}

\title{
Higher-order Fermi-liquid corrections for an Anderson impurity away from half-filling  III: 
non-equilibrium transport
}

 \author{Akira Oguri}
 \affiliation{
 Department of Physics, Osaka City University, Sumiyoshi-ku, 
 Osaka 558-8585, Japan }

  \author{A.\ C.\ Hewson}
 \affiliation{
 Department of Mathematics, Imperial College London, London SW7 2AZ, 
  United Kingdom }

\date{\today}

\begin{abstract}

We extend the microscopic Fermi-liquid theory for the Anderson impurity  
[Phys.\ Rev.\ B {\bf  64}, 153305  (2001)]  
to explore  non-equilibrium transport at finite magnetic fields.
Using the Ward identities in the Keldysh formalism with  
the analytic and anti-symmetric properties of the vertex function,
the  spin-dependent Fermi-liquid corrections of order  $T^2$ and $(eV)^2$ 
are determined at low temperatures $T$ and low bias voltages $eV$. 
Away from half-filling, these corrections  can be expressed in terms of  
the linear and non-linear static susceptibilities 
which represent  the two-body and three-body fluctuations, respectively. 
We calculate the non-linear susceptibilities 
using  the numerical renormalization group, 
to explore  the differential conductance $dI/dV$ through a quantum dot.   
We find that the two-body fluctuations  
dominate the corrections  in the Kondo regime at zero magnetic field.  
The contribution of  the three-body fluctuations become significant far away from half-filling, 
especially in the valence-fluctuation  regime and empty-orbital regimes.  
In finite magnetic fields, 
 the three-body contributions  become comparable to  the two-body contributions, 
and play an essential role in the splitting of  the zero-bias conductance peak  
 occurring at a magnetic field of the order of  the Kondo energy scale.
We  also apply  our microscopic formulation  
to  the magneto-resistance  and  thermal conductivity 
of dilute magnetic alloys away from half-filling.

\end{abstract}


\maketitle

\section{Introduction}
\label{sec:introduction}

It has already been more than forty years since   
Nozi\`{e}res' phenomenological Fermi-liquid  theory 
for the Kondo system\cite{NozieresFermiLiquid} 
and the corresponding microscopic description of Yamada-Yosida
\cite{YamadaYosida2,YamadaYosida4,ShibaKorringa,Yoshimori}
 successfully  explained  the universal low-energy behavior,   
which had been clarified by  Wilson's  
numerical normalization group (NRG).\cite{WilsonRMP,KWW1,KWW2}
Recently, there has been a significant breakthrough,   
which extends Nozi\`{e}res' phenomenological description 
and reveals  higher-order Fermi-liquid corrections in the 
particle-hole asymmetric case.\cite{MoraMocaVonDelftZarand,FilipponeMocaVonDelftMora} 
Specifically,  Filippone, Moca, von Delft and Mora (FMvDM) 
have presented  the low-energy asymptotic form 
of the Green's function $G_{\sigma}(\omega)$ 
up to terms of order  $\omega^2$,  $T^2$, and $(eV)^2$, 
at finite temperatures   $T$, bias voltages $V$, and 
magnetic fields.\cite{FilipponeMocaVonDelftMora} 
 It shed light on a long standing problem 
in the Kondo physics away from half-filling,    
which has been studied  for dilute magnetic alloys\cite{NozieresFermiLiquid,YamadaYosida2,YamadaYosida4,ShibaKorringa,Yoshimori} 
and  quantum dots.\cite{Hershfield1,ao2001PRB,Aligia,Munoz}

In the previous two papers,\cite{ao2017_1,ao2017_2_PRB} 
 we provided a microscopic description for 
 the higher-order Fermi-liquid corrections  away from half-filling, 
 extending the  approach of Yamada-Yosida using Ward identities.
We have shown that the next-leading Fermi-liquid corrections, 
which cannot be neglected away from half-filling,  
are deduced from one of the 
 key features of the vertex function for parallel spins: 
the $\omega$-linear term of 
 $\Gamma_{\sigma\sigma;\sigma\sigma} (\omega, 0; 0, \omega) $ 
becomes  pure imaginary with no real part at $T=0$ and $eV=0$.  
The additional Fermi-liquid parameters  
can be expressed in terms of the static three-body correlation functions 
of the impurity occupations, i.e.,  $n_{d\sigma}$'s. 
The first paper is a letter, in which have described  
an overview of the results that  follow  from this property.\cite{ao2017_1}
It has been proved in the second paper, hereafter referred to as {\it paper II},   
that the fermionic anti-symmetry property causes 
the absence of an $\omega$-linear term in the real part 
of $\Gamma_{\sigma\sigma;\sigma\sigma} (\omega, 0; 0, \omega)$.\cite{ao2017_2_PRB} 
In addition,  we have also calculated 
the $\omega^2$ and $T^2$ real part of the self-energy at equilibrium  
using the Matsubara imaginary-time Green's function.\cite{FilipponeMocaVonDelftMora}

In the present paper, 
we continue the precise discussion started in {\it paper II}.
We microscopically derive the 
low-energy asymptotic form of the Keldysh Green's function, 
extending the non-equilibrium Ward identities for finite magnetic fields.\cite{ao2001PRB}
We also calculate the Fermi-liquid corrections to transport through a quantum dot
\cite{GlazmanRaikh,NgLee,Hershfield1,WingreenMeir,MeirWingreen} 
 and also  thermoelectric transport\cite{CostiThermo,ao1990PRB} 
in dilute magnetic alloys  away from half-filling.    
In addition,  we apply the microscopic description 
to the multi-orbital case with $N$ impurity components, 
 and present  the precise form of 
the expansion coefficients for the self-energy.
The result  of  the order $\omega^2$ real part of the self-energy, 
which has been deduced from the Ward identity,  
completely agrees  with the FMvDM's formula.\cite{FilipponeMocaVonDelftMora}

In order to see how the higher-order Fermi-liquid parameters 
evolve as the system deviates from the particle-hole symmetric point, 
we also explore some typical cases using the NRG.
The corrections away from the symmetric case 
are determined  not only by the two-body fluctuations 
which enter through the linear susceptibilities $\chi_{\sigma\sigma'}$ 
but the three-body fluctuations described 
by the static nonlinear susceptibilities $\chi_{\sigma_1\sigma_2\sigma_3}^{[3]}$. 
Specifically, we see that each of these two types of the fluctuations  
contributes to  the $T^2$ and $(eV)^2$ Fermi-liquid corrections 
for the conductance through a quantum dot 
 away from half-filling, and also at finite magnetic fields. 
The result  shows that at zero field  
the contributions of the two-body fluctuations  dominate in the Kondo regime, 
whereas the three-body fluctuations are significant 
in valence fluctuation and empty-orbital regimes. 
In contrast, in the case where a magnetic field is applied to the Kondo regime, 
both the two-body and three-body fluctuations give comparable contributions 
to the $T^2$ and $(eV)^2$ corrections.  
We also discuss how these two types of  fluctuations contribute 
to the  $T^2$ corrections of the electric resistance and thermal conductivity 
of the dilute magnetic alloys.

The paper is organized as follows. 
In Sec.\ \ref{sec:formulation}, 
static non-linear susceptibilities and 
the Ward identities which have been described in  
{\it paper II}  are summarized.
The non-equilibrium Ward identities for finite magnetic fields 
are derived in Sec.\ \ref{sec:nonlinear_self_energy_at_h}. 
The results for the asymptotic form of the retarded self-energy is 
described  in  Sec.\ \ref{sec:result_self_energy}. 
Then, in Sec.\ \ref{sec:transport_dot}, 
differential conductance of quantum dot  is discussed 
at symmetric tunneling couplings.
In Sec.\ \ref{sec:magnetic_alloy}, 
we apply the microscopic Fermi-liquid description 
to  thermoelectric transport of dilute magnetic alloy away from half-filling.
A summary is given in Sec.\ \ref{sec:summary_III}.

\section{Formulation and summary of equilibrium properties}
\label{sec:formulation}

We study the transport properties in the 
Fermi-liquid regime away from half-filling in this paper.
We consider  the single Anderson impurity 
coupled to two noninteracting leads:   
$\mathcal{H} =
\mathcal{H}_d + \mathcal{H}_c  + \mathcal{H}_\mathrm{T}$,  
\begin{align}
\mathcal{H}_d =& \  
 \sum_{\sigma}
 \epsilon_{d\sigma}^{}\, n_{d\sigma}  
+
   U\,n_{d\uparrow}\,n_{d\downarrow} ,
 \label{eq:H_U_spin}
\\ 
\mathcal{H}_c =& \   
\sum_{\lambda=L,R} \sum_{\sigma} 
\int_{-D}^D  \!\! d\epsilon\,  \epsilon\, 
 c^{\dagger}_{\epsilon \lambda \sigma} c_{\epsilon \lambda \sigma}^{},
 \label{eq:Ham_cond}
\\
 \mathcal{H}_\mathrm{T} =& \   
   \sum_{\lambda=L,R} \sum_{\sigma}  v_{\lambda}^{}
 \left( \psi_{\lambda,\sigma}^\dag d_{\sigma}^{} + 
  d_{\sigma}^{\dag} \psi_{\lambda,\sigma}^{} \right) .
 \label{eq:Ham_mix}
\end{align}
Here, 
 $d^{\dag}_{\sigma}$ creates an impurity electron 
with spin $\sigma$ 
 in the impurity level of energy $\epsilon _{d\sigma}$, and  
 $n_{d\sigma} = d^{\dag}_{\sigma} d^{}_{\sigma}$. 
 $U$ is the Coulomb interaction between  electrons occupying the impurity level. 
Conduction electrons in the two leads at  $\lambda=L$ and $R$ 
obey  the  anti-commutation relation 
$
\{ c^{\phantom{\dagger}}_{\epsilon\lambda\sigma}, 
c^{\dagger}_{\epsilon'\lambda'\sigma'}
\} = \delta_{\lambda\lambda'} \,\delta_{\sigma\sigma'}   
\delta(\epsilon-\epsilon')$. 
The linear combination of the conduction electrons, 
$\psi^{}_{\lambda \sigma} \equiv  \int_{-D}^D d\epsilon \sqrt{\rho_c^{}} 
\, c^{\phantom{\dagger}}_{\epsilon\lambda \sigma}$  with $\rho_c^{}=1/(2D)$, 
 couples to the impurity level. The  bare width  is given by 
 $\Delta \equiv \Gamma_L + \Gamma_R$ with 
 $\Gamma_{\lambda} = \pi \rho_c^{} v_{\lambda}^2$.  
We consider the parameter region, where 
 the  half band-width 
 $D$ is much greater than the other energy scales,    
$D \gg \max( U, \Delta, |\epsilon_{d\sigma}^{}|, |\omega|, T, eV)$. 
For  finite magnetic fields $h$, the impurity energy takes the form
 $\epsilon_{d\sigma}^{} = \epsilon_{d}^{} - \sigma h$, 
 where  $\sigma = +1$ (-1) for $\uparrow$ ($\downarrow$) spin.
The relation between the differentiations is 
\begin{align}
&
\frac{\partial}{\partial \epsilon_{d}} 
= 
\frac{\partial}{\partial \epsilon_{d\uparrow}} 
+
\frac{\partial}{\partial \epsilon_{d\downarrow}} 
,  
\qquad
\frac{\partial}{\partial h} 
= 
-\frac{\partial}{\partial \epsilon_{d\uparrow}} 
+
\frac{\partial}{\partial \epsilon_{d\downarrow}} 
,
\end{align}
and 
$\frac{\partial}{\partial \epsilon_{d\sigma}^{}} 
 = 
 \frac{1}{2}\left(
 \frac{\partial}{\partial \epsilon_{d}} 
 -\sigma \, 
 \frac{\partial}{\partial h} 
 \right)$.

\subsection{Local Fermi-liquid parameters in equilibrium}

\subsubsection{Free energy $\Omega$ and Green's function at $T=0$}

The low-bias behavior of the self-energy 
can be deduced from the equilibrium quantities. 
Specifically, at $T =0$ and $eV = 0$, 
the usual zero-temperature formalism  
is applicable to the causal Green's function  
defined with respect to the equilibrium ground state,
\begin{align}
G_{\mathrm{eq},\sigma}^{--}(\omega)
\ = & \ 
 -i\, 
\int_{-\infty}^{\infty} dt \, e^{i \omega  t}
\langle \,\mbox{T}\,
d^{\phantom{\dagger}}_{\sigma}(t)\, d^{\dagger}_{\sigma}(0)
\,\rangle 
\nonumber \\  
 =& \     
\frac{1}{\omega -\epsilon_{d\sigma} +i\Delta \,\mathrm{sgn}\,\omega 
- \Sigma_{\mathrm{eq},\sigma}^{--}(\omega)  } \;. 
 \label{eq:G--_eq}
\end{align}
The corresponding $T=0$ retarded Green's function is given by
$G_{\mathrm{eq},\sigma}^{r}(\omega)
= \ 
\theta(\omega) G_{\mathrm{eq},\sigma}^{--}(\omega)
+\theta(-\omega) \left\{G_{\mathrm{eq},\sigma}^{--}(\omega)\right\}^*
$, where $\theta(\omega)$ is the Heaviside step function. 
The density of states for impurity electrons is defined by 
\begin{align}
 \rho_{d\sigma}^{}(\omega) \,\equiv\, -\,\frac{1}{\pi} \,
\mathrm{Im}\,  G_{\mathrm{eq},\sigma}^{r}(\omega)  \;.
\end{align}
We will write the density of states at the Fermi energy $\omega=0$ in the following 
way, suppressing the frequency argument  
$\rho_{d\sigma}^{} \equiv   \rho_{d\sigma}^{}(0)
= {\sin^2 \delta_{\sigma}}/{\pi \Delta}$, where 
\begin{align}
\delta_{\sigma} 
\,=\,    
\cot^{-1} \left[
\frac{\epsilon_{d\sigma}^{} + \Sigma_{\mathrm{eq},\sigma}^{r}(0)}{\Delta} \right].
\end{align}
The phase shift $\delta_{\sigma}$  is a primary parameter 
which characterizes the Fermi-liquid ground state.
The Friedel sum rule relates  the phase shift to  the occupation number 
which also corresponds to the first derivative of the free energy 
  $\Omega \equiv - T \log \left[\mathrm{Tr}\,e^{-\mathcal{H}/T}\right]$,  
\begin{align}
&\langle n_{d\sigma} \rangle \,=\,
\frac{\partial \Omega}{\partial \epsilon_{d\sigma}^{}} 
 \,\xrightarrow{\,T\to 0\,} \,  
\frac{\delta_{\sigma}}{\pi} \;.
\end{align}

\subsubsection{Second derivative of $\Omega$}

The  leading Fermi-liquid corrections 
can be described by  
the static susceptibilities  
following Yamada-Yosida:\cite{YamadaYosida2}
\begin{align}
\chi_{\sigma\sigma'}^{} \equiv  
- \frac{\partial^2 \Omega}
{\partial \epsilon_{d\sigma'}^{} \partial \epsilon_{d\sigma}^{}} 
 =
- \,\frac{\partial \langle n_{d\sigma} \rangle }{\partial \epsilon_{d\sigma'}} 
\,\xrightarrow{\,T\to 0\,} \,  
 \rho_{d\sigma}^{} \, \widetilde{\chi}_{\sigma\sigma'}^{}  .
\!\!
\label{eq:chi_org}
\end{align}
Note that $\chi_{\uparrow\downarrow}^{}=\chi_{\downarrow\uparrow}^{}$. 
The renormalization factors are  defined by 
\begin{align}
\!\!\!
\widetilde{\chi}_{\sigma\sigma'}\! 
\equiv  
\delta_{\sigma\sigma'}  \!+\!
\frac{\partial   \Sigma_{\mathrm{eq},\sigma}^{r}(0)}{\partial \epsilon_{d\sigma'}}
,  \ \ \,  
\frac{1}{z_{\sigma}^{}}\! 
\equiv      
1 \!- \! \left.
 \frac{\partial  \Sigma_{\mathrm{eq},\sigma}^{r}(\omega)}{\partial \omega}
\right|_{\omega= 0} 
\!\!   . \!\!\!  
\label{eq:chi_tilde}
\end{align}
The susceptibility can be written as a static 2-body correlation function
\begin{align}
\chi_{\sigma\sigma'}^{}   =   & \ 
\int_0^{1/T}  \!\! d \tau \, 
\left\langle  \delta n_{d\sigma}(\tau)\,\delta  n_{d\sigma'}\right\rangle    , 
\end{align}
where $\delta n_{d\sigma} \equiv   n_{d\sigma} - \langle n_{d\sigma}  \rangle$. 
The  usual spin and charge susceptibilities  are  given by 
\begin{subequations}
\begin{align}
\!\!\!
\chi_{c}  \,\equiv&  \    - \frac{\partial^2 \Omega}{\partial \epsilon_{d}^2} \,=\, 
 \chi_{\uparrow\uparrow}^{} +\chi_{\downarrow\downarrow}^{} 
+  \chi_{\uparrow\downarrow}^{} +  \chi_{\downarrow\uparrow}^{}
\;, \\ 
\!\!\!
\chi_{s}  \,\equiv&  \   
- \frac{1}{4} \frac{\partial^2 \Omega}{\partial h^2} \, = \,  
\frac{1}{4} \left(\,  
 \chi_{\uparrow\uparrow}^{} +\chi_{\downarrow\downarrow}^{} 
-  \chi_{\uparrow\downarrow}^{} -  \chi_{\downarrow\uparrow}^{}
\,\right) . 
\label{eq:chi_s}
\end{align}
\end{subequations}

The free energy $\Omega$ is an even function of the field $h$.
Therefore, $\chi_{c}$  and   $\chi_{s}$ are also even functions of $h$. 
Furthermore, $\chi_{\uparrow\downarrow}$ is an even function of $h$,  
\begin{align}
\chi_{\uparrow\downarrow} 
=\chi_{\downarrow\uparrow} 
 =  
-\frac{\partial^2 \Omega}
{\partial \epsilon_{d\uparrow}\partial \epsilon_{d\downarrow}} 
= -  
\frac{1}{4}
\left( 
\frac{\partial^2 \Omega}{\partial \epsilon_{d}^2} 
- \frac{\partial^2\Omega}{\partial h^2} 
 \right)  
 . 
\end{align}
Similarly, 
$\chi_{\uparrow\uparrow}+\chi_{\downarrow\downarrow}$ is an even function of $h$,  
and $\chi_{\uparrow\uparrow}-\chi_{\downarrow\downarrow}$ 
is an odd function of $h$:   
\begin{subequations}
\begin{align}
\chi_{\uparrow\uparrow} +\chi_{\downarrow\downarrow}
\, = &  \ 
 -\frac{1}{2}
\left( 
\frac{\partial^2 \Omega}{\partial \epsilon_{d}^2} 
+\frac{\partial^2 \Omega}{\partial h^2} 
\right) 
\;,
\\
\chi_{\uparrow\uparrow} -\chi_{\downarrow\downarrow}
\, = &  \ 
\frac{\partial}{\partial \epsilon_{d}} 
\left(\frac{\partial \Omega}{\partial h}\right)  
\;.
\end{align}
\end{subequations}
Therefore, 
 $\chi_{\uparrow\uparrow} =\chi_{\downarrow\downarrow}$ at zero field $h=0$.

\subsubsection{Third derivative of $\Omega$}

The next leading  Fermi-liquid corrections 
 are determined by  the static nonlinear  susceptibilities, as we will describe later,
\begin{align}
\chi_{\sigma_1\sigma_2\sigma_3}^{[3]} \equiv \  
- \,
\frac{\partial^3 \Omega }{\partial \epsilon_{d\sigma_1}\partial \epsilon_{d\sigma_2}\partial \epsilon_{d\sigma_3}} 
\,=\,  \frac{\partial \chi_{\sigma_2\sigma_3}}
{\partial \epsilon_{d\sigma_1}} .
\label{eq:canonical_correlation_3_dif}
\end{align}
It also corresponds to the thee-body correlations 
of the impurity occupation 
\begin{align}
\chi_{\sigma_1\sigma_2\sigma_3}^{[3]} 
\!  = 
- \!
\int_{0}^{\frac{1}{T}} \!\!\! d\tau_3 \!\! 
\int_{0}^{\frac{1}{T}} \!\!\! d\tau_2\, 
\langle T_\tau 
\delta n_{d\sigma_3} (\tau_3) \,
\delta n_{d\sigma_2} (\tau_2) \,
\delta n_{d\sigma_1}
\rangle .
\label{eq:canonical_correlation_3}
\end{align}
Similarly,  the  $n$-th derivative of $\Omega$ 
for $n=4,5,6\cdots$ corresponds to 
 the $n$-body correlation function 
$\chi_{\sigma_1\sigma_2\sigma_3\cdots}^{[n]}$. 
The Fermi-liquid corrections can be classified according to $n$,  
and  the derivative of the Ward identity reveals 
 a hierarchy of Fermi-liquid relations, as described in the next 
subsection. 

The $n$-body correlation function 
have permutation symmetry for the spin indexes  
$\chi_{\sigma_1\sigma_2\sigma_3 \cdots}^{[n]} 
 = 
\chi_{\sigma_2\sigma_1\sigma_3\cdots}^{[n]} 
 =  
\chi_{\sigma_3\sigma_2\sigma_1 \cdots}^{[n]} 
 =  \cdots$, and thus it has $n+1$ independent components 
at finite magnetic fields.  
%
There are four independent components for the  $n=3$ case: 
\begin{align}
\frac{\partial \chi_{\uparrow\downarrow}}{\partial \epsilon_{d\sigma}^{}} 
\,   =&   \    
\frac{1}{2}
\left(
\frac{\partial \chi_{\uparrow\downarrow}}{\partial \epsilon_{d}^{}} 
-\sigma 
\frac{\partial \chi_{\uparrow\downarrow}}{\partial h} \right) 
 \label{eq:eV2_real_part_mag_again_move}
\, \xrightarrow{\,h\to 0\,}   \, 
\frac{1}{2}
\frac{\partial \chi_{\uparrow\downarrow}}{\partial \epsilon_{d}^{}} \;,
\\
\frac{\partial \chi_{\sigma\sigma}}{\partial \epsilon_{d\sigma}^{}} 
\   =  &\ 
\frac{\partial \chi_{\sigma\sigma}}{\partial \epsilon_{d}^{}} 
- \frac{\partial \chi_{\uparrow\downarrow}}{\partial \epsilon_{d\sigma}^{}} 
 \xrightarrow{\,h\to 0\,} \, 
\frac{\partial \chi_{\uparrow\uparrow}}{\partial \epsilon_{d}^{}} 
- \frac{1}{2} 
\frac{\partial \chi_{\uparrow\downarrow}}{\partial \epsilon_{d}^{}} 
.
 \label{eq:essential_real_part_mag_again_move}
\end{align}
for $\sigma = \uparrow, \downarrow$.
At zero field $h = 0$, only two components are independent because  
$
\chi_{\uparrow\uparrow\uparrow}^{[3]}
=\chi_{\downarrow\downarrow\downarrow}^{[3]} $ and 
$\chi_{\uparrow\uparrow\downarrow}^{[3]}
= \chi_{\uparrow\downarrow\downarrow}^{[3]}$   
due to the spin rotation symmetry, 
and  $\partial \chi_{\uparrow\downarrow}/\partial h$ vanishes  
as  $\chi_{\uparrow\downarrow}$ is an even function of $h$. 
Furthermore, in the particle-hole symmetric case 
for which  $\xi_d \equiv \epsilon_d + U/2 \to 0$,
the phase shift reaches the unitary limit value $\delta_{\sigma} \to \frac{\pi}{2}$. 
Then the charge susceptibility $\chi_c$ and spin susceptibility  $\chi_s$
 take a minimum and a maximum, respectively, and thus 
\begin{align}
\left.
\frac{\partial \chi_{\uparrow\uparrow}}{\partial \epsilon_{d}^{}} 
\right|_{h=0  \atop \xi_d=0}^{} =0\;,
\qquad  \ 
\left.
\frac{\partial \chi_{\uparrow\downarrow}}{\partial \epsilon_{d}^{}} 
\right|_{h=0  \atop \xi_d=0}^{} =0\;.
\end{align}

The derivative of the renormalization factors  $\widetilde{\chi}_{\sigma\sigma'}$ 
can also be written in terms of  the susceptibilities, 
\begin{align}
& 
\!\!\!\!
\frac{\partial \widetilde{\chi}_{\sigma_1\sigma_2}^{}}
{\partial \epsilon_{d\sigma_3}^{}} 
= 
\frac{1}{\rho_{d\sigma_1}^{}}\! 
\left( 
\!
\frac{\partial \chi_{\sigma_1\sigma_2}^{}}{\partial \epsilon_{d\sigma_3}^{}} 
+
2 \pi \cot \delta_{\sigma_1}\,
\chi_{\sigma_1\sigma_3}^{} \chi_{\sigma_1\sigma_2}^{} 
\! \right)  \!  .
\label{eq:Dren_to_Dsus_org}
\end{align}
Note that the derivative of the density of states with respect to 
the frequency and that with respect to the impurity level  can also be written as 
\begin{align}
\! 
\rho_{d\sigma}'  \equiv  \left.
\frac{\partial \rho_{d\sigma}^{}(\omega)}{\partial \omega} \right|_{\omega=0}^{} 
\!\!
=   -  
\frac{\partial \rho_{d\sigma}^{}}{\partial \epsilon_{d\sigma}^{}} 
 =   
2 \pi  \cot \delta_{\sigma} \,\chi_{\sigma\sigma} \rho_{d\sigma}^{} 
.  \!
\label{eq:rho_d_omega_2}
\end{align}
Furthermore, the derivative of  
$\widetilde{\chi}_{\sigma\sigma'}$ 
has the permutation symmetry  for the spin indexes in a constrained way  
\begin{align}
\frac{\partial^2 \Sigma_{\mathrm{eq},\sigma}^r(0)}
{\partial \epsilon_{d\sigma_2}^{}\partial \epsilon_{d\sigma_1}^{}}
\,=& \  
\frac{\partial \widetilde{\chi}_{\sigma\sigma_1}}
{\partial \epsilon_{d\sigma_2}^{}} 
\,=\, 
\frac{\partial \widetilde{\chi}_{\sigma\sigma_2}}
{\partial \epsilon_{d\sigma_1}^{}}  \;,
\end{align}
namely,  
the spin indexes other than $\sigma$ can be exchanged.

\subsection{Ward identities at equilibrium ground state}

The Ward identity for the causal Green's function 
for  the equilibrium  ground state, at  $T=0$,  
follows  from  the local current conservation for each spin component 
$\sigma$,\cite{YamadaYosida2,Yoshimori} 
\begin{align}
\delta_{\sigma\sigma'} \frac{\partial \Sigma_{\mathrm{eq},\sigma}^{--}(\omega) }{\partial \omega} 
+ \frac{\partial \Sigma_{\mathrm{eq},\sigma}^{--}(\omega) }{\partial \epsilon_{d\sigma'}^{}} 
 = \, 
- 
\Gamma_{\sigma\sigma';\sigma'\sigma}(\omega,0;0,\omega) 
\,
\rho_{d\sigma'}^{} ,
\label{eq:YYY_T0_causal} 
\end{align}
where $\Gamma_{\sigma\sigma';\sigma'\sigma}(\omega,\omega';\omega',\omega)$ 
is the vertex function for the causal Green's function
in the $T=0$ formalism.
The Ward identity describes a relation between 
the vertex function and the differential coefficients of the self-energy.

\subsubsection{Leading Fermi-liquid corrections}

At the Fermi energy  $\omega=0$,   
 the Ward identity represents the Fermi-liquid relation 
of Yamada-Yosida,\cite{YamadaYosida2,Yoshimori}  
i.e.,  the anti-parallel $\sigma'=-\sigma$  
and the parallel $\sigma'=\sigma$ spin components 
of Eq.\ \eqref{eq:YYY_T0_causal}  can be written as
\begin{align}
\chi_{\uparrow\downarrow} =
 -
 \rho_{d\uparrow}^{}  \rho_{d\downarrow}^{}
 \,\Gamma_{\uparrow\downarrow;\downarrow\uparrow}(0, 0; 0, 0) 
, \quad \ \ 
\frac{1}{z_{\sigma}^{}}
 =
\widetilde{\chi}_{\sigma\sigma} .
\label{eq:YY2_results}
\end{align}
Note that  
$\Gamma_{\uparrow\downarrow;\downarrow\uparrow}(0 , 0; 0 , 0)=
\Gamma_{\downarrow\uparrow;\uparrow\downarrow}(0 , 0; 0 , 0)$, 
and  $\Gamma_{\sigma\sigma;\sigma\sigma}(0 , 0; 0 , 0)=0$. 
These parameters also determine low-energy properties of quasi-particles.
The residual interaction  $\widetilde{U}$ 
and renormalized density 
of states $\widetilde{\rho}_{d\sigma}^{}$ are given by,\cite{HewsonRPT2001} 
\begin{align}
\widetilde{U} \equiv z_{\uparrow}^{}z_{\downarrow}^{}  
 \Gamma_{\uparrow\downarrow;\downarrow\uparrow}(0, 0; 0, 0) , 
\quad \ \ 
\widetilde{\rho}_{d\sigma}^{} \equiv \frac{\rho_{d\sigma}^{}}{z_{\sigma}^{} } 
=  \chi_{\sigma\sigma} 
. 
  \end{align}
In addition, the Wilson ratio  $R_{W}^{}$  
and characteristic energy scale $T^*$ which 
at zero field  corresponds  to the Kondo temperature can be defined in the form
\begin{subequations}
\begin{align}
R_W^{} \,\equiv & \   
1+\sqrt{\widetilde{\rho}_{d\uparrow}^{}\widetilde{\rho}_{d\downarrow}^{}} 
\ \widetilde{U} 
\ = \    1 - 4T^* \chi_{\uparrow\downarrow}^{}, 
\\
T^* \equiv & \ 
\frac{1}{4\sqrt{\chi_{\uparrow\uparrow}^{}  \chi_{\downarrow\downarrow}^{}}} \;.
 \label{eq:Kondo_scale_general} 
\end{align}
\end{subequations}

\subsubsection{Higher-order Fermi-liquid correction at $T=0$}

Most of  our recent results for  the higher-order Fermi-liquid 
corrections follow from an important property of    
the vertex function for the parallel spins 
$\Gamma_{\sigma\sigma;\sigma\sigma}(\omega,0;0,\omega)$, i.e.,  
its $\omega$-linear part does not have an analytic real component 
but has a pure imaginary non-analytic $|\omega|$ component, 
\begin{align}
\Gamma_{\sigma\sigma;\sigma\sigma}^{}(\omega , 0; 0, \omega) 
 \rho_{d\sigma}^{2}
= 
 i \pi  \chi_{\uparrow\downarrow}^2
\,  \omega \,\,  \mathrm{sgn} (\omega) 
+  O(\omega^2) . 
\label{eq:GammaUU_general}
\end{align}
From this property of the vertex correction, 
the order $\omega^2$ term of the self-energy can also be deduced,  
taking a derivative of  Eq.\ \eqref{eq:YYY_T0_causal} with respect to $\omega$,  
\begin{align}
\left.
\frac{\partial^2 \Sigma_{\mathrm{eq},\sigma}^{--}(\omega)
}{\partial \omega^2}\, 
\right|_{\omega \to 0}^{} 
= \ 
\frac{\partial \widetilde{\chi}_{\sigma\sigma}}{\partial \epsilon_{d\sigma}^{}}
\,  
- i  \pi  \,\frac{\chi_{\uparrow\downarrow}^2}{\rho_{d\sigma}^{}}
\ \mathrm{sgn} (\omega) \;.
\label{eq:self_w2}
\end{align}
Furthermore, the vertex function for the anti-parallel spins 
can be calculated up to the  $\omega^2$  contributions, 
using the Ward identity Eq.\ \eqref{eq:YYY_T0_causal} again,  
\begin{align}
& \Gamma_{\sigma -\sigma;-\sigma \sigma}^{}
(\omega, 0; 0 , \omega) 
\,\rho_{d\sigma}^{}\rho_{d,-\sigma}^{}
\ \,  = \,   
-\chi_{\uparrow\downarrow} \,+\, 
\rho_{d\sigma}^{}\frac{\partial \widetilde{\chi}_{\sigma,-\sigma}}
{\partial \epsilon_{d\sigma}^{}} \  \omega 
\nonumber \\
& \qquad  - \frac{\rho_{d\sigma}^{}}{2} \, 
\frac{\partial}{\partial \epsilon_{d,-\sigma}^{}} 
\! \left[
\frac{\partial \widetilde{\chi}_{\sigma\sigma}}{\partial \epsilon_{d\sigma}^{}} 
 - i \pi  \frac{\chi_{\uparrow\downarrow}^2}{\rho_{d\sigma}^{}} 
 \, \mathrm{sgn} (\omega) 
\right]  \omega^2
  +  O(\omega^3) .
\label{eq:GammaUD_general}
\end{align}

\subsection{Asymptotic form of $\Gamma_{\sigma\sigma';\sigma'\sigma}
(\omega, \omega'; \omega',\omega )$ and   $T^2$ corrections}
\label{sec:T2_ward}

We have shown in  {\it paper II} 
that the low-frequency behavior of   
the vertex corrections with two independent frequencies
$\Gamma_{\sigma\sigma';\sigma'\sigma}
(i\omega, i\omega'; i\omega',i\omega )$ 
can also be described by the Fermi-liquid theory 
 up to the linear terms in $i\omega$ and $i\omega'$. 
The results which were described using the Matsubara formalism    
can be converted into the real-frequency expressions 
in terms of the  $T=0$ causal Green's functions:   
\begin{align}
 & 
\Gamma_{\sigma\sigma;\sigma\sigma}(\omega , \omega'; \omega', \omega) 
\,\rho_{d\sigma}^{2}
\ =  \ 
 i \pi \chi_{\uparrow\downarrow}^2
\,\bigl|\omega-\omega' \bigr| 
+ \cdots ,
\label{eq:GammaUU_general_omega_dash}
\\
& \Gamma_{\sigma, -\sigma;-\sigma, \sigma}(\omega, \omega'; \omega' ,\omega) 
\,\rho_{d\sigma}^{}\rho_{d,-\sigma}^{}
 \nonumber \\
& \qquad  = \,  
-\chi_{\uparrow\downarrow} + 
\rho_{d\sigma}^{}
\frac{\partial \widetilde{\chi}_{\sigma,-\sigma}}
{\partial \epsilon_{d\sigma}^{}} \, \omega  
+ 
\rho_{d,-\sigma}^{}
\frac{\partial \widetilde{\chi}_{-\sigma,\sigma}}
{\partial \epsilon_{d,-\sigma}^{}} \, \omega'   
 \nonumber \\
& \qquad \ \ 
+ 
i \pi \,\chi_{\uparrow\downarrow}^2
\Bigl(
\,\bigl|  \omega - \omega'\bigr| 
-
\,\bigl| \omega + \omega' \bigr| 
\Bigr)
+ \cdots
.
\label{eq:GammaUD_general_omega_dash}
\end{align}
This asymptotically exact result 
captures the essential features of the Fermi liquid, 
and is analogous to Landau's quasi-particle interaction $f(\bf{p}\, \sigma,\bf{p}' \sigma')$ 
and Nozi\`{e}res' function $\phi_{\sigma\sigma'}(\varepsilon,\varepsilon')$. 
\cite{AGD,NozieresFermiLiquid}  One important difference is that 
the vertex function also has  the non-analytic imaginary part which 
directly determines  the damping of the quasi-particles.

We have also reexamined  the finite-temperature corrections 
in  {\it paper II}.  We  have obtain a simplified formula, 
with which the leading  $T^2$ contribution of the retarded self-energy 
$\Sigma_{\mathrm{eq},\sigma}^r(\omega, T)$ 
can be deduced from the derivative 
of  $\Gamma_{\sigma\sigma';\sigma'\sigma}
(\omega, \omega'; \omega',\omega )$
with respect to the intermediate frequency  $\omega'$:  
\begin{align}
\Sigma_{\mathrm{eq},\sigma}^r(\omega, T) 
-
\Sigma_{\mathrm{eq},\sigma}^r(\omega, 0) 
\, = & 
\ \frac{(\pi  T)^2}{6}\, 
\Psi_{\sigma}^{r}(\omega)   +  O(T^4)\;.
\label{eq:Psi_result_T2}
\end{align}
 Here, $\Psi_{\sigma}^{r}(\omega)$ is a retarded function,  
the corresponding causal function of which is given  by
\begin{align}
\Psi_{\sigma}^{--}(\omega) 
\,\equiv & \ 
 \lim_{\omega' \to 0}
\frac{\partial}{\partial \omega'} \, 
 \sum_{\sigma'}\,
\Gamma_{\sigma \sigma';\sigma' \sigma}(\omega , \omega'; \omega' , \omega) 
\rho_{d\sigma'}^{}(\omega') \;.
\label{eq:Psi_T0}
 \end{align}
Equation \eqref{eq:Psi_result_T2}  shows that this 
function determines the $T^2$ corrections as 
 $\Psi_{\sigma}^{r}(\omega) = \Psi_{\sigma}^{--}(\omega)$ for $\omega >0$, 
and  $\Psi_{\sigma}^{r}(\omega) = \left\{\Psi_{\sigma}^{--}(\omega)\right\}^*$ 
for $\omega <0$. 
The zero-frequency limit can be calculated,  
substituting the double-frequency expansion of the vertex functions  
 Eqs.\ \eqref{eq:GammaUU_general_omega_dash} and 
\eqref{eq:GammaUD_general_omega_dash}
 into Eq.\ \eqref{eq:Psi_T0}: 
 \begin{align}
 \lim_{\omega \to 0} 
\Psi_{\sigma}^{--}(\omega)  
\, =&  \ 
 \frac{1}{\rho_{d\sigma}^{}} 
\frac{\partial \chi_{\uparrow\downarrow}}{\partial \epsilon_{d,-\sigma}} 
\, - i \,3\,\pi \, \frac{\chi_{\uparrow\downarrow}^2}{\rho_{d\sigma}^{}}
\, \mbox{sgn}(\omega) 
\;.
\label{eq:Psi_result_+}
\end{align}
In Appendix  \ref{sec:D2_Psi_causal_detail},  
we provide an alternative derivation 
which is also applicable to the multi-orbital case.  
We will discuss in the next section  an exact relation 
  between the $T^2$  and  the $(eV)^2$ contributions 
which was first pointed out by FMvMD.\cite{FilipponeMocaVonDelftMora} 
In our formulation, it follows from an identity  
$\Psi_{\sigma}^{--}(\omega) \equiv  
\widehat{D}^2  \Sigma_{\mathrm{eq},\sigma}^{--}(\omega)$, 
given  in Eq.\ \eqref{eq:d2_T0}.

\section{Non-equilibrium Fermi-liquid relations at finite magnetic fields}
\label{sec:nonlinear_self_energy_at_h}

The higher-order Fermi-liquid corrections, 
 summarized in the previous section for thermal equilibrium,  
are described in terms of the differential coefficients which are taken  
with respect to the spin-dependent impurity level  $\epsilon_{d\sigma}^{}$.
The non-equilibrium Ward identities were previously obtained 
for the spin SU(2) symmetric case,  and were used to calculate  
non-linear conductance  through a quantum dot at low bias voltages.\cite{ao2001PRB} 
In the formulation, the impurity-level derivative  
$
{\partial}/{\partial \epsilon_d^{}} 
  =   
 {\partial}/{\partial \epsilon_{d\uparrow}^{}} +  
 {\partial}/{\partial \epsilon_{d\downarrow}^{}} $,  
which does not  distinguish the two spin components, were taken. 
In this section, we describe how the previous formulation 
can be extended at finite magnetic fields.   Using the extended identities,  
we  calculate the Fermi-liquid corrections 
to  magneto-conductance through a quantum dot, 
and also provide transport coefficients for  the thermoelectric transport  
of  dilute magnetic alloys.

We use  the Keldysh Green's function\cite{Keldysh}  for  impurity  electrons,
\begin{subequations}
\begin{align}
G^{--}_{\sigma} (t_1, t_2) \, \equiv&  
\, -i\, \langle \,\mbox{T}\,
d^{\phantom{\dagger}}_{\sigma}(t_1)\, d^{\dagger}_{\sigma}(t_2)
\,\rangle
\,,
\\
G^{++}_{\sigma} (t_1, t_2) \, \equiv&  
\,-i\, 
\langle \,\widetilde{\mbox{T}}\,
d^{\phantom{\dagger}}_{\sigma}(t_1)\, d^{\dagger}_{\sigma}(t_2)
\,\rangle
\,,
 \\
G^{+-}_{\sigma} (t_1, t_2) \, \equiv& 
\,-i\, 
\langle 
d^{\phantom{\dagger}}_{\sigma}(t_1)\, d^{\dagger}_{\sigma}(t_2)
\,\rangle 
\;,
\\
G^{-+}_{\sigma} (t_1, t_2) \equiv&  
\ \  i\, 
\langle 
d^{\dagger}_{\sigma}(t_2) \, d^{\phantom{\dagger}}_{\sigma}(t_1) 
\,\rangle 
\label{eq:G<}
\,.
\end{align}
\end{subequations}
Non-equilibrium steady state driven by the  bias voltage $eV$ can be  
described using the noninteracting Green's function,   
the Fourier transform of which is given by\cite{Caroli} 
\begin{subequations}
\begin{align}
&
\!\!\!\!\!\!
G_{0\sigma}^{--}(\omega) = 
\bigl[1-f_\mathrm{eff}(\omega)\bigr] G_{0\sigma}^{r}(\omega) 
+f_\mathrm{eff}(\omega)
        \,G_{0\sigma}^{a}(\omega) ,
\label{eq:G0_00^--}
\\ 
&
\!\!\!\!\!\!
G_{0\sigma}^{++}(\omega) = 
-f_\mathrm{eff}(\omega)
\,G_{0\sigma}^{r}(\omega) 
-\bigl[1-f_\mathrm{eff}(\omega)\bigr]
        G_{0\sigma}^{a}(\omega) ,  \!
 \label{eq:G0_00^++}
\\ 
&
\!\!\!\!\!\!
G_{0\sigma}^{-+}(\omega) =  
-f_\mathrm{eff}(\omega)
        \bigl[G_{0\sigma}^{r}(\omega)- G_{0\sigma}^{a}(\omega)\bigr] , 
\label{eq:G0_00^-+}
\\ 
&
\!\!\!\!\!\!
G_{0\sigma}^{+-}(\omega) = 
\bigl[1-f_\mathrm{eff}(\omega)\bigr]
        \bigl[G_{0\sigma}^{r}(\omega)- G_{0\sigma}^{a}(\omega)\bigr] .
\label{eq:G0_00^+-}  
\end{align}
\end{subequations}
The retarded and the advanced Green's functions are  
written  explicitly in the following  form
\begin{align}
 G_{0\sigma}^{r}(\omega) \,=\,
\frac{1}{\omega-\epsilon_{d\sigma}^{}  +i (\Gamma_L+\Gamma_R)} \;,
\end{align}
and 
 $G_{0\sigma}^{a}(\omega) = \left\{G_{0\sigma}^{r}(\omega)\right\}^*$. 
Similarly, the Fourier transform of 
the causal Green's function $G_{0\sigma}^{--}$ and 
its time-reversal counter part $G_{0\sigma}^{++}$ are related to each other through 
  $G_{0\sigma}^{++}(\omega) = -\left\{G_{0\sigma}^{--}(\omega)\right\}^*$.  
One of the most important  properties of these Green functions is  that 
both the bias voltage $eV$ and temperature $T$ enter through 
a local distribution function for impurity electrons,
\begin{align}
f_\mathrm{eff}(\omega) \,=\,
\frac{  f_L(\omega) \,\Gamma_L 
  + f_R(\omega) \,\Gamma_R}{ 
 \Gamma_L +\Gamma_R } \;. 
\label{eq:f_0}
\end{align}
Here,  $f_{L/R}(\omega) \equiv  f(\omega-\mu_{L/R}^{})$ 
and  $f(\omega)=[e^{\omega/T+1}]^{-1}$  is the Fermi function. 
We choose the chemical potentials such that 
$\mu_L= \alpha_L eV$, 
$\mu_R= -\alpha_R eV$, and $\alpha_L + \alpha_R =1$.  
The parameters $\alpha_L$ and $\alpha_R$ specify how the bias is applied 
relative to the Fermi level at equilibrium  $\omega=0$.

\subsection{Ward identities for the  Keldysh  Green's functions at finite magnetic fields}
\label{subsec:Ward_eV_mag}

The corresponding self-energy satisfies  the Dyson equation 
of a matrix form,  $\bm{G}_{\sigma}^{-1}  = 
\bm{G}_{0\sigma}^{-1} - \bm{\Sigma}_{\sigma}^{} $,    
\begin{align}
\bm{G}_{\sigma}^{}  =  \left[ 
\begin{matrix}
G_{\sigma}^{--} & G_{\sigma}^{-+}   \cr
           G_{\sigma}^{+-} & G_{\sigma}^{++}  \cr  
\end{matrix}
                             \right] 
, \qquad 
\bm{\Sigma}_{\sigma}^{}  =  \left[ 
\begin{matrix}
\Sigma_{\sigma}^{--} & \Sigma_{\sigma}^{-+}   \cr
           \Sigma_{\sigma}^{+-} & \Sigma_{\sigma}^{++}  \cr  
\end{matrix}
                             \right] .
\end{align}
In the Keldysh formalism, 
the dependence of $\bm{\Sigma}_{\sigma}^{}$  on 
the bias voltage and temperature enters  through  
the internal $\bm{G}_{0\sigma}$'s, each of which accompanies 
the non-equilibrium distribution  $f_\mathrm{eff}(\omega)$. 
Therefore, 
 the first few differential coefficients  of this function 
play a central role in low-energy properties,  
\begin{subequations}
\begin{align}
\frac{\partial  f_\mathrm{eff}(\omega)}{\partial (eV)}  
\,=&\  
\frac{\alpha_R\Gamma_R\,f_R'(\omega) - \alpha_L\Gamma_L\,f_L'(\omega)
}{\Gamma_R + \Gamma_L} \;,
\\
\frac{\partial^2  f_\mathrm{eff}(\omega)}{\partial (eV)^2}  
\,=& \   
\frac{\alpha_R^2\Gamma_R\,f_R''(\omega) + \alpha_L^2\Gamma_L\,f_L''(\omega)
}{\Gamma_R + \Gamma_L}
\;. 
\label{eq:dif_feff}
\end{align} 
\end{subequations}
Note that the low-energy limit of these  two derivatives  do  not depend on 
the order to take the  limits   $eV \to 0$ and $\omega \to 0$, 
\begin{subequations}
\begin{align}
& \lim_{\omega\to 0 \atop eV\to 0}
\frac{\partial  f_\mathrm{eff}(\varepsilon + \omega)}{\partial (eV)}  
\,=\,   
-\alpha \, \frac{\partial f(\varepsilon)}{\partial \varepsilon}\;,
\\
& \lim_{\omega \to 0 \atop eV\to 0}
\frac{\partial^2  f_\mathrm{eff}(\varepsilon + \omega)}{\partial (eV)^2}  
\,=\,    
\kappa \,  \frac{\partial^2 f(\varepsilon)}{\partial \varepsilon^2}
\;,
\end{align}
\end{subequations}
Here,  $\varepsilon$ is  an arbitrary frequency argument,  
which for our purpose can be regarded as an internal frequency of a Feynman diagram.
The coefficients are defined by 
\begin{align}
&\alpha  \equiv   
\frac{\alpha_L \Gamma_L - \alpha_R \Gamma_R}{\Gamma_L+ \Gamma_R}
, \quad  
\kappa \equiv 
   \frac{\alpha_L^2 \Gamma_L + \alpha_R^2 \Gamma_R}
{\Gamma_L+ \Gamma_R} ,
\end{align}
and thus 
$\kappa -\alpha^2 = 
{\Gamma_L\,\Gamma_R}/{(\Gamma_L+ \Gamma_R)^2}$.

The differential coefficients of   $\Sigma_{\sigma}^{\nu\nu'}(\omega)$ 
 with respect to $eV$ can be calculated by taking derivatives of 
the  internal Green's functions  in the Feynman diagrams for the self-energy.  
To be specific,  we  assign  the internal frequencies  $\varepsilon$'s  
in a way  such that every internal  propagator carries 
the external frequency $\omega$. 
Then  the $eV$ derivative of the noninteracting Green's function   
 can be rewritten as a linear combination of 
the $\omega$ derivative and 
the $\epsilon_{d}^{}$ derivative which  includes both spin components,  
$
{\partial}/{\partial \epsilon_d^{}} 
  =   
 {\partial}/{\partial \epsilon_{d\uparrow}^{}} +  
 {\partial}/{\partial \epsilon_{d\downarrow}^{}} $,  
\begin{subequations}
\begin{align}
 \left.
  \frac{\partial  
 G_{0\sigma}^{\nu\nu'}(\varepsilon + \omega)}{
 \partial (eV)} 
 \right|_{eV=0}
 =& \ 
  - \alpha 
 \left(
 \frac{\partial}{\partial \omega}
 + \frac{\partial}{\partial \epsilon_{d}^{}}
 \right)
 G_{0:\text{eq},\sigma}^{\nu\nu'}(\varepsilon + \omega),
\label{eq:derivative_1b} 
\\
\left.
\frac{\partial^2 G_{0,\sigma}^{\nu \nu' }(\varepsilon + \omega)}{
\partial ({\sl e}V)^{2}}\right|_{eV=0}
 =& \ 
\kappa 
\left(
\frac{\partial}{\partial \omega}
+ \frac{\partial}{\partial \epsilon_{d}^{}}
\right)^{2}
G_{0:\text{eq},\sigma}^{\nu \nu'}(\varepsilon + \omega) .  
\label{eq:derivative_2b}
\end{align}
\end{subequations}
Here,  the label \lq\lq eq" represents the  \lq\lq equilibrium" limit, 
 $ G_{0:\text{eq},\sigma}^{\nu\nu'}
 \equiv 
 \bigl.G_{0\sigma}^{\nu\nu'}\bigr|_{eV=0}$. 
The right-hand side of 
Eqs.\ \eqref{eq:derivative_1b} and \eqref{eq:derivative_2b}
have been expressed in terms of  the equilibrium Green's functions,  
which can  be calculated  further, as 
 \begin{subequations}
\begin{align}
 & 
\left(
\frac{\partial}{\partial \omega}
+ \frac{\partial}{\partial \epsilon_{d}^{}}
\right)
G_{0:\text{eq},\sigma}^{\nu \nu'}(\varepsilon + \omega) 
\nonumber \\
 =&  \ 
 - \frac{\partial f(\varepsilon + \omega) }{\partial \omega}
 \bigl[
G_{0:\text{eq},\sigma}^{r}(\varepsilon + \omega) -
G_{0:\text{eq},\sigma}^{a}(\varepsilon + \omega) 
\bigr] , 
\label{eq:derivative_1a}
\\
& 
\left(
\frac{\partial}{\partial \omega}
+ \frac{\partial}{\partial \epsilon_{d}^{}}
\right)^{2}
G_{0:\text{eq},\sigma}^{\nu \nu'}(\varepsilon + \omega) 
\nonumber \\
=&  \ -
 \frac{\partial^2 f(\varepsilon + \omega)}{\partial \omega^2}
 \bigl[
G_{0:\text{eq},\sigma}^{r}(\varepsilon + \omega) -
G_{0:\text{eq},\sigma}^{a}(\varepsilon + \omega) 
\bigr].
\label{eq:derivative_2a}
 \end{align}
\end{subequations}
These relations between the derivatives of 
 the noninteracting Green's functions at  finite magnetic fields   
keep the same form as those at  $h=0$.\cite{ao2001PRB}
Nevertheless,  
it is necessary  for taking  a variational derivative with respect 
to the internal Green's functions to keep track of the spin index $\sigma$.

The first two differential coefficients  of $\bm{\Sigma}_\sigma(\omega)$ 
with respect to $eV$  can be expressed in the following form,  
using  Eqs.\ \eqref{eq:derivative_1b} and \eqref{eq:derivative_2b} 
for the derivatives of internal lines  in  the self-energy diagrams
\begin{subequations}
\begin{align}
\left.
\frac{\partial \bm{\Sigma}_\sigma(\omega) }{\partial (eV)}\right|_{eV=0} 
 &= \ 
 -\,
 \alpha  
\left(
\frac{\partial}{\partial \omega}
+ \frac{\partial}{\partial \epsilon_{d}^{}}
\right)
\bm{\Sigma}_{\text{eq},\sigma}(\omega) 
,
\label{eq:derivative_self_1}
\\
\left.
\frac{\partial^2  
\bm{\Sigma}_\sigma(\omega)}{
\partial (eV)^2}\right|_{eV=0} 
& \, = 
\  \alpha^2
\left(
\frac{\partial}{\partial \omega}
 + \frac{\partial}{\partial \epsilon_{d}^{}}
\right)^2 
\bm{\Sigma}_{\text{eq},\sigma}(\omega) 
 \nonumber \\
 & \quad  \ 
+   
\frac{ \Gamma_L\,\Gamma_R}{ \left( \Gamma_L+ \Gamma_R \right)^2}
   \  \widehat{D}^2 
\bm{\Sigma}_{\text{eq},\sigma}(\omega) 
.
\label{eq:derivative_self_2_with_CII}
\end{align}
\end{subequations}
Here,   $\bm{\Sigma}_{\text{eq},\sigma}(\omega) 
\equiv \left. \bm{\Sigma}_{\sigma}(\omega) \right|_{eV=0}$, 
and thus  the right-hand side of 
Eqs.\  \eqref{eq:derivative_self_1} and \eqref{eq:derivative_self_2_with_CII} 
are written in terms of the equilibrium self-energy. 
The operator $\widehat{D}^2$ takes 
the second derivative 
$(\partial /\partial \omega  + \partial/ \partial \epsilon_{d}^{} )^2$  
for  each single  internal Green's function of the  Feynman  diagrams 
for $\bm{\Sigma}_{\text{eq},\sigma}(\omega)$.\cite{ao2001PRB}

Specifically at  zero temperature,  the standard  $T=0$  diagrammatic formulation  
which only needs the causal Green's function  is applicable, 
and the right-hand side of 
Eqs.\ \eqref{eq:derivative_self_1} and \eqref{eq:derivative_self_2_with_CII} 
can be calculated further.  
Taking  the variational derivative of  
 $\Sigma_{\text{eq},\sigma}^{--}$  component with respect 
to the internal Green's functions  and then    
 using Eqs.\ \eqref{eq:derivative_1a} and \eqref{eq:derivative_2a}, 
we obtain the following two identities, 
\begin{subequations}
\begin{align}
& \left(
\frac{\partial}{\partial \omega}
+ \frac{\partial}{\partial \epsilon_{d}^{}}
\right)
  \Sigma_{\mathrm{eq},\sigma}^{--}(\omega) 
\nonumber \\
 &=   
 -\sum_{\sigma'}\int d\omega'\, 
 \Gamma_{\sigma\sigma';\sigma'\sigma}(\omega,\omega';\omega',\omega)
 \,\rho_{d\sigma'}^{}(\omega')
 \left\{-\frac{\partial f(\omega') }{\partial {\omega'}} \right\}
%
 \nonumber
 \\
&= 
- \sum_{\sigma'}  
\Gamma_{\sigma\sigma';\sigma'\sigma}(\omega,0;0,\omega)
\,\rho_{d\sigma'}^{}(0) \;,
\label{eq:d1_T0}
\\
 &\widehat{D}^2  \Sigma_{\mathrm{eq},\sigma}^{--}(\omega) 
\nonumber \\
 &=      
 -\sum_{\sigma'}\int d\omega'\, 
 \Gamma_{\sigma\sigma';\sigma'\sigma}(\omega,\omega';\omega',\omega)
 \,\rho_{d\sigma'}^{}(\omega')
 \left\{-\frac{\partial^2 f(\omega') }{\partial {\omega'}^2} \right\} 
%
 \nonumber
 \\
&=  
\sum_{\sigma'}  
\left.
\frac{\partial}{\partial \omega'}\,  
\Gamma_{\sigma\sigma';\sigma'\sigma}(\omega,\omega';\omega',\omega)
\,
\rho_{d\sigma'}^{}(\omega')\right|_{\omega'=0} .
\end{align}
\end{subequations}
The first one corresponds  to the Ward identity given 
 in Eq.\ \eqref{eq:YYY_T0_causal}.  
The second identity shows that 
 $\widehat{D}^2  \Sigma_{\mathrm{eq},\sigma}^{--}(\omega)$
is identical to  the correlation function 
 $\Psi_{\sigma}^{--}(\omega)$:  
\begin{align}
 \widehat{D}^2  \Sigma_{\mathrm{eq},\sigma}^{--}(\omega)  
\, \equiv  \,      \Psi_{\sigma}^{--}(\omega) . 
\label{eq:d2_T0}
\end{align}
Thus, the $(eV)^2$  contribution emerging  
through the second term of Eq.\ \eqref{eq:derivative_self_2_with_CII}  
and  the  $T^2$  contribution determined by  Eq.\ \eqref{eq:Psi_result_T2} 
appear in the self-energy as   a linear combination,  
\begin{align}
 \frac{\Gamma_L \Gamma_R}{\left( \Gamma_L + \Gamma_R \right)^2} 
\,\frac{(eV)^2}{2}  
+\frac{(\pi  T)^2}{6} \;.
\label{eq:single_variable_T_eV}
\end{align}
Although this was known for the imaginary part,\cite{Hershfield1,ao2001PRB}
it has not been recognized until recently that  the $T^2$ and $(eV)^2$ 
contributions of the real part of the self-energy are determined by the same processes. 
This was first pointed out by FMvDM, using  the   
Nozi\`{e}res' phenomenological description.\cite{FilipponeMocaVonDelftMora} 
Our description provides an alternative microscopic proof.

The common coefficient  for the set of the  $(eV)^2$ and $T^2$ contributions 
can be calculated taking the $\omega \to 0$  limit  for Eq.\ \eqref{eq:d2_T0},
and the result corresponding to Eq.\ \eqref{eq:Psi_result_+}  is given by    
\begin{align}
\!\!\!
 \lim_{\omega \to 0} 
 \widehat{D}^2  \Sigma_{\mathrm{eq},\sigma}^{--}(\omega) 
 = 
 \frac{1}{\rho_{d\sigma}^{}} 
\frac{\partial \chi_{\uparrow\downarrow}}{\partial \epsilon_{d,-\sigma}} 
\, - i \,3\pi  \frac{\chi_{\uparrow\downarrow}^2}{\rho_{d\sigma}^{}}
\, \mathrm{sgn}(\omega) 
.
\label{eq:Psi_result_D2}
\end{align}
See  Appendix \ref{sec:D2_Psi_causal_detail} for the details, 
where a general proof  applied to multi-orbital Anderson impurity 
with $N$  components  $\sigma=1,2,\ldots, N$ is given using 
the $T=0$ causal-Green's-function formulation. 
The non-analytic $\mathrm{sgn}(\omega)$ dependence   
in the imaginary part of Eq.\ \eqref{eq:Psi_result_D2} 
reflects the behavior caused by the branch cuts of  
the vertex function 
 $\Gamma_{\sigma\sigma';\sigma'\sigma}(\omega,\omega';\omega',\omega)$
along 
 $\omega-\omega'=0$ and  $\omega+\omega'=0$.
\cite{Eliashberg,EliashbergJETP15,Yoshimori,ao2017_2_PRB} 
This imaginary part  generalizes the previous result\cite{ao2001PRB} 
 obtained at $h=0$ to finite magnetic fields.  
It also agrees with the corresponding FMvDM's formula,\cite{FilipponeMocaVonDelftMora}
and with the second-order-renormalized-perturbation result 
 as well.\cite{HewsonOguriMeyer}

 Note that   
  Eq.\ \eqref{eq:Psi_result_D2} has been deduced  from  Eq.\ \eqref{eq:d2_T0}. 
 The antisymmetry property of the vertex function 
 imposes a strong restriction on the intermediate states, 
 i.e.,   in  the summation over $\sigma'$  in  Eq.\ \eqref{eq:d2_T0} 
 the contribution of  $\sigma' = \sigma$ component  vanishes  because of  
  $\Gamma_{\sigma\sigma;\sigma\sigma}(0,0;0,0) = 0$ and 
 $\mathrm{Re}\, \partial 
 \Gamma_{\sigma\sigma;\sigma\sigma}(0,\omega';\omega',0)
 / \partial \omega'|_{\omega'=0}=0$,  
 as shown in Appendix \ref{sec:D2_Psi_causal_detail}.
 Thus, for the  $N=2$ spin Anderson model,  the intermediate state must be unique, 
 i.e.,  the spin  $\sigma'=-\sigma$ state,   
 and it  gives  a finite contribution   $(1/\rho_{d\sigma}^{}) \,
 {\partial \chi_{\sigma\sigma'}}/{\partial \epsilon_{d\sigma'}}$.


\subsection{Additional  $eV$,  $\omega eV$, and  $(eV)^2$  contributions 
emerging  for the case of  $\alpha \neq  0$ 
}

In the situation where  $\alpha \neq 0$, 
the self-energy also captures the terms of order  $eV$,  $\omega eV$, 
and an additional  $(eV)^2$  contribution 
emerging through the first term in the right-hand 
side of Eq.\ \eqref{eq:derivative_self_2_with_CII}. 
We calculate the coefficients for these terms in the following, 
using the low-energy asymptotic form of 
$\Gamma_{\sigma \sigma';\sigma' \sigma}(\omega, 0; 0 , \omega)$,  
given in Eqs.\ \eqref{eq:GammaUU_general} and  \eqref{eq:GammaUD_general}.

The order  $eV$  contribution is determined by   
 Eqs.\ \eqref{eq:derivative_self_1} and \eqref{eq:d1_T0}. 
Using the explicit form of the vertex function given 
in Eqs.\ \eqref{eq:GammaUU_general} and  \eqref{eq:GammaUD_general}, 
we obtain 
\begin{align}
\lim_{\omega \to 0} 
\left.
\frac{\partial \Sigma^{--}_\sigma(\omega)}{\partial (eV)}\,\right|_{eV=0} 
 &= \, 
 -
 \alpha \, 
\lim_{\omega \to 0} 
\left(
\frac{\partial}{\partial \omega}
+ \frac{\partial}{\partial \epsilon_{d}^{}}
\right) \Sigma^{--}_{\text{eq},_\sigma}(\omega) 
\nonumber \\
  &= \,   
 \alpha 
\sum_{\sigma'} 
\Gamma_{\sigma \sigma';\sigma' \sigma}(0, 0; 0 , 0) 
\,\rho_{d\sigma'}^{}  
\nonumber \\
& = \, 
- \alpha\, \widetilde{\chi}_{\sigma,-\sigma}^{}
 \, .
\label{eq:self_eV_linear}
\end{align}
The order $\omega\, eV$ contribution can also be deduced from 
 Eqs.\ \eqref{eq:derivative_self_1} and \eqref{eq:d1_T0},  
using Eqs.\ \eqref{eq:GammaUU_general} and  \eqref{eq:GammaUD_general}, 
\begin{align}
&
\!\!\!\!
\lim_{\omega \to 0} 
\frac{\partial}{\partial \omega} \left[
\frac{\partial \Sigma^{--}_\sigma(\omega)}{\partial (eV)}\right]_{eV=0} 
\nonumber \\
 &
=  
 -  \alpha 
\lim_{\omega \to 0} 
\frac{\partial}{\partial \omega}
\left(
\frac{\partial}{\partial \omega}
+ \frac{\partial}{\partial \epsilon_{d}^{}}
\right) \Sigma^{--}_{\text{eq},_\sigma}(\omega) 
\nonumber \\
 &= 
 \alpha  
\lim_{\omega \to 0} 
\frac{\partial}{\partial \omega}
\sum_{\sigma'} 
\Gamma_{\sigma \sigma';\sigma' \sigma}(\omega, 0; 0 , \omega) 
\,\rho_{d\sigma'}^{}  
 \nonumber  \\
&=  
\alpha 
\left[ \,
\frac{\partial \widetilde{\chi}_{\sigma,-\sigma}}{\partial \epsilon_{d\sigma}^{}} 
+i\pi \, \frac{\chi_{\uparrow\downarrow}^2}{\rho_{d\sigma}^{}} 
\mathrm{sgn}(\omega) 
\,\right] .
\label{eq:self_w_eV}
\end{align}

The additional  $(eV)^2$ contribution, 
which enters through the  $\alpha^2$ term 
 in  Eq.\ \eqref{eq:derivative_self_2_with_CII},
 can be deduced from Eqs.\ \eqref{eq:d1_T0}   
using Eqs.\ \eqref{eq:GammaUU_general} and  \eqref{eq:GammaUD_general}: 
\begin{align}
&
\alpha^2
\lim_{\omega \to 0} 
\left(
\frac{\partial}{\partial \omega}
 + \frac{\partial}{\partial \epsilon_{d}^{}}
\right)^2 
\Sigma^{--}_{\text{eq},_\sigma}(\omega) 
\,
\nonumber \\
&=  \ 
- \alpha^2
\lim_{\omega \to 0}
\left(
\frac{\partial}{\partial \omega}
 + \frac{\partial}{\partial \epsilon_{d}^{}}
\right) 
\sum_{\sigma'}
\Gamma_{\sigma \sigma';\sigma' \sigma}(\omega , 0; 0 , \omega) 
\,\rho_{d\sigma'}^{} 
\nonumber 
\\
&= \ 
\alpha^2 \left[\,
\frac{\partial\widetilde{\chi}_{\sigma,-\sigma}}{\partial\epsilon_{d,-\sigma}^{}}
-  
i\pi \, \frac{\chi_{\uparrow\downarrow}^2}{\rho_{d\sigma}^{}} \, 
\mathrm{sgn}(\omega)  \, \right] .
\label{eq:self_w2_alpha2}
\end{align}

\section{Low-energy asymptotic form of self-energy}
\label{sec:result_self_energy}

\begin{figure}[t]
 \leavevmode

\begin{minipage}{1\linewidth}
\includegraphics[width=0.6\linewidth]{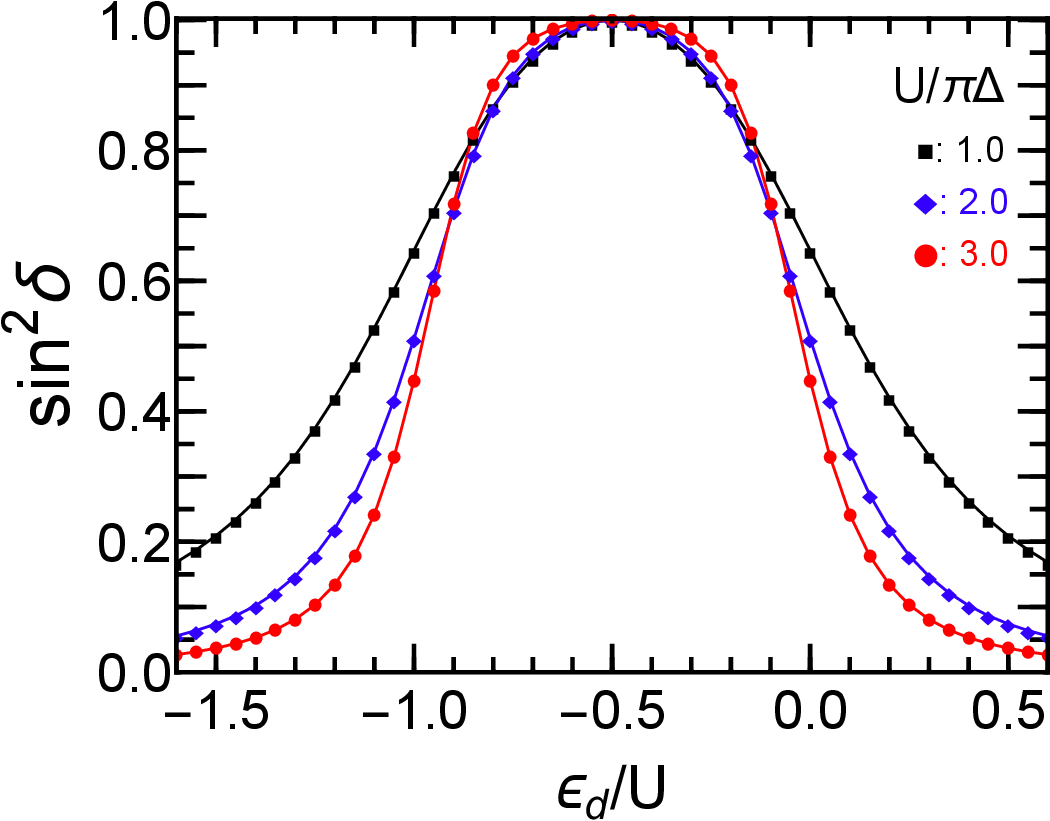}
\\
\includegraphics[width=0.6\linewidth]{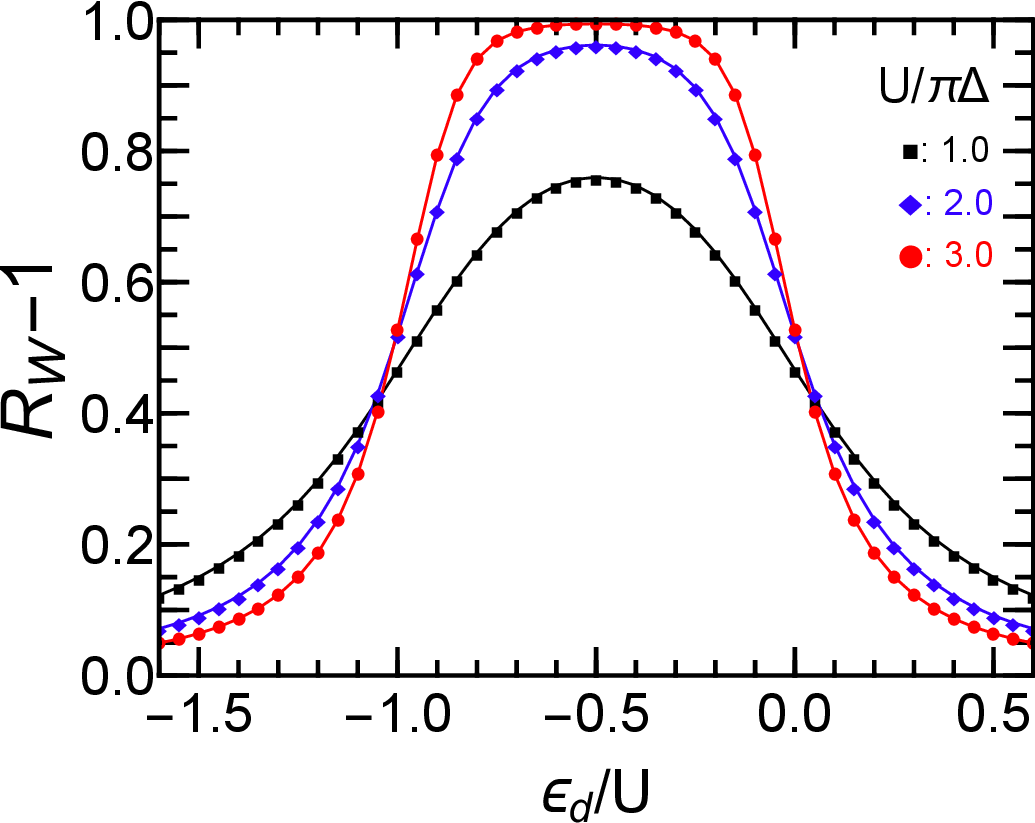}
\\
\includegraphics[width=0.6\linewidth]{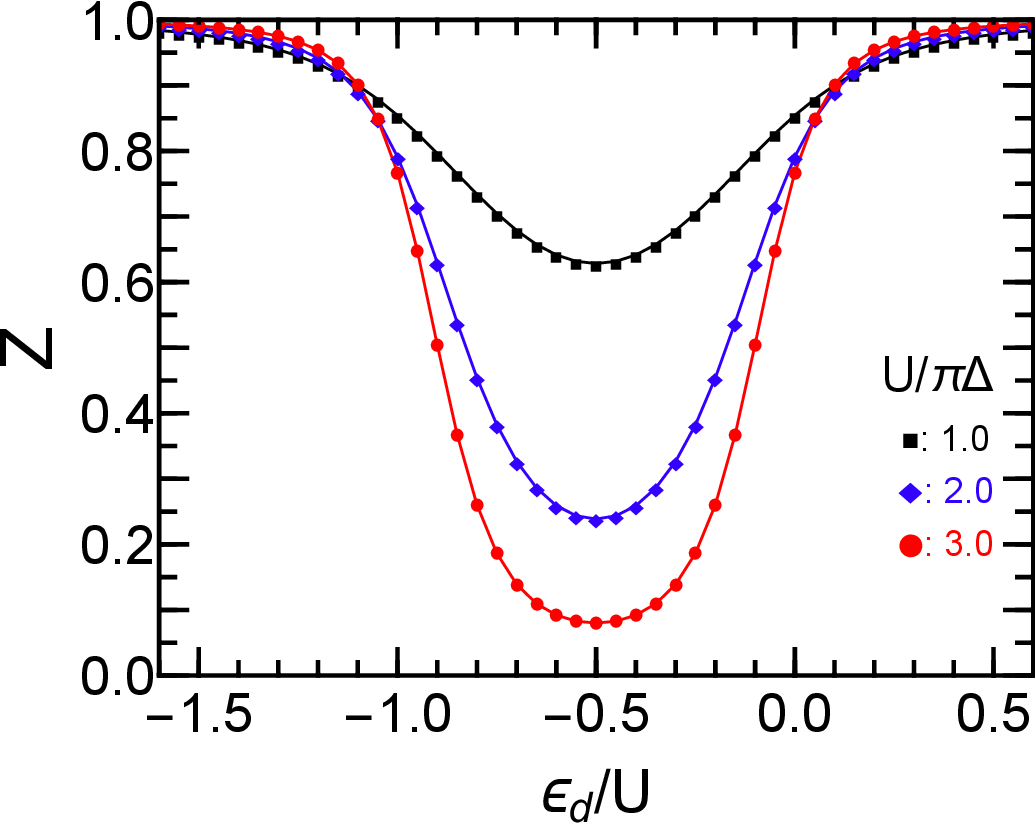}
\end{minipage}
 \caption{
(Color online) 
NRG results of $\sin^2 \delta$  ($= \pi \Delta \rho_{d}^{})$, 
$R_{W}-1$, and $z$ at zero magnetic field  $h=0$ 
are plotted vs  $\epsilon_{d}^{}/U$ for $U/(\pi \Delta) = 1.0,  2.0,$, and $3.0$.
}
 \label{fig:FL_parameter_1}
\end{figure}

\begin{figure}[t]
\begin{minipage}{1\linewidth}
\includegraphics[width=0.6\linewidth]{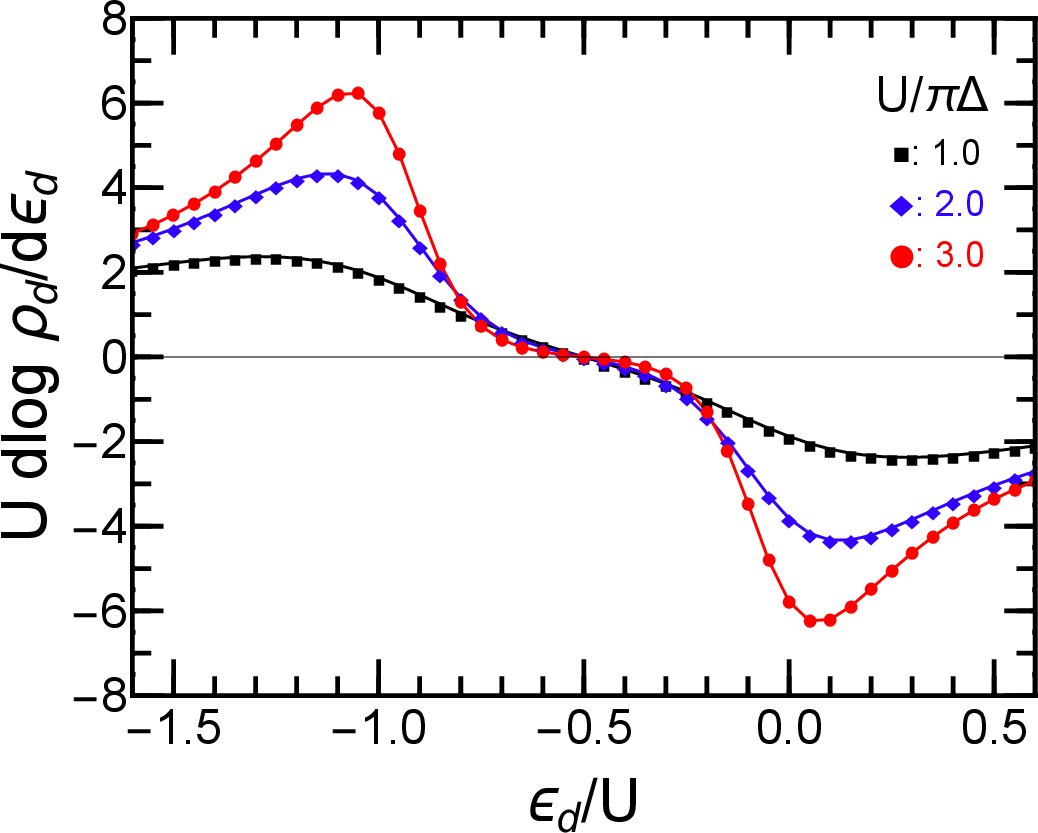}
\\
\includegraphics[width=0.6\linewidth]{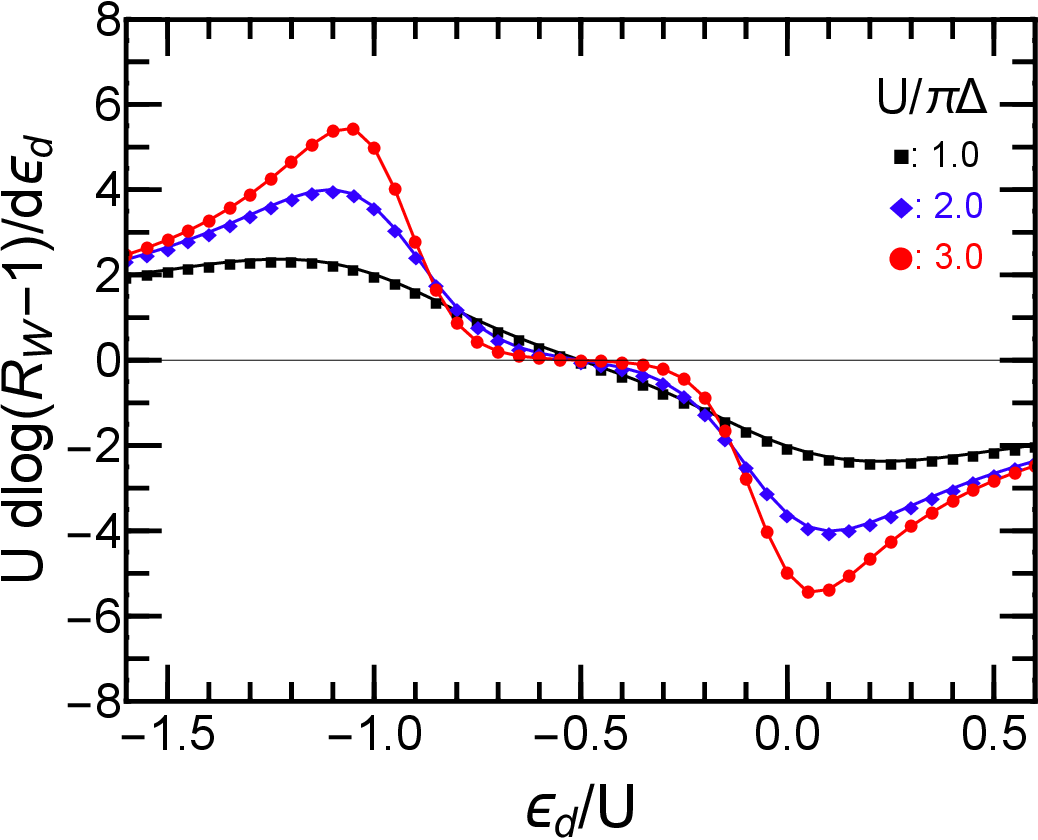}
\\
\includegraphics[width=0.6\linewidth]{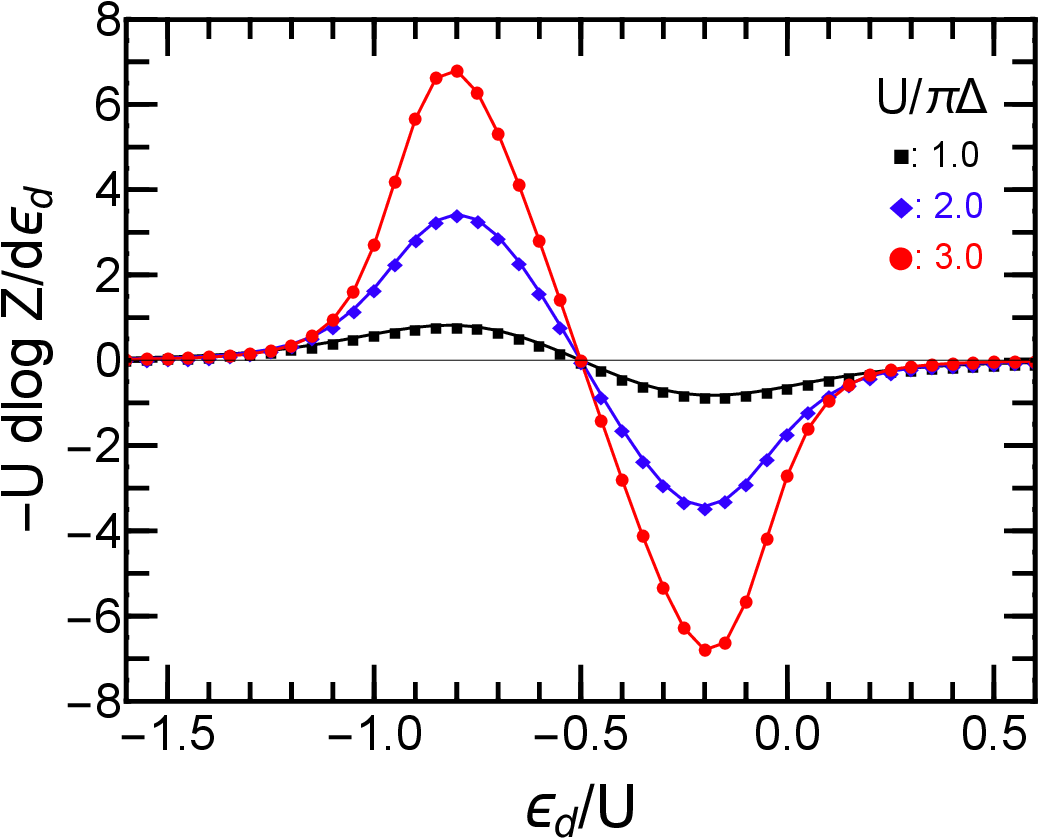}
\end{minipage}
 \caption{
(Color online) 
Logarithmic derivatives  of $\rho_{d}^{}$ ($=\sin^2 \delta/\pi\Delta $), 
$R_{W}-1$, and $z$ with respect to  $\epsilon_{d}^{}$ at zero magnetic field  $h=0$ 
are  plotted   vs  $\epsilon_{d}^{}/U$  for $U/(\pi \Delta) = 1.0,  2.0, $, and $3.0$.
}
 \label{fig:FL_parameter_2}
\end{figure}

The low-energy behavior of  the retarded self-energy 
for finite magnetic field  $\Sigma_{\sigma}^r(\omega,T,eV)$ 
can be deduced exactly  up to terms of order $\omega^2$, $T^2$  and  $(eV)^2$,  
from the results given  in  Eqs.\ \eqref{eq:self_w2} and  
\eqref{eq:Psi_result_T2}--\eqref{eq:Psi_result_+}  for equilibrium,   
and Eqs.\ \eqref{eq:d2_T0}  and \eqref{eq:self_eV_linear}--\eqref{eq:self_w2_alpha2} 
for finite bias voltages.  
\begin{widetext}
The imaginary part can be expressed in the form
\begin{align}
\mathrm{Im}\, \Sigma_{\sigma}^r(\omega, T, eV) 
\,  = & \ -\,   \frac{\pi}{2}\,   
\frac{\chi_{\uparrow\downarrow}^2}{\rho_{d\sigma}^{}}
\,
   \left[\,
    \left(\,\omega -   \alpha\, eV\,  \right)^2 
 + \frac{ 3\,\Gamma_L \Gamma_R}{\left( \Gamma_L + \Gamma_R \right)^2} \,(eV)^2 
 +(\pi T)^2  
    \,\right] \ + \, \cdots .
\label{eq:self_imaginary}
\end{align}
The spin dependence enters  through  
the density of states $\rho_{d\sigma}^{}$ in the prefactor.

 Owing to the recent knowledge about the double derivative 
$\mathrm{Re}\,\partial^2 \Sigma_{\mathrm{eq},\sigma}^{--}/\partial \omega^2$ 
 described in Eq.\ \eqref{eq:self_w2}, 
the real part of the self-energy can also be expressed 
in terms of the susceptibilities, or renormalized parameters for the quasi-particles,  
\begin{align}
\epsilon_{d\sigma}+ 
\mathrm{Re}\, \Sigma_{\sigma}^r(\omega,T,eV) 
\,  = & \ \ 
\Delta\, \cot \delta_{\sigma}
\,+ \bigl( 1-\widetilde{\chi}_{\sigma\sigma} \bigr)\, \omega 
\, + \frac{1}{2}\,\frac{\partial \widetilde{\chi}_{\sigma\sigma}}
{\partial \epsilon_{d\sigma}^{}}\, \omega^2 
 \, +  \frac{1}{6}\,
\frac{1}{\rho_{d\sigma}^{}} 
\frac{\partial \chi_{\uparrow\downarrow}}{\partial \epsilon_{d,-\sigma}} 
\left[
\frac{3\Gamma_L \Gamma_R}{\left( \Gamma_L + \Gamma_R \right)^2} 
 \,(eV)^2 
+
 \left( \pi T\right)^2 
\right] 
\nonumber \\ 
 & 
\ -\widetilde{\chi}_{\sigma,-\sigma}^{} \,  \alpha\,eV 
\,+ 
 \frac{\partial \widetilde{\chi}_{\sigma,-\sigma}}{\partial \epsilon_{d\sigma}^{}} \,\alpha\,eV \, \omega
\, +  \frac{1}{2}\,
\frac{\partial\widetilde{\chi}_{\sigma,-\sigma}}{\partial\epsilon_{d,-\sigma}}
\,\alpha^2 (eV)^2 
\ + \,\cdots
\label{eq:self_real_ev_mag}
\;.
\end{align}


At zero magnetic field $h=0$, 
the real part can be rewritten in the following form,  
using Eqs.\  \eqref{eq:eV2_real_part_mag_again_move}--\eqref{eq:Dren_to_Dsus_org}; 
\begin{align}
&\epsilon_{d}^{}+ 
\mathrm{Re}\, \Sigma_{}^r(\omega,T,eV) 
\,   \xrightarrow{\,h\to 0\,}   \, 
\nonumber 
\\
& 
\ \quad  
\Delta\, \cot \delta
\,+\, 
\bigl( 1-\widetilde{\chi}_{\uparrow\uparrow} \bigr)\, \omega 
\,+\, 
\frac{1}{2\rho_{d}^{}}
\left(
\frac{\partial  \chi_{\uparrow\uparrow}^{}}{\partial \epsilon_{d}} 
- 
\frac{1}{2}
\frac{\partial \chi_{\uparrow\downarrow}^{}}{\partial \epsilon_{d}} 
+ 
2 \pi \cot \delta\   \chi_{\uparrow\uparrow}^{2} 
\right) 
 \omega^2 
+ 
\frac{1}{12}  \frac{1}{\rho_{d}}
\frac{\partial \chi_{\uparrow\downarrow}}{\partial \epsilon_{d}} 
\left[
\frac{3\Gamma_L \Gamma_R}{\left( \Gamma_L + \Gamma_R \right)^2} \, 
 (eV)^2  +\left(\pi T\right)^2 \right]
\nonumber \\ 
& \  \ 
-\,\widetilde{\chi}_{\uparrow\downarrow}^{} \,  \alpha\,eV 
\,+\, 
\frac{1}{\rho_{d}^{}}\left(
\frac{1}{2}\frac{\partial \chi_{\uparrow\downarrow}^{}}{\partial \epsilon_{d}} 
\,+\, 
2 \pi \cot \delta\,
\chi_{\uparrow\uparrow} \chi_{\uparrow\downarrow}^{} 
\right)
 \alpha\,eV \, \omega
\, + \, 
\frac{1}{2\rho_{d}^{}}\left(  
\frac{1}{2}\frac{\partial \chi_{\uparrow\downarrow}^{}}{\partial \epsilon_{d}} 
\,+\, 
2 \pi \cot \delta\, \chi_{\uparrow\downarrow}^{2} 
\right) \alpha^2 (eV)^2 .
\label{eq:self_real_ev_zero_field}
\end{align}
\end{widetext}
This expression agrees with the previous result, Eq.\  (19) of Ref.\ \onlinecite{ao2001PRB} 
as shown in Appendix \ref{sec:h=0_previous}.
The higher-order fluctuations emerging away from half-filling enter through 
$\partial \chi_{\uparrow\uparrow}^{}/\partial \epsilon_{d}^{} $ and 
$\partial \chi_{\uparrow\downarrow}^{}/\partial \epsilon_{d}^{}$ at zero-magnetic field, and  these two parameters can also be written in terms 
of the wave-function renormalization factor  $z=1/\widetilde{\chi}_{\uparrow\uparrow}$  
and the Wilson ratio; 
\begin{align}
\frac{\partial \log  \chi_{\uparrow\uparrow}}{\partial  \epsilon_{d}^{}}
\,= & \  -\frac{\partial \log z}{\partial  \epsilon_{d}^{}}  + 
\frac{\partial \log \rho_{d}^{}}{\partial \epsilon_{d}^{}}  , 
\label{eq:dlog_chi_UU}
\\
\frac{\partial \log (-\chi_{\uparrow\downarrow})}{\partial  \epsilon_{d}^{}}
\,= & \ 
\frac{\partial \log \chi_{\uparrow\uparrow}}{\partial  \epsilon_{d}^{}} 
\,+\,
\frac{\partial \log (R_{W}^{}-1)}{\partial \epsilon_{d}^{}}  ,
\label{eq:dlog_chi_UD}
 \\
\frac{\partial \log \rho_{d}^{}}{\partial \epsilon_{d}^{}}  
\, = &\ 
\,-\,2\pi \left(2-R_{W}^{}\right)\, \chi_{\uparrow\uparrow} \cot \delta\, .
\label{eq:dlog_rho_d}
\end{align}
Figure \ref{fig:FL_parameter_1} shows the $\epsilon_{d}^{}$ dependence 
of $\sin^2 \delta$,  $R_{W}^{}-1$, and $z$  at zero field $h=0$ obtained with 
the NRG.\cite{HewsonOguriMeyer}  
Correspondingly, their logarithmic derivatives with respect to $\epsilon_{d}^{}$ 
are shown in Fig.\ \ref{fig:FL_parameter_2}. 
The derivative of   the density of states 
$\rho_{d}^{}=\sin^2 \delta/\pi\Delta$  is obtained using Eq.\ \eqref{eq:dlog_rho_d}
while the derivatives  
 $\partial \log z/\partial  \epsilon_{d}^{}$ and 
 $\partial \log \left(R_{W}^{} -1\right)/\partial \epsilon_{d}^{}$ 
are numerically evaluated from  the discrete NRG data for $z$ and $R_{W}^{}$.   
These derivatives with respect to  $\epsilon_{d}^{}$ 
 are enhanced near the two valence-fluctuation regions  
at $\epsilon_d^{} \simeq 0$ and at $\epsilon_d^{} \simeq -U$. 
Note that the logarithmic derivatives 
can be  related to the $\beta$-functions for 
 renormalization group equations.\cite{Amit}

%


\section{
Non-equilibrium transport through a quantum dot 
}
\label{sec:transport_dot}

We apply  the low-energy asymptotic form of the self-energy obtained in the above 
to the non-equilibrium current $I$ through 
quantum dots.\cite{GlazmanRaikh,NgLee,Hershfield1,WingreenMeir}
The retarded Green's function $G_{\sigma}^{r}(\omega,T,eV)$ 
and the spectral function $A_{\sigma}^{}(\omega,T,eV)$ 
can be obtained from Eqs.\ \eqref{eq:self_imaginary} and \eqref{eq:self_real_ev_mag}: 
\begin{align}
\left\{
G_{\sigma}^{r}(\omega,T,eV) \right\}^{-1} =&  \ \,
\omega - \bigl[\, \epsilon_{d\sigma}^{}+ 
\mathrm{Re}\, \Sigma_{\sigma}^r(\omega,T,eV) 
\,\bigr] 
\nonumber \\
&  +i \,\bigl[\, \Delta - \mathrm{Im}\, \Sigma_{\sigma}^r(\omega,T,eV) 
 \,\bigr]
 \;,\\
A_{\sigma}^{}(\omega,T,eV) \,\equiv & \  
-\frac{1}{\pi} \,\mathrm{Im}\,G_{\sigma}^{r}(\omega,T,eV)\;.
 \end{align}
Note that  $\rho_{d\sigma}^{}(\omega) \equiv A_{\sigma}^{}(\omega,0,0)$.
Then, the current  $I$ can be calculated 
using the  Meir-Wingreen formula,\cite{MeirWingreen,Hershfield1}
\begin{align}
I \,= & \  \frac{e}{2\pi\hbar}
\sum_{\sigma}  
\frac{4\Gamma_L \Gamma_R}{\Gamma_L + \Gamma_R } 
\nonumber \\
& \ \  \times 
\int_{-\infty}^{\infty} \! d\omega\,  
\bigl[\,f_L(\omega)-f_R(\omega) \,\bigr]\, 
\pi A_{\sigma}^{}(\omega,T,eV) 
.
\label{eq:current_formula}
\end{align}
Thus,  $T$ and $eV$ enter through  
the distribution function $f_L-f_R$ 
and the spectral function  $ A_{\sigma}^{}$.

\subsection{
Conductance formula for $\Gamma_L=\Gamma_R$ and $\alpha=0$}

In the following,  we consider the situation in which  $\alpha=0$, 
taking the tunneling couplings and the bias voltages such that 
 $\Gamma_L =  \Gamma_R = \Delta/2$ and  $\alpha_L=\alpha_R=1/2$.
We obtain the spectral function 
up to terms of order $\omega^2$, $(eV)^2$, and $T^2$,  
\begin{align}
& \pi\Delta \, A_{\sigma}(\omega, eV, T) 
\  =  \ \sin^2 \delta_{\sigma}
+ \pi \sin 2\delta_{\sigma}\, \chi_{\sigma\sigma} \ \omega
\nonumber \\
& +  
\pi^2
\left[
\cos 2\delta_{\sigma}
\left(
\chi_{\sigma\sigma}^2
+ 
\frac{1}{2} \chi_{\uparrow\downarrow}^2
\right)
- 
\frac{\sin 2\delta_{\sigma}}{2\pi}
\, \frac{\partial \chi_{\sigma\sigma}}{\partial \epsilon_{d\sigma}^{}}
\right]  \omega^2 
\nonumber \\
& 
+ 
\frac{\pi^2}{3} \! 
\left(
\frac{3}{2}\cos 2 \delta_{\sigma}\,\chi_{\uparrow\downarrow}^2
- 
\frac{\sin 2\delta_{\sigma}}{2\pi}
\frac{\partial \chi_{\uparrow\downarrow}}{\partial \epsilon_{d,-\sigma}} 
\right) 
\! 
\left[\frac{3}{4}\left(eV\right)^2   \! + \left(\pi T\right)^2
\right]  
\nonumber \\
& +  \cdots .
\label{eq:A_including_T_eV}
\end{align}
%
The contribution of the non-linear fluctuation,   
 $\partial \chi_{\uparrow\downarrow}/\partial \epsilon_{d,-\sigma}$, 
enters  in the coefficient for $(\pi T)^2+(3/4) (eV)^2$  through 
 Eq.\ \eqref{eq:self_real_ev_mag}.
We calculate the current  $I$  up to order $(eV)^3$ 
using  Eqs.\ \eqref{eq:current_formula} and \eqref{eq:A_including_T_eV}, 
and  obtain the  differential conductance, 
\begin{align}
\frac{dI}{dV}  =
\frac{e^2}{2\pi\hbar}  \sum_{\sigma} 
\left[
\sin^2 \delta_{\sigma}  
-  c_{T,\sigma}^{}\left(\pi T\right)^2
-  c_{V,\sigma}^{} \left(eV \right)^2  +  \cdots 
\right] .
\end{align}
The coefficients $c_{T,\sigma}^{}$ and $c_{V,\sigma}^{}$ are given by  
\begin{align}
c_{T,\sigma}^{} =  & \ 
%
\frac{\pi^2}{3} 
\Biggl[\,
 -\cos 2 \delta_{\sigma}
\left(
\chi_{\sigma\sigma}^2
+ 
2
\chi_{\uparrow\downarrow}^2 
\right) 
\nonumber \\
& \qquad \ 
 + \,
\frac{\sin 2\delta_{\sigma}}{2\pi}\,
\left(
 \frac{\partial \chi_{\sigma\sigma}}{\partial \epsilon_{d\sigma}^{}}
+\frac{\partial \chi_{\uparrow\downarrow}}{\partial \epsilon_{d,-\sigma}} 
\right)
\Biggr] \;, 
\label{eq:c_T_last}
\\
c_{V,\sigma}^{} = & \ 
\frac{\pi^2}{4}
\Biggl[
\,
-\cos 2 \delta_{\sigma} 
\left( \chi_{\sigma\sigma}^2 +  5\,\chi_{\uparrow\downarrow}^2
\right)
\nonumber \\
& \qquad \ 
+\,
\frac{\sin 2\delta_{\sigma}}{2\pi} 
\left(
\, \frac{\partial \chi_{\sigma\sigma}}{\partial \epsilon_{d\sigma}^{}}
+
3\, \frac{\partial \chi_{\uparrow\downarrow}}{\partial \epsilon_{d,-\sigma}} 
\right) 
\Biggr]
\label{eq:c_V_last}
\;. 
\end{align} 
We note that the derivatives, for which  $\sin 2\delta_{\sigma}$ are multiplied, 
 can be rewritten in terms  of the derivatives  with respect to $\epsilon_{d}^{}$ and $h$,
\begin{align}
 \frac{\partial \chi_{\sigma\sigma}}{\partial \epsilon_{d\sigma}^{}}
+\frac{\partial \chi_{\uparrow\downarrow}}{\partial \epsilon_{d,-\sigma}} 
\,=& \ 
\frac{\partial \chi_{\sigma\sigma}}{\partial \epsilon_{d}}
+ \sigma \frac{\partial \chi_{\uparrow\downarrow}}{\partial h} 
\\
\, \frac{\partial \chi_{\sigma\sigma}}{\partial \epsilon_{d\sigma}^{}}
\,+\,
3\, \frac{\partial \chi_{\uparrow\downarrow}}{\partial \epsilon_{d,-\sigma}} 
\,= & \ 
\frac{\partial \chi_{\sigma\sigma}}{\partial \epsilon_{d}^{}} 
+
\frac{\partial \chi_{\uparrow\downarrow}}{\partial \epsilon_{d}} 
\,+\,
\sigma\, 2\, \frac{\partial \chi_{\uparrow\downarrow}}{\partial h} .
\end{align}
In  the particle-hole symmetric case  
at which $\epsilon_d^{}=-U/2$ and  $h=0$, 
the previous result is also reproduced\cite{ao2001PRB} 
\begin{align}
\!\!\!\!\!
c_{T,\sigma}^{}  
\xrightarrow{h\to 0 \atop \xi_d \to 0}& \  
\frac{
 \widetilde{\chi}_{\uparrow\uparrow}^2
+2 \widetilde{\chi}_{\uparrow\downarrow}^2
}{3\Delta^2},  
\ \ 
c_{V,\sigma}^{}  
\xrightarrow{h\to 0 \atop \xi_d \to 0}\,
\frac{
 \widetilde{\chi}_{\uparrow\uparrow}^2
+5 \widetilde{\chi}_{\uparrow\downarrow}^2
}{4 \Delta^2} , 
\!\!
\end{align}
since  $\delta_{\sigma} = \pi/2$ and $\rho_{d\sigma}^{} = 1/(\pi\Delta)$.

The last line of Eq.\ \eqref{eq:c_T_last}  and  
that of Eq.\ \eqref{eq:c_V_last}  
are expressed in terms of the derivative 
with respect to the center of the impurity levels  $\epsilon_{d}^{}$ 
and the magnetic field $h$.
We may also express these coefficients in a dimensionless way such that 
$c_{T,\sigma}^{} ( T^*)^2$ and $c_{V,\sigma}^{} ( T^*)^2$,  
scaling the quadratic  $(\pi T)^2$ and $(eV)^2$  parts  
by the characteristic energy 
 $T^*=1/4\sqrt{\chi_{\uparrow\uparrow}\chi_{\downarrow\downarrow}}{}$ 
that is introduced in Eq.\ \eqref{eq:Kondo_scale_general} and 
is a function of  $\epsilon_{d}^{}$ and $h$: 
\begin{align}
\frac{dI}{dV}  =& \ 
\frac{2e^2}{2\pi\hbar} 
\Biggl[
\frac{1}{2}
  \sum_{\sigma} 
\sin^2 \delta_{\sigma}
\nonumber \\  
& \qquad \ 
-  C_{T}^{}\left(\frac{\pi T}{T^*}\right)^2
-  C_{V}^{} \left(\frac{eV}{T^*} \right)^2  +  \cdots 
\Biggr] ,
\\
C_{T}^{} \equiv & \ 
\frac{(T^*)^2}{2} 
\sum_{\sigma} \,c_{T,\sigma}^{} ,
\quad  
C_{V}^{} \equiv 
\frac{(T^*)^2}{2} 
\sum_{\sigma} \,c_{V,\sigma}^{} .
\label{eq:dimensionless_coefficients}
\end{align}

\begin{figure}[t]
\begin{minipage}{1\linewidth}
\includegraphics[width=0.6\linewidth]{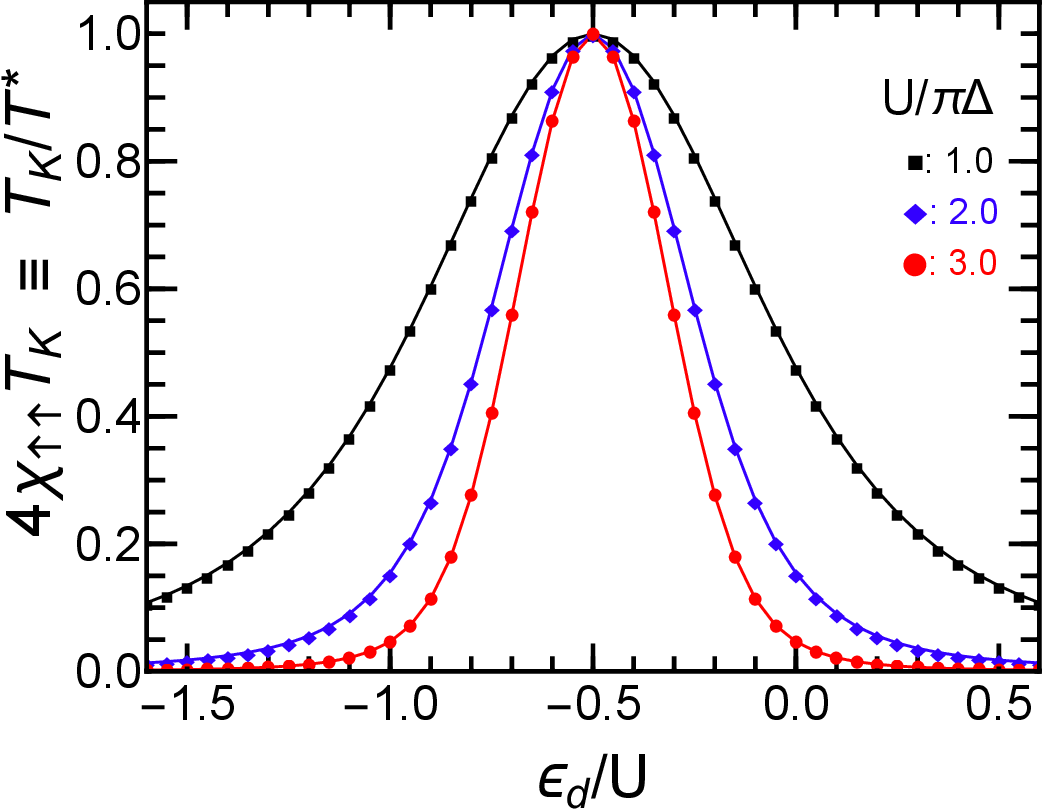}
\end{minipage}
 \caption{
(Color online) 
Susceptibility  $\chi_{\uparrow\uparrow}$ 
is plotted vs  $\epsilon_{d}/U$ at zero magnetic field $h=0$.
The reciprocal of it is proportional to the characteristic energy  
 $T^* \equiv 1/4\chi_{\uparrow\uparrow}$.
Here, the Kondo temperature  $T_K \equiv z_0 \pi \Delta/4$  
is defined  at half-filling $\epsilon_d^{}/U = -0.5$ with 
the renormalization factor $z_0$ which depends on $U$:     
$z_0 \simeq 0.63$,  $0.24$, and $0.08$, 
for $U/\pi \Delta = 1.0$, $2.0$, and $3.0$, 
respectively.
}
 \label{fig:T_scale}
\end{figure}

\subsection{Conductance away from half-filling at zero field  
}

The coefficients given in 
Eqs.\ \eqref{eq:c_T_last} and \eqref{eq:c_V_last}
take a much simpler form  at  zero magnetic field  $h=0$. 
Since  $ {\partial \chi_{\uparrow\downarrow}}/{\partial h}|_{h=0}=0$ 
as  $\chi_{\uparrow\downarrow}$ is an even function of $h$,  we obtain 
\begin{align}
c_{T,\sigma}^{} \, 
& \xrightarrow{\,h\to 0\,} \,   
\frac{\pi^2}{3} 
\left[\,
-
\left(\chi_{\uparrow\uparrow}^2
+ 2 \chi_{\uparrow\downarrow}^2 
\right)  \cos 2 \delta
+ 
\frac{\sin 2\delta}{2\pi}\,
\frac{\partial \chi_{\uparrow\uparrow}}{\partial \epsilon_{d}}
\,\right] , 
\\
c_{V,\sigma}^{} \,& 
\xrightarrow{\,h\to 0\,} \, 
\frac{\pi^2}{4}
\biggl[
\,
-
\left( \chi_{\uparrow\uparrow}^2 +  5\,\chi_{\uparrow\downarrow}^2
\right) \cos 2 \delta
\nonumber \\
& \qquad \qquad \qquad 
+ \, 
 \frac{\sin 2\delta}{2\pi}\left(
\frac{\partial \chi_{\uparrow\uparrow}^{}}{\partial \epsilon_{d}^{}} 
+\frac{\partial \chi_{\uparrow\downarrow}^{}}{\partial \epsilon_{d}^{}}
\right) 
\biggr] .
\end{align} 
These coefficients  $c_{T,\sigma}^{}$ and  $c_{V,\sigma}^{}$ 
for $h=0$ coincide with  those of FMvDM's,\cite{FilipponeMocaVonDelftMora} 
which were first  presented  in Ref.\  \onlinecite{MoraMocaVonDelftZarand} 
by Mora, Moca, von Delft, and Zar\'{a}nd (MMvDZ) away from half-filling 
at zero magnetic field.

\begin{figure}[t]
\begin{minipage}{1\linewidth}
\includegraphics[width=0.6\linewidth]{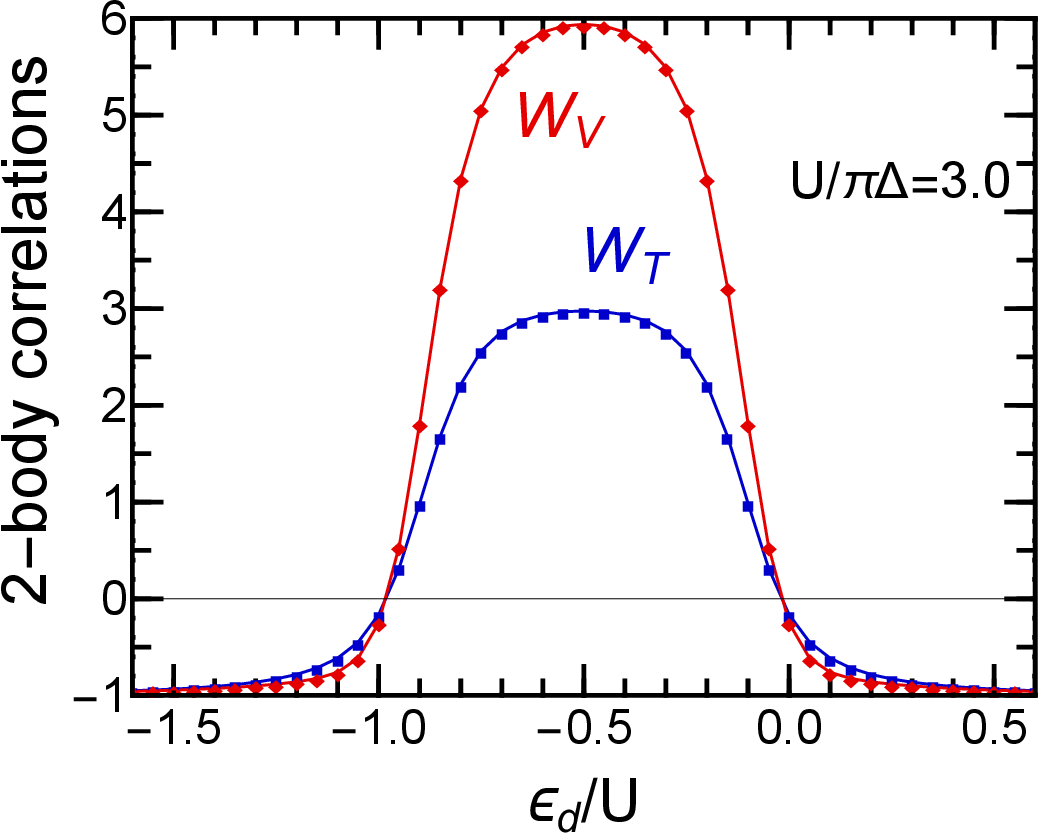}
\\
\includegraphics[width=0.6\linewidth]{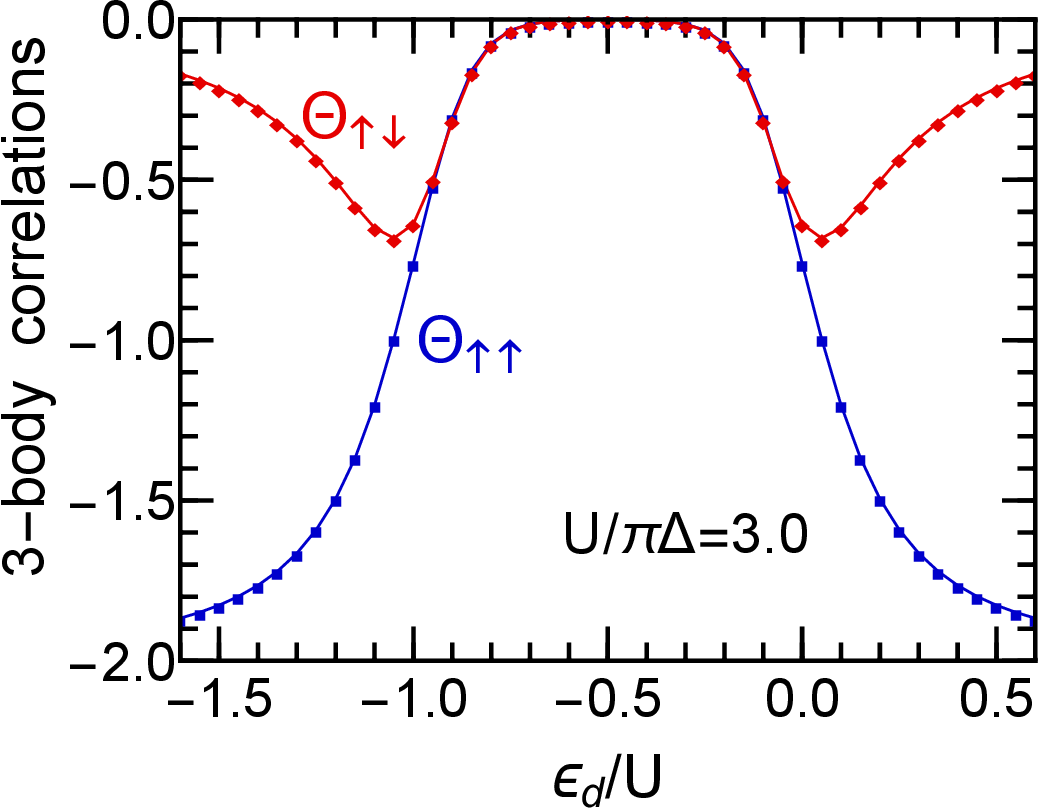}
\end{minipage}
 \caption{
(Color online) 
Upper panel: contributions of the two-body fluctuations parts  
$W_{T}^{} = -\left[1+  2\left(R_W^{}-1\right)^2 \right]\cos 2 \delta$, 
and 
$W_{V}^{} = -\left[1+  5\left(R_W^{}-1\right)^2 \right]\cos 2 \delta$ 
are plotted vs $\epsilon_{d}^{}$ for  $U/\pi \Delta = 3.0$ at $h=0$.
Lower panel: contributions of the three-body  fluctuations 
$\Theta_{\uparrow\uparrow}^{} 
 \equiv  
\frac{\sin 2\delta}{2\pi}\,
\frac{1}{\chi_{\uparrow\uparrow}^2}
\frac{\partial \chi_{\uparrow\uparrow}^{}}{\partial \epsilon_{d}^{}}$, 
and 
$\Theta_{\uparrow\downarrow}^{} 
 \equiv
-\frac{\sin 2\delta}{2\pi}\,
\frac{1}{\chi_{\uparrow\uparrow}^2}
\frac{\partial \chi_{\uparrow\downarrow}^{}}{\partial \epsilon_{d}^{}} 
$. 
In the limit of $|\epsilon_{d}^{}| \to \infty$, 
these parameters converge towards  
$W_{T}^{}\to -1$, $W_{V}^{}\to -1$,
$\Theta_{\uparrow\uparrow}^{} \to -2$, and  
$\Theta_{\uparrow\downarrow}^{} \to 0$. 
%
%
}
 \label{fig:Wt_Wv_Theta_for_u3}
\end{figure}

\begin{figure}[t]
\begin{minipage}{1\linewidth}
\includegraphics[width=0.6\linewidth]{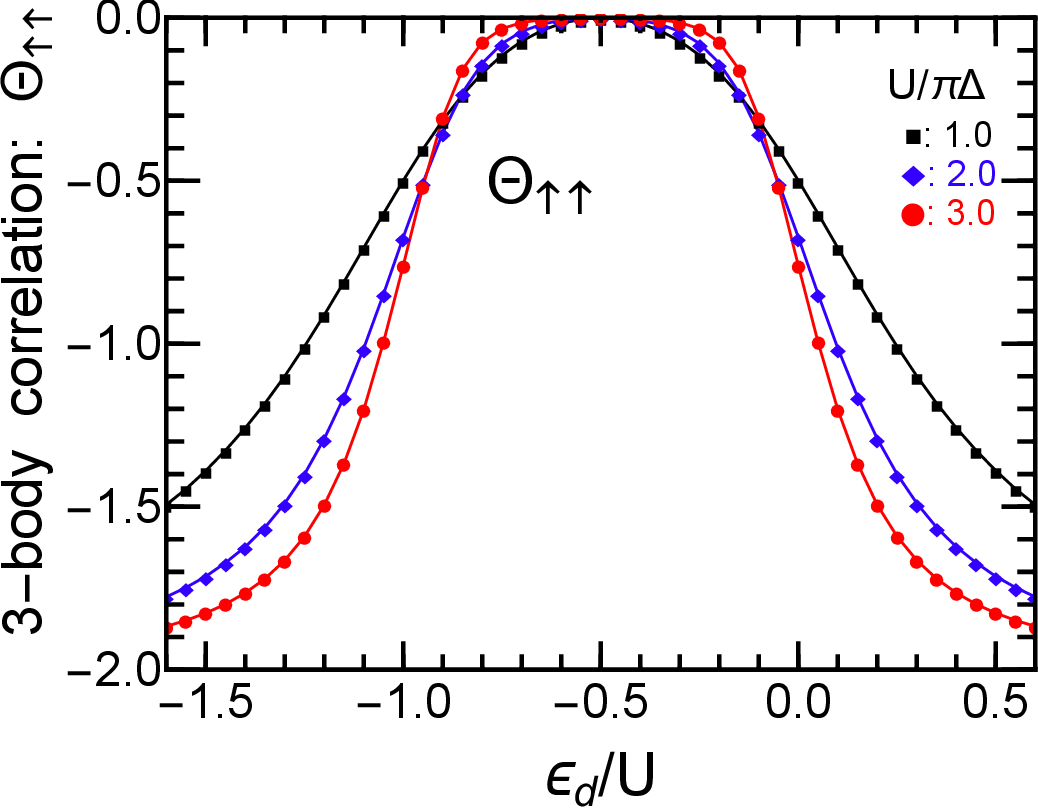}
\\
\includegraphics[width=0.6\linewidth]{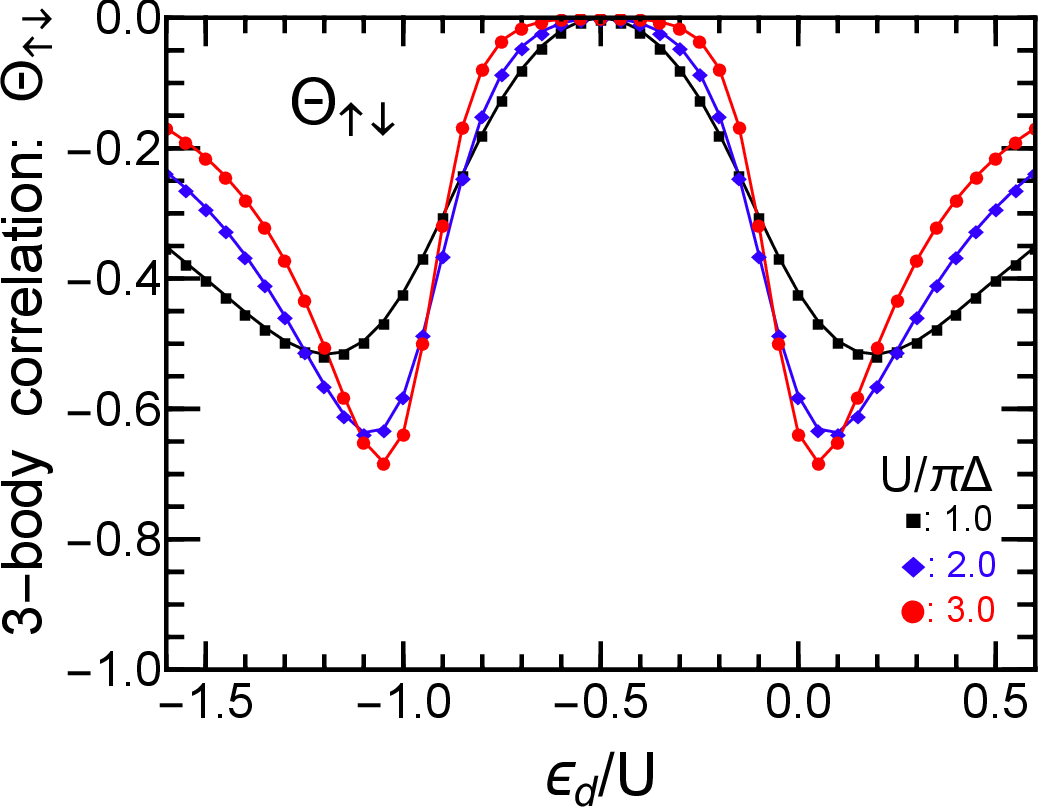}
\end{minipage}
 \caption{
(Color online) 
Contributions of the three-body fluctuations 
$\Theta_{\uparrow\uparrow}^{}$ and 
$\Theta_{\uparrow\downarrow}^{}$ 
are plotted for several different values of $U/\pi \Delta$ 
($= 1.0,  2.0$ and $3.0$).  
 As $U$ increases, both of these two significantly vary 
at the crossover region from the Kondo regime to 
the empty  (fully-occupied) orbital regime 
seen at $\epsilon_{d}^{}/U \simeq 0.0$ 
($\epsilon_{d}^{}/U \simeq -1.0$). 
}
 \label{fig:FL_UU_UD_several_U}
\end{figure}

Corresponding dimensionless parameters in this case  
are scaled by  the characteristic energy   $T^* =  1/4\chi_{\uparrow\uparrow}$ 
which increases as  $\epsilon_{d}^{}$ deviates 
from the particle-hole symmetric point as  shown in Fig.\ \ref{fig:T_scale}:    
\begin{align} 
C_{T}^{}  
\,=& \ 
\frac{\pi^2}{48} 
\,\bigl(\,
W_{T}^{} 
\,+\, 
\Theta_{\uparrow\uparrow}^{} 
\,\bigr)
\;, 
\label{eq:dimensionless_Ct_h=0}
\\
C_{V}^{}  
\, = & \  \frac{\pi^2}{64}
 \,\bigl(
\,
W_{V}^{} 
\,+ \,
\Theta_{\uparrow\uparrow}^{} 
-
\Theta_{\uparrow\downarrow}^{} 
 \bigr) 
\;.
\label{eq:dimensionless_Cv_h=0}
\end{align}
Here, 
$W_{T}^{}$ and $W_{V}^{}$ represents 
contributions of two-body fluctuations  determined by 
 the spin and charge susceptibilities,  or the Wilson ratio $R_W^{}$:
\begin{align}
W_{T}^{} \,=& \ 
-\left[\,1+  2\left(R_W^{}-1\right)^2 \,\right]\cos 2 \delta
\;, \\
W_{V}^{} \,=& \ 
-\left[ \,1+  5\left(R_W^{}-1\right)^2 \,\right]\cos 2 \delta \;.
\end{align}
The other parts, 
$\Theta_{\uparrow\uparrow}^{}$  and $\Theta_{\uparrow\downarrow}^{}$,   
represent  contributions of three-body fluctuations 
which can also  be described in terms  of the non-linear susceptibilities 
$\chi_{\sigma_1\sigma_2\sigma_3}^{[3]}$ defined in 
Eq.\ \eqref{eq:canonical_correlation_3_dif}: 
\begin{align}
\Theta_{\uparrow\uparrow}^{} 
\, \equiv & \  
\frac{\sin 2\delta}{2\pi}\,
\frac{1}{\chi_{\uparrow\uparrow}^2}
\frac{\partial \chi_{\uparrow\uparrow}^{}}{\partial \epsilon_{d}^{}}  \;,
\label{eq:Theta_UU}
\\
\Theta_{\uparrow\downarrow}^{} 
\, \equiv & \ 
-\frac{\sin 2\delta}{2\pi}\,
\frac{1}{\chi_{\uparrow\uparrow}^2}
\frac{\partial \chi_{\uparrow\downarrow}^{}}{\partial \epsilon_{d}^{}} 
\label{eq:Theta_UD} 
\;.
\end{align}
At half-filling  $\delta = \pi/2$,  
the Wilson ratio approaches $R_{W}^{} \to 2$ 
for the Kondo regime  $U\gtrsim 2\Delta$,  
and then $W_{T}^{} \to 3$ and  $W_{V}^{} \to 6$,  
whereas the  contribution of the three-body fluctuations vanish 
 $\Theta_{\uparrow\uparrow}^{}\to 0 $ and 
$\Theta_{\uparrow\downarrow}^{}\to 0$
as charge fluctuation is minimized and spin fluctuation is maximized.\cite{YamadaYosida2} 
We discuss  in the following how the two-body and three-body contributions 
vary as  $\epsilon_{d}^{}$ deviates  away from  the particle-hole symmetric point.

\begin{figure}[t]
\begin{minipage}{1\linewidth}
\includegraphics[width=0.6\linewidth]{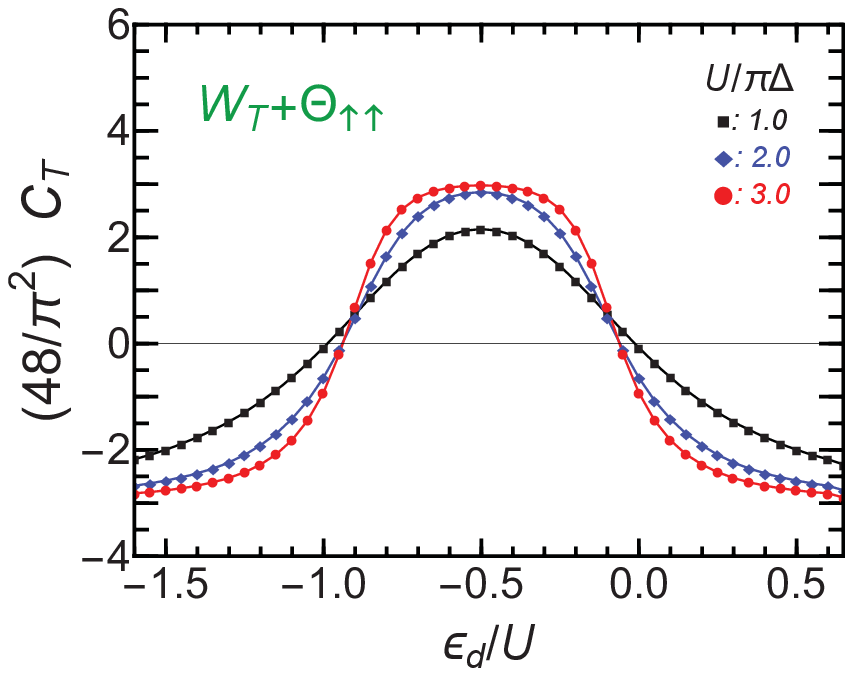}
\\
\includegraphics[width=0.6\linewidth]{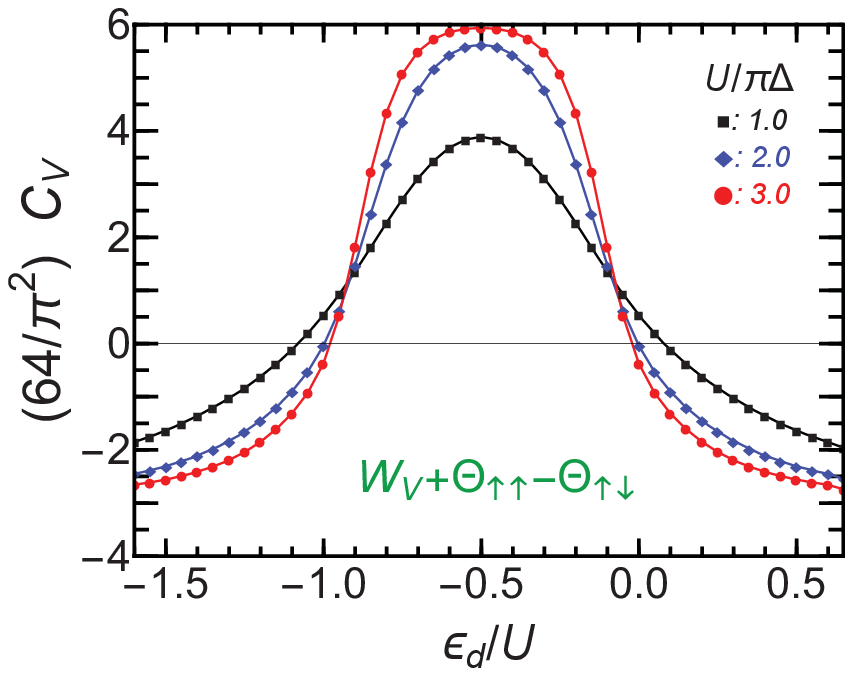}
\end{minipage}
 \caption{
(Color online) 
Dimensionless coefficients  
$C_T$ and $C_V$ 
are plotted vs  $\epsilon_{d}^{}/U$  
for $U/\pi \Delta = 1.0$, $2.0$, and $3.0$  at  $h=0$. 
Note that numerical factor has been introduced such that 
$(48/\pi^2)\,C_T^{} 
= W_{T}^{} + \Theta_{\uparrow\uparrow}^{}$ 
and 
$(64/\pi^2)\,C_V^{} = W_{V}^{} 
+ \Theta_{\uparrow\uparrow}^{} 
-  \Theta_{\uparrow\downarrow}^{}$.
}
 \label{fig:CV_CT_h=0}
\end{figure}

The behavior in the other limit at  $\epsilon_d \gg \max(U, \Delta)$ 
corresponds to the empty-orbital regime 
as already examined by MMvDZ.\cite{MoraMocaVonDelftZarand} 
In the empty-orbital regime,  the interaction can be neglected at low energies 
and thus for $\epsilon_{d}^{} \to \infty$ 
 the parameters asymptotically behave such that 
$R_{W} \to 1$, 
$\delta \simeq \Delta/\epsilon_{d}^{}$, 
$\chi_{\uparrow\uparrow} \simeq \Delta/(\pi \epsilon_{d}^{2})$,
and $\chi_{\uparrow\downarrow} \simeq 0$.  
Therefore,  
\begin{align}
&
\lim_{|\epsilon_{d}^{}|\to \infty}
W_{T}^{} \,=\, -1,
\qquad 
\lim_{|\epsilon_{d}^{}|\to \infty}
W_{V}^{} \,=\, -1,
\\
&
\lim_{|\epsilon_{d}^{}|\to \infty}
\Theta_{\uparrow\uparrow}^{} 
\,=\, -2 ,
\qquad 
\lim_{|\epsilon_{d}^{}|\to \infty}
\Theta_{\uparrow\downarrow}^{} 
\,=\, 0. 
\end{align}
The opposite limit  $\epsilon_{d}^{} \to -\infty$,  corresponding to 
a fully-filled orbital, links to the empty-orbital regime 
through the particle-hole transformation. 
The behavior  of the two-body-fluctuation and three-body-fluctuation parts 
at intermediate  $\epsilon_{d}^{}$ can be explored  using the NRG. 
Figure \ref{fig:Wt_Wv_Theta_for_u3} shows a typical result 
obtained  for $U=3\pi \Delta$. We see in the right panel explicitly 
the contributions of the three-body fluctuation,  $\Theta_{\uparrow\uparrow}^{}$ and 
$\Theta_{\uparrow\downarrow}^{}$, are suppressed in the Kondo 
regime  $-1.0 \lesssim \epsilon_{d}^{}/U \lesssim 0.0$.
It also shows  that the three-body fluctuations become important 
outside the Kondo regime. 
The anti-parallel component  $\Theta_{\uparrow\downarrow}^{}$ shows 
a  minimum 
in the valence fluctuation region near  $\epsilon_{d}^{}/U \simeq -1.0$ and $0.0$,  
whereas the parallel component  $\Theta_{\uparrow\uparrow}^{}$ 
does not have an extremal point. 
Figure \ref{fig:FL_UU_UD_several_U} shows the three-body contributions 
for several values of  the  interaction;  $U/\pi \Delta=1.0, 2.0$ and $3.0$. 
The crossover between the Kondo and empty (or fully-occupied) orbital  
regimes becomes sharp as $U$ increases, 
and correspondingly the transient region becomes very narrow for large $U$.
The dependence of  $C_{T}^{}$  and $C_{V}^{}$ on 
$\epsilon_{d}^{}$ was already discussed by MMvDZ.\cite{MoraMocaVonDelftZarand}
We also provide similar results in Fig.\ \ref{fig:CV_CT_h=0} 
in order to explicitly show how the sum of  two-body and  three-body 
fluctuations  determines these coefficients. 
The contributions of the two-body fluctuations which enter through 
$W_{T}^{}$  and $W_{V}^{}$ dominate in the Kondo regime, 
whereas the three-body fluctuation give significant 
contributions for $|\epsilon_{d}+U/2| \gtrsim U/2$.
In the  $|\epsilon_{d}^{}|\to \infty$ limit 
of  the  empty (or fully-occupied) orbital regime,  
the coefficients converge towards 
 $(48/\pi^2)C_{T}^{} \to -3$ 
and  $(64/\pi^2)C_{V}^{} \to -3$,\cite{MoraMocaVonDelftZarand}
while those in the Kondo regime are given by  
 $(48/\pi^2)C_{T}^{} \to 3$ and $(64/\pi^2)C_{V}^{} \to 6$ 
for $U\gtrsim 2 \pi \Delta$.

\begin{figure}[t]
 \leavevmode
\begin{minipage}{1\linewidth}
\includegraphics[width=0.6\linewidth]{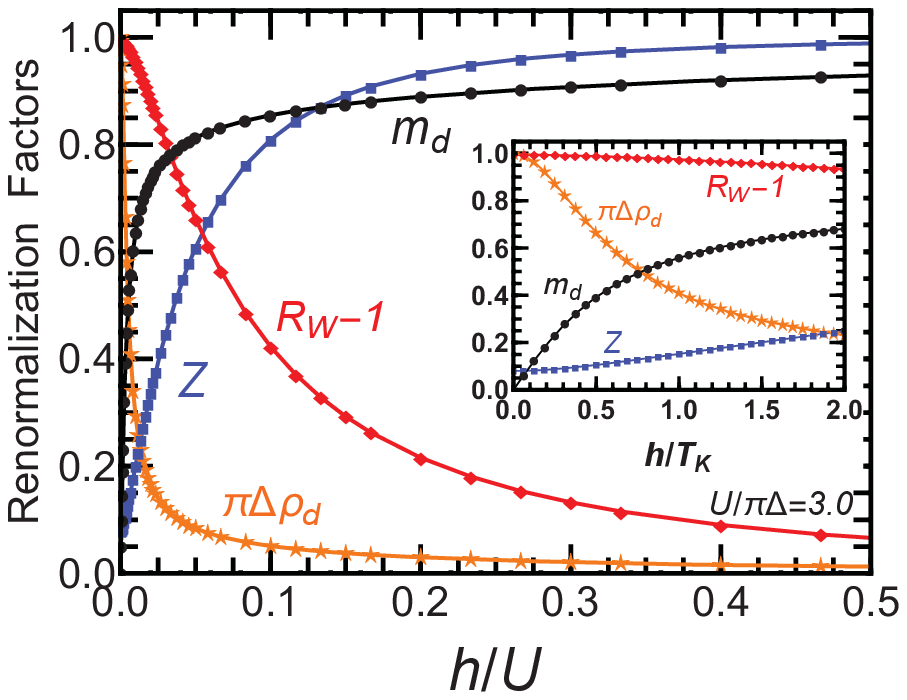}
\\
 \includegraphics[width=0.6\linewidth]{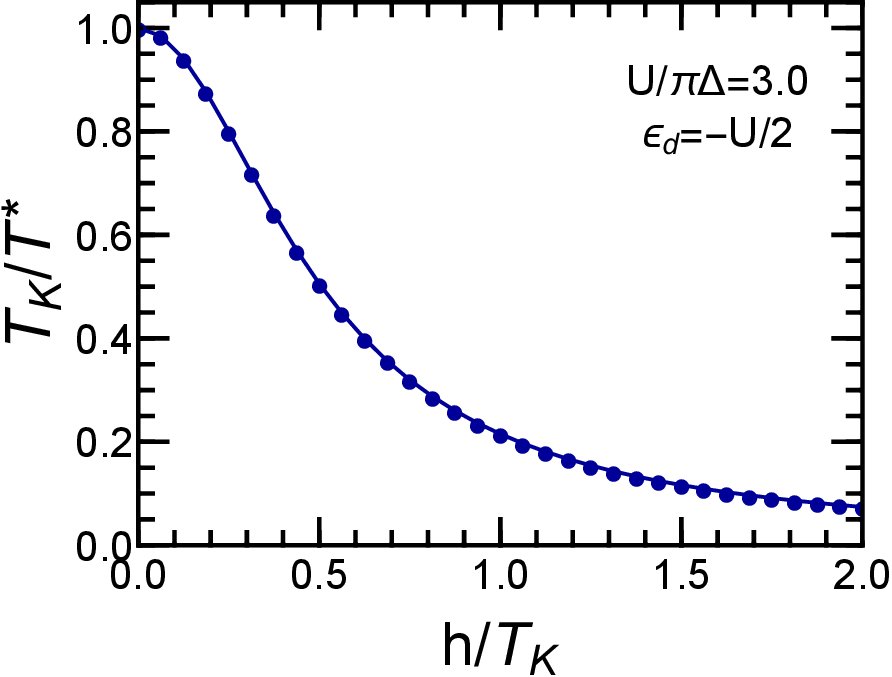}
\end{minipage}
 \caption{
(Color online) 
Magnetic field dependence of Fermi-liquid  parameters 
 at half-filling  $\epsilon_d^{}=-U/2$ for  $U/\pi \Delta =3.0$  plotted vs $h/U$.  
 Inset shows an enlarged view of a small $h$ region, 
for which the horizontal axis is scaled by $T_K = 0.02 \pi \Delta$ determined at  $h=0$. 
Upper panel shows  $Z$, $R_{W}^{}-1$, 
$\pi\Delta \rho_{d}^{} = \cos^2 (\pi m_d/2)$,  
and  $m_d^{} \equiv n_{d\uparrow}^{}-n_{d\downarrow}^{}$.
Using this definition of  $T_K$,  
the reciprocal of the field-dependent characteristic energy  $T^*$ 
is plotted vs $h/T_K$ in the right panel.
} 
 \label{fig:field_dependence_half_filling}
\end{figure}

\begin{figure}[t]
\begin{minipage}{1\linewidth}
 \includegraphics[width=0.6\linewidth]{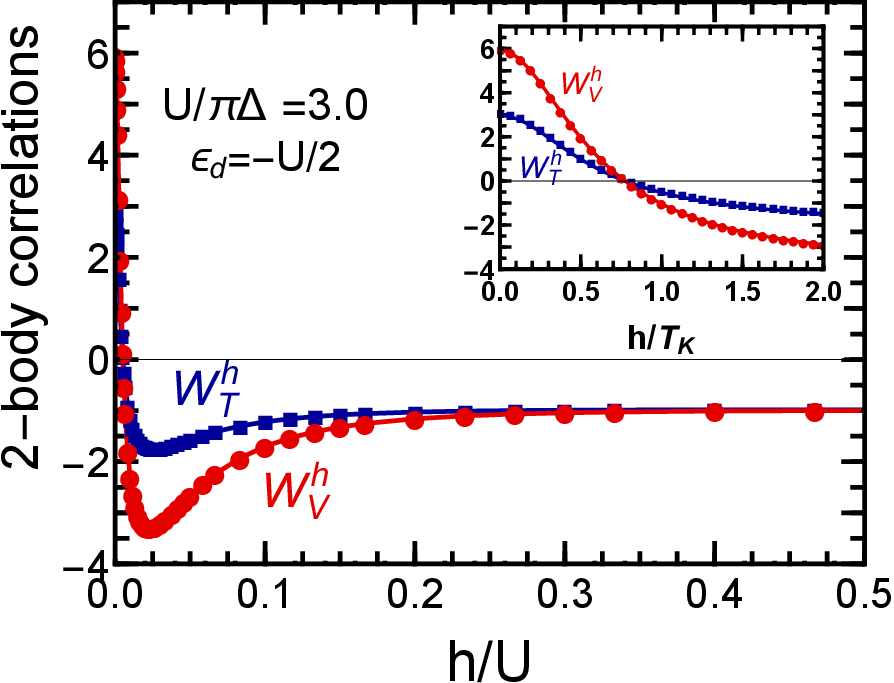}
\\
\includegraphics[width=0.6\linewidth]{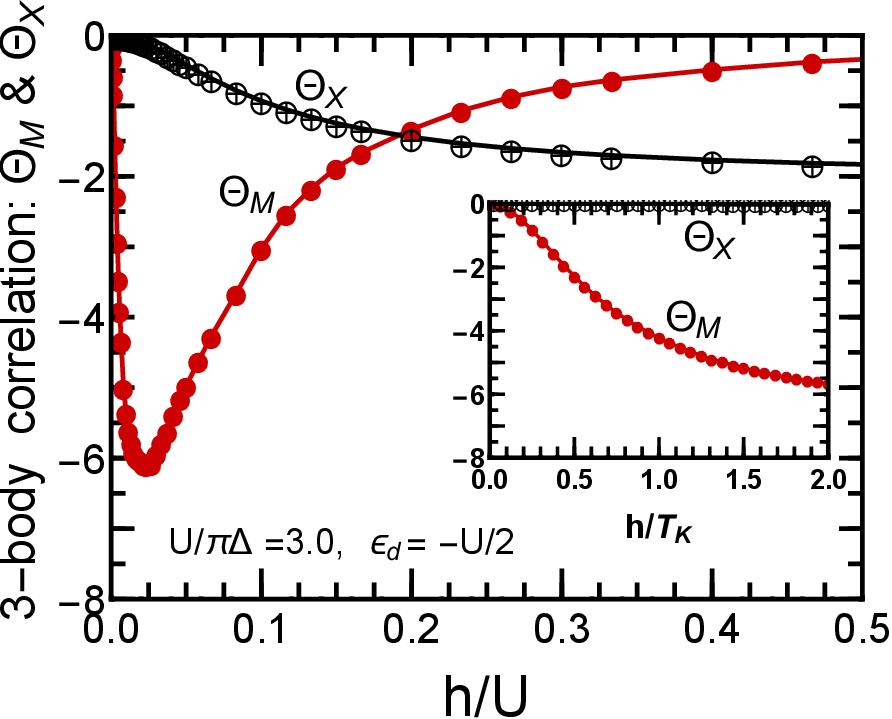}
\end{minipage}
 \caption{
(Color online) 
Two-body and three-body correlations which determine 
$C_{T}^{h}$ and $C_{V}^{h}$ are plotted vs $h/U$ 
 at half-filling $\epsilon_{d}^{}=-U/2$ for  $U/\pi \Delta = 3.0$.   
 Inset shows an enlarged view of the small $h$ region, 
for which the horizontal axis is scaled by $T_K = 0.02 \pi \Delta$ ($=0.0066U$)  
determined at  $h=0$. 
Upper panel shows the contribution of  two-body fluctuations  
$W_{T}^{h} = \left[1+  2\left(R_W^{}-1\right)^2 \right]\cos (\pi m_d)$, 
and 
$W_{V}^{h} = \left[1+  5\left(R_W^{}-1\right)^2 \right]\cos (\pi m_d)$. 
Lower panel shows the  contribution of three-body fluctuations 
 $\Theta_{M}^{}$ and   $\Theta_{X}^{}$, defined 
in Eqs.\ \eqref{eq:Theta_M} and \eqref{eq:Theta_X}.
}
 \label{fig:FL_mag}
\end{figure}

\begin{figure}[t]
 \leavevmode
\begin{minipage}{1\linewidth}
 \includegraphics[width=0.6\linewidth]{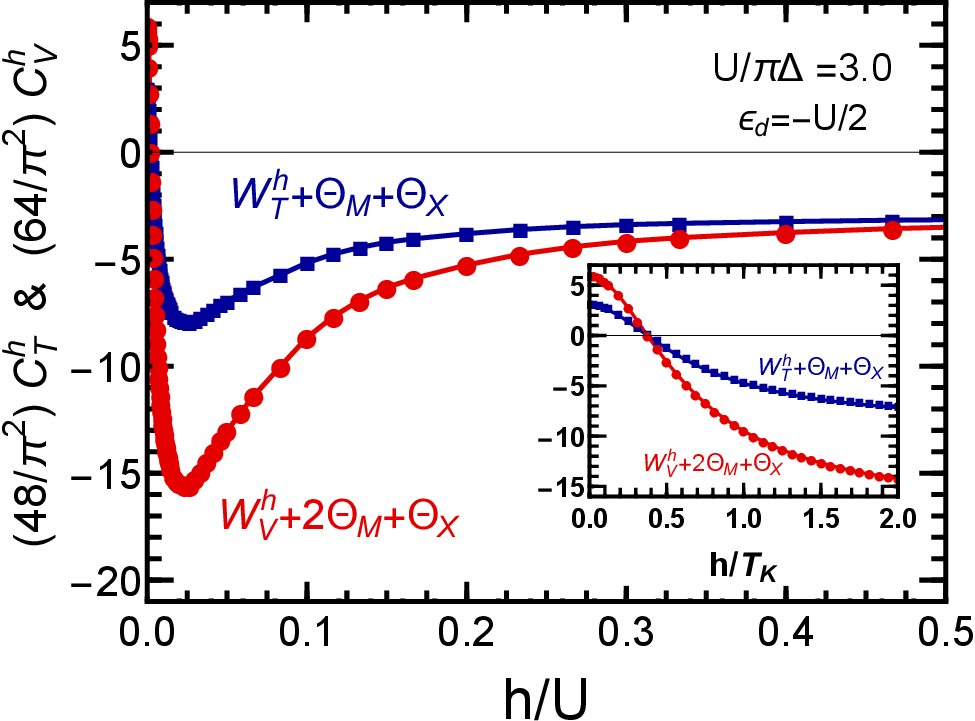}
\\
 \includegraphics[width=0.6\linewidth]{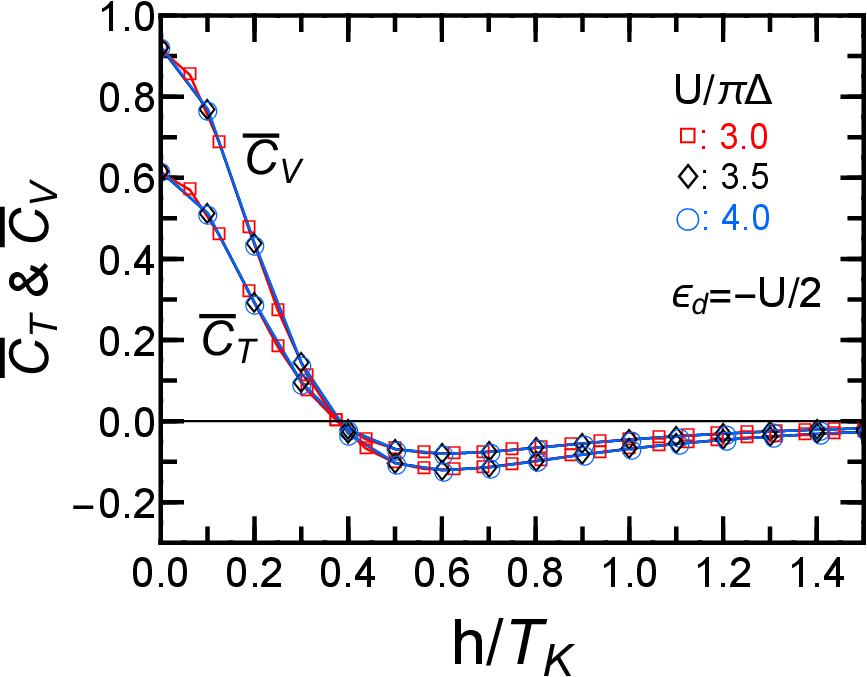}
\end{minipage}
 \caption{
(Color online)
Magnetic-field  dependence of the $dI/dV$ coefficients: 
Upper panel shows 
$(48/\pi^2)\, C_{T}^{h}  = W_{T}^{h} 
 + \Theta_{M}^{} + \Theta_{X}^{} $ 
and  $(64/\pi^2)\,C_{V}^{h}= W_{V}^{h}  
+ 2\Theta_{M}^{}  + \Theta_{X}^{}$.  
Inset describes an enlarged view of the small $h$ region.
Lower panel shows rescaled coefficients 
$\overline{C}_{T}^{} = (T_K/T^*)^2 C_{T}^{h}$ and 
$\overline{C}_{V}^{} = (T_K/T^*)^2 C_{V}^{h}$ 
defined in  Eq.\ \eqref{eq:dimensionless_coefficients_2},
using the  $h$-independent $T_K$. 
The Kondo temperature  $T_K =  z_0 \pi \Delta/4$  is
determined  at  $h=0$ with the renormalization 
 factor   $z_0 \simeq 0.08$,  $0.05$, and  $0.03$  
for $U/\pi \Delta = 3.0$, $3.5$, and $4.0$, respectively
} 
 \label{fig:field_dependence_CT_CV}
\end{figure}

\subsection{Conductance at finite magnetic fields  
for $\epsilon_{d}^{}=-U/2$}

We next consider the conductance at finite magnetic fields $h\neq 0$,  
applied  at half-filling $\xi_d=0$. In this case, 
  the average  of  total occupation number 
for both spin components is fixed at  $n_{d\uparrow}+n_{d\downarrow}=1$, 
and  thus the phase shift for each spin component can be expressed in the form  
 $\delta_\sigma = \pi( 1 + \sigma m_d)/2$,  
with  $m_d= n_{d\uparrow}-n_{d\downarrow}$  the induced magnetization.
Furthermore, since $\rho_{d\uparrow}^{}=\rho_{d\downarrow}^{}$, 
$\chi_{\uparrow\uparrow}^{}=\chi_{\downarrow\downarrow}^{}$, 
 $z_{\uparrow}=z_{\downarrow}$($\equiv z$),  
and the coefficients  for  the  $T^2$ and $(eV)^2$ terms 
defined in Eqs.\ \eqref{eq:c_T_last} and \eqref{eq:c_V_last} 
are  simplified, 
\begin{align}
& \frac{1}{2} \sum_{\sigma} c_{T,\sigma}^{}
\nonumber  \\   
& \xrightarrow{\xi_d \to 0} \frac{\pi^2}{3} 
\Biggl[
\left(\chi_{\uparrow\uparrow}^2
+ 2 \chi_{\uparrow\downarrow}^2 \right) 
\cos (\pi m_d)
- 
\frac{\sin (\pi m_d)}{2\pi}
 \frac{\partial \chi_{\uparrow\downarrow}}{\partial h} 
\nonumber \\
& 
\qquad \qquad \quad
-\, \frac{\sin (\pi m_d)}{2\pi} 
\frac{\partial }{\partial \epsilon_d} 
\left(
\frac{\chi_{\uparrow\uparrow} 
- \chi_{\downarrow\downarrow}}{2}
\right)
\Biggr],  
\label{eq:c_T_mag_half}
\\
&  \frac{1}{2} \sum_{\sigma} c_{V,\sigma}^{}
\nonumber  \\   
& 
 \xrightarrow{\xi_d \to 0} 
\frac{\pi^2}{4}
\Biggl[
\left( \chi_{\uparrow\uparrow}^2 +  5\,\chi_{\uparrow\downarrow}^2
\right)
\cos (\pi m_d)
- 
\frac{\sin (\pi m_d)}{\pi} 
 \, \frac{\partial \chi_{\uparrow\downarrow}}{\partial h} 
\nonumber \\
& 
\qquad \qquad \quad 
-\, \frac{\sin (\pi m_d)}{2\pi} 
\frac{\partial }{\partial \epsilon_d} 
\left(
\frac{\chi_{\uparrow\uparrow} 
- \chi_{\downarrow\downarrow}}{2}
\right)
\Biggr] .
\label{eq:c_V_mag_half}
 \end{align}
The three-body contribution that  enters through  
$\partial \chi_{\uparrow\downarrow}/\partial \epsilon_{d}^{}$ 
has vanished 
because the contributions of $\uparrow$ and $\downarrow$ spin components 
cancel each other out. 
The characteristic energy  $T^*=1/4\chi_{\uparrow\uparrow}$  
in the present case depends on $h$ 
as shown in Fig.\ $\ref{fig:field_dependence_half_filling}$.

Multiplying   Eqs. \eqref{eq:c_T_mag_half}--\eqref{eq:c_V_mag_half} 
by $(T^*)^2$,  we obtain the dimensionless coefficients 
\begin{subequations}
\label{allequations}
 \begin{align}
C_{T}^{h}
\equiv & \ 
\frac{\pi^2}{48} 
\left(
W_{T}^{h}
+ \Theta_{M}^{} 
+ \Theta_{X}^{} 
\right) ,
\label{eq:CT_mag_rescale}
\\
C_{V}^{h}  \equiv  & \  
\frac{\pi^2}{64}
\left(
W_{V}^{h}
+ 2\Theta_{M}^{} 
+ \Theta_{X}^{} 
\right)
.  \! 
\label{eq:CV_mag_rescale}
\end{align} 
\end{subequations}
Here,  $W_{T}^{h}$ and  $W_{V}^{h}$ represent 
contributions of the two-body fluctuations, 
\begin{align}
W_{T}^{h} \,\equiv & \ 
\left[\,1+ 2\left(R_{W}^{} -1 \right)^2 \,\right] \cos (\pi m_d) ,  \\
W_{V}^{h} \,\equiv & \ 
\left[\,1+ 5\left(R_{W}^{} -1 \right)^2 \,\right] \cos (\pi m_d) . 
 \end{align}
The remaining contribution of the  three-body fluctuations 
are described by  $\Theta_{M}^{}$  and $\Theta_{X}^{}$, 
\begin{subequations}
\label{allequations}
\begin{align}
\Theta_{M}^{} 
\, \equiv & \   
-\, \frac{\sin (\pi m_d)}{2\pi} 
\frac{1}{\chi_{\uparrow\uparrow}^2}
 \, \frac{\partial \chi_{\uparrow\downarrow}}{\partial h} \,,
\label{eq:Theta_M} 
\\
\Theta_{X}^{} 
\, \equiv & \     
-\, \frac{\sin (\pi m_d)}{2\pi} 
\frac{1}{\chi_{\uparrow\uparrow}^2}
\frac{\partial }{\partial \epsilon_d} 
\left(
\frac{\chi_{\uparrow\uparrow} 
- \chi_{\downarrow\downarrow}}{2}
\right)
\;.
\label{eq:Theta_X} 
\end{align}
\end{subequations}
The contribution of this three-body correlation at finite magnetic fields 
can also be decomposed into  the logarithmic derivatives of the 
renormalization factor  and the Wilson ratio, similarly  
to Eqs.\ \eqref{eq:dlog_chi_UD} and \eqref{eq:dlog_rho_d}, 
\begin{align}
&\frac{\partial  \log (- \chi_{\uparrow\downarrow})}{\partial  h}
\,= \, 
\frac{\partial \log \chi_{\uparrow\uparrow}}{\partial  h} 
+ \frac{\partial \log (R_{W}^{}-1)}{\partial h}  ,
\label{eq:dlog_chi_UD_dh}
\\
&\frac{\partial \log \chi_{\uparrow\uparrow}}{\partial  h}
=    -\frac{\partial \log z}{\partial  h} 
-2\pi R_{W}^{} \, \chi_{\uparrow\uparrow} \tan \left(\frac{\pi m_d}{2}\right) .  
\label{eq:dlog_chi_UU_dh}
\end{align}

Figure \ref{fig:field_dependence_half_filling} shows the 
magnetic-field dependence of  the renormalized parameters,  
obtained with the NRG.\cite{HewsonOguriMeyer}
It indicates that  the induced magnetization $m_d$ and 
the density of states  $\sin^2 \delta= \pi \Delta \rho_{\sigma}^{}$ 
rapidly vary  at small fields $h \lesssim T_K$ as the Kondo resonance  
goes away from the Fermi level.  
In contrast, the wavefunction renormalization factor  $z$ and $R_{W}^{}$  
vary more slowly than $m_d$ and $\sin^2 \delta_{\sigma}$ with 
the energy scale of the Coulomb interaction $U$.  
Figure \ref{fig:FL_mag} shows 
the magnetic-field dependence of the contributions 
of two-body fluctuations  and three-body fluctuations on the coefficients 
 $C_{T}^{h}$ and  $ C_{V}^{h}$.
The two-body correlations are given by 
 $W_{T}^{h}= 3$ and  $W_{T}^{h}= 6$ 
at zero field  for large interactions ($U \gtrsim 2 \pi \Delta$)  
as $m_d=0$  and $R_{W}\to 2$. 
As $h$ increases, these two-body  contributions change  sign near $h=0.8T_K$ with  
$T_K=0.02\pi \Delta = 0.0066U$ that is determined at $h=0$ for $U=3.0 \pi \Delta$.
Both of these two correlations show a minimum near  $h\simeq 0.02U$, 
and then  approach  
 $\lim_{h\to \infty} W_{T}^{h}= -2$ and 
 $\lim_{h\to \infty}W_{V}^{h} =-2$ for large magnetic  fields 
 where  $m_d \to 1$ and $R_{W}^{} \to 1$.
 The three-body contribution  $\Theta_{M}^{}$ vanishes at $h=0$, 
and in the large-field limit  $\lim_{|h|\to \infty}
\Theta_{M}^{} =0$  as $\chi_{\uparrow\downarrow}$ 
decreases faster than $\chi_{\uparrow\uparrow}$.
It also has a deep minimum  of  $\Theta_{M}^{}\simeq -6.0$,  
which is deeper than that of  $W_{V}^{h} \simeq -3.7$,  
at an intermediate field  $h \simeq 0.02U$ for the case $U=3.0 \pi \Delta$.
We also see that  $\Theta_{M}^{}$  gives a comparable contribution 
 with that of $W_{T}^{h}$ and $W_{V}^{h}$  at small fields $h \lesssim T_K$.
 The other three-body term $\Theta_{X}^{}$ also vanishes at  $h=0$,  
and at  $h \lesssim T_K$ it takes a very small negative value 
and does not contribute  to  $C_{T}^{h}$ and $C_{V}^{h}$ very much. 
However, it becomes comparable to  $\Theta_{M}^{}$ 
 at  high fields  $h \gtrsim 0.1 U$,  
and then approaches $\lim_{|h|\to \infty}\Theta_{X}^{}  =  -2$ 
which corresponds to the value in the noninteracting case. 

The lower panel of Fig.\ \ref{fig:field_dependence_CT_CV} shows the 
total contributions:   $(48/\pi^2)\,C_{T}^{h} =W_{T}^{h} 
+ \Theta_{M}^{}   + \Theta_{X}^{}$,  
and  $(64/\pi^2)\,C_{V}^{h} =W_{V}^{h}  
+ 2\Theta_{M}^{} + \Theta_{X}^{}$ for the same interaction  $U=3.0\pi \Delta$.  
For $h \gtrsim 0.8T_K =0.0053U$, both the two-body and three-body correlations 
give negative contributions, and thus the minimum of $C_{T}^{h}$ 
and  also that of $C_{V}^{h}$  become deeper than the minimum 
of the individual contributions alone.
It indicates that the three-body correlation  $\Theta_{M}^{}$ 
dominates  the contribution on the  $T^2$ and $(eV)^2$ part 
of $dI/dV$ near the minimum  $0.01U \lesssim h \lesssim 0.1U$. 
So far, we have used the field-dependent energy $T^*$ to 
scale the $T^2$ and $(eV)^2$ dependences. 
In order to examine the universal Kondo-scaling  behavior for small magnetic fields, 
however, we use $T_K$ determined at $h=0$ as an $h$-independent 
characteristic  energy and rescale $dI/dV$ such that 
\begin{align}
\frac{dI}{dV}  =& \ 
\frac{2e^2}{2\pi\hbar} 
\Biggl[\,
\frac{1}{2}
  \sum_{\sigma} 
\sin^2 \delta_{\sigma}  
\nonumber \\
& 
\qquad 
-  \overline{C}_{T}^{}\left(\frac{\pi T}{T_K}\right)^2
-  \overline{C}_{V}^{} \left(\frac{eV}{T_K} \right)^2 +  \cdots 
\Biggr] ,
\\
\overline{C}_{T}^{} 
\, \equiv&  \  
\left(\frac{T_K}{T^*}\right)^2 C_{T}^{h} \,,
 \qquad 
\overline{C}_{V}^{} \,\equiv  \,  
\left(\frac{T_K}{T^*}\right)^2  C_{K}^{h} . 
\label{eq:dimensionless_coefficients_2}
\end{align} 
In the right panel of  Fig.\  \ref{fig:field_dependence_CT_CV},  
 $\overline{C}_{T}^{}$ and  $ \overline{C}_{V}^{}$ are plotted vs $h/T_K$,
using $T_K$  for each $U/\pi \Delta = 3.0, 3.5, 4.0$. 
We see that  both the  coefficients   $\overline{C}_{T}^{}$ and  $\overline{C}_{V}^{}$  
show universal Kondo behavior.   
This is mainly caused by the fact that the Wilson ratio  is almost saturated 
 $R_W \simeq 2$  for strong  interactions $U$.
These two coefficients, 
 $\overline{C}_{T}^{}$ and  $ \overline{C}_{V}^{}$,  also 
show a similar $h$ dependence, 
especially they both change sign at  finite magnetic field 
$h \simeq 0.38 T_K$  of the order of the Kondo temperature.
Therefore, the zero-bias peak of  
the conductance splits  for large magnetic fields $h\gtrsim 0.38 T_K$   
as  $dI/dV$  increases from the zero-bias value  as  $eV$  or  
 $T$ increases.\cite{HewsonBauerKoller}  
These observations  are also consistent with 
the result of the 
  second-order renormalized perturbation theory.
\cite{HewsonBauerKoller,HewsonBauerOguri}

\section{Thermoelectric transport of  dilute magnetic alloy}
\label{sec:magnetic_alloy}

The Kondo effect in dilute magnetic alloy  (MA) has been studied  for 
a wide variety of  $3d$, $4f$, and  $5f$ electron systems. 
Our formulation  can also  be applied to these original Kondo systems. 
In this subsection,  
we provide the microscopic description of  the  Fermi-liquid corrections for 
 magneto-transport properties of  dilute magnetic alloys away from half-filling. 
Specifically, we calculate 
the electric resistance $R_\mathrm{MA}^{}$, 
thermoelectric power $S$, and thermal conductivity  $\kappa$   
using the  linear-response formulas,\cite{CostiThermo,ao1990PRB}
\begin{align}
\frac{1}{R_\mathrm{MA}^{}} 
=   & \  
\frac{1}{2R_\mathrm{MA}^{0}} \sum_\sigma \mathcal{L}_{0,\sigma}^{} \,, 
 \quad \  \ 
\mathcal{S}\, =\, \frac{-1}{|e|T} 
\frac{\sum_{\sigma} \mathcal{L}_{1,\sigma}^{}}
{\sum_{\sigma} \mathcal{L}_{0,\sigma}^{}} \,,
\\ 
\kappa \,= & \  
\frac{\eta_0^{}}{T}
\left(  
\sum_{\sigma}
 \mathcal{L}_{2,\sigma}^{} 
- 
\frac{ \left(
\sum_{\sigma}
\mathcal{L}_{1,\sigma}\right)^2}{\sum_{\sigma}
 \mathcal{L}_{0,\sigma}^{}} \right) .
\label{eq:thermal_coefficients}
\end{align}
The coefficients are defined by 
\begin{align}
\mathcal{L}_{n,\sigma} = 
\int_{-\infty}^{\infty}  
d\omega\, 
\frac{\omega^n}{\pi \Delta A_{\sigma}(\omega,T)}
\left( -\frac{\partial f(\omega)}{\partial \omega}\right)  .  
\label{eq:L_thermal}
\end{align}
The factor  $R_\mathrm{MA}^{0}$  is 
the unitary-limit value of the electric resistance at zero field. 
Similarly,  $\eta_0^{}$ is defined  such that 
 the  $T$-linear thermal conductivity should take  the following form  in the unitary limit, 
\begin{align}
\kappa_0 =  \frac{2\pi^2\eta_0^{}}{3} \, T\;.
\end{align}
Note that thermoelectric transport through quantum dots can 
also be determined in a similar way\cite{GuttmanBergman}.   
\footnote{
Thermoelectric response of quantum dots is described by the coefficients: 
$\mathcal{L}_{n,\sigma}^\mathrm{QD} = 
-\int_{-\infty}^{\infty}  
d\omega\, 
\omega^n
 A_{\sigma}(\omega,T) \,
\frac{\partial f(\omega)}{\partial \omega}
$}

\subsection{Coefficients  $\mathcal{L}_{n,\sigma}$ for finite magnetic fields}

The coefficients  $\mathcal{L}_{n,\sigma}$, 
defined in   Eqs.\ \eqref{eq:L_thermal}, 
are written in terms of  the inverse  spectral function 
which physically represents the relaxation time due to 
 the many-body scattering by the impurity at equilibrium  $eV=0$. 
For this spectral function, we use the low-energy 
asymptotic form given in Eq.\ \eqref{eq:A_including_T_eV}, 
\begin{align}
& \frac{A_{\sigma}(0, 0, 0)}{A_{\sigma}(\omega, T, eV=0)} \, 
\nonumber \\
 =   &  \ 
1 
-
\frac{\pi}{3 \Delta \rho_{d\sigma}^{}} 
\left(
 \frac{3}{2} \cos 2 \delta_{\sigma}\,\chi_{\uparrow\downarrow}^2
- 
\frac{\sin 2\delta_{\sigma}}{2\pi}\,
\frac{\partial \chi_{\uparrow\downarrow}}{\partial \epsilon_{d,-\sigma}} 
 \right)   \left(\pi T\right)^2 
\nonumber \\
& 
- \frac{\sin 2\delta_{\sigma}\, \chi_{\sigma\sigma}}{\Delta \rho_{d\sigma}^{}} 
\,\omega
+ 
\frac{\pi}{ \Delta\rho_{d\sigma}^{}}
\Biggl[
\bigl( 2+\cos 2\delta_{\sigma} \bigr)
\, \chi_{\sigma\sigma}^2
\nonumber \\
& 
-
\frac{1}{2}\cos 2\delta_{\sigma}
 \chi_{\uparrow\downarrow}^2
+  
\frac{\sin 2\delta_{\sigma}}{2\pi}
\, \frac{\partial \chi_{\sigma\sigma}}{\partial \epsilon_{d\sigma}^{}} 
\Biggr] 
\,\omega^2  
+ \cdots .
\label{eq:current_int3_invert}
\end{align}
Note that $\pi \Delta \rho_{d\sigma}^{} = \sin^2 \delta_{\sigma}$. 
Using also  the integration formulas, 
\begin{subequations}
\begin{align}
 \int_{-\infty}^{\infty} d\omega\,  \omega^2
\left(- \frac{\partial f(\omega)}{\partial \omega} \right) 
= & \ \frac{1}{3} \,(\pi T)^2,
\\
 \int_{-\infty}^{\infty} d\omega\,  \omega^4
\left(- \frac{\partial f(\omega)}{\partial \omega} \right) 
=& \  \frac{7}{15}\,(\pi T)^4 \;,
\end{align}
\end{subequations}
we  obtain   $\mathcal{L}_{n,\sigma}$ for  $n=0,1,$ and $2$: 
\begin{align}
&\mathcal{L}_{0,\sigma}  =  
\frac{1}{\pi \Delta \rho_{d\sigma}^{}}
\Biggl[
1+
\frac{\pi}{ 3\Delta \,\rho_{d\sigma}^{}} 
\biggl\{
\bigl(2+\cos 2\delta_{\sigma}  \bigr)\,
\chi_{\sigma\sigma}^2
\nonumber \\
& 
\quad  - 
2\cos 2 \delta_{\sigma}\,
\chi_{\uparrow\downarrow}^2
+ 
\frac{\sin 2\delta_{\sigma}}{2\pi}\,
\left(
 \frac{\partial \chi_{\sigma\sigma}}{\partial \epsilon_{d\sigma}^{}} 
 + \frac{\partial \chi_{\uparrow\downarrow}}{\partial \epsilon_{d,-\sigma}} 
\right)
\, \biggr\}
  \left(\pi T\right)^2 
\Biggr] 
\nonumber \\
&  \quad +  O(T^4) ,
\label{eq:L0_result}
\\
\nonumber \\
&\mathcal{L}_{1,\sigma}  = 
- \frac{2\pi}{3} 
\frac{\cot \delta_{\sigma}}{\pi\Delta \rho_{d\sigma}^{}} 
\, \chi_{\sigma\sigma}\,(\pi T)^2  +  O(T^4) \;, 
\label{eq:L1_result}
\\
\nonumber \\
& \mathcal{L}_{2,\sigma} 
 =    
\frac{\left(\pi T\right)^2}{3\pi \Delta \rho_{d\sigma}^{}}
\Biggl[
1 
+ 
\frac{7\pi}{ 5\Delta\rho_{d\sigma}^{}}
\biggl\{
\bigl( 2+\cos 2\delta_{\sigma} \bigr) \, \chi_{\sigma\sigma}^2
\nonumber \\
& \quad 
- \frac{6}{7}\cos 2\delta_{\sigma}  \chi_{\uparrow\downarrow}^2
+  
\frac{\sin 2\delta_{\sigma}}{2\pi}
\left(
  \frac{\partial \chi_{\sigma\sigma}}{\partial \epsilon_{d\sigma}^{}} 
 +
 \frac{5}{21}
 \frac{\partial \chi_{\uparrow\downarrow}}{\partial \epsilon_{d,-\sigma}} 
\right)
\biggr\}\left(\pi T\right)^2 
\Biggr]  
\nonumber \\
& \quad +  O(T^6).
\label{eq:L2_result}
\end{align}
The derivatives in the last part of  $\mathcal{L}_{2,\sigma}$ can also be written as 
\begin{align}
  \frac{\partial \chi_{\sigma\sigma}}{\partial \epsilon_{d\sigma}^{}} 
 +
 \frac{5}{21}
 \frac{\partial \chi_{\uparrow\downarrow}}{\partial \epsilon_{d,-\sigma}} 
=
  \frac{\partial \chi_{\sigma\sigma}}{\partial \epsilon_{d}^{}} 
-\frac{8}{21}  \frac{\partial \chi_{\uparrow\downarrow}}{\partial \epsilon_{d}^{}} 
+ \sigma \,\frac{13}{21}  
\frac{\partial \chi_{\uparrow\downarrow}}{\partial h} 
.  
\end{align}
The asymptotic exact low-temperature form of  
the  transport coefficients  $R_\mathrm{MA}^{}$,   $S$, and  $\kappa$ 
for finite magnetic fields can be explicitly written down  using  
 Eqs.\ \eqref{eq:L0_result}--\eqref{eq:L2_result} 
for Eq.\ \eqref{eq:thermal_coefficients}.  
As those general Fermi-liquid expressions  become rather lengthy  for  $h\neq 0$, 
we explicitly write  in the following the transport coefficients of dilute 
magnetic alloys  at  zero magnetic field.

\subsection{Thermoelectric transport coefficients at zero magnetic field}

The electric resistance takes the following form  at zero magnetic field $h=0$ 
away from half-filling,  
\begin{align}
&\frac{R_\mathrm{MA}^{}}{{R_\mathrm{MA}^{0}}}
\, =    \,  
\sin^2 \delta - 
c_{R}^\mathrm{MA}
%
 \left(\pi T\right)^2    +  O(T^4)\;, 
\\ 
%
&c_{R}^\mathrm{MA} 
=   
\frac{\pi^2}{3} 
\left[
\bigl(2+\cos 2\delta  \bigr)\, \chi_{\uparrow\uparrow}^2
-  2\cos 2 \delta\, \chi_{\uparrow\downarrow}^2
+ 
\frac{\sin 2\delta}{2\pi}\,
\frac{\partial \chi_{\uparrow\uparrow}}{\partial \epsilon_{d}^{}} 
 \right]  .
 \end{align}
Note that it  reproduces the results of Yamada-Yosida in the particle-hole symmetric case,
\cite{YamadaYosida2,YamadaYosida4}
\begin{align}
\frac{R_\mathrm{MA}^{}}{{R_\mathrm{MA}^{0}}}
\, \xrightarrow{\,\xi_d\to 0 \,}  & \   
1\, - \, 
\frac{\widetilde{\chi}_{\uparrow\uparrow}^2 
+ 2 \widetilde{\chi}_{\uparrow\downarrow}^2}{3}
  \left(\frac{\pi T}{\Delta}\right)^2    +  O(T^4)\;.
\end{align}

We introduce the dimensionless coefficient $C_{R}^\mathrm{MA}$ 
which is scaled by  $T^*=1/(4 \chi_{\uparrow\uparrow})$, 
the characteristic energy at $h=0$;
 \begin{align}
C_{R}^\mathrm{MA} \, \equiv& \    c_{R}^\mathrm{MA} (T^*)^2
\,=\, 
\frac{\pi^2}{48} 
\left(
\, 
W_{R}^\mathrm{MA} 
\,+\, 
\Theta_{\uparrow\uparrow}^{} 
\right)
\label{eq:C_R_rescale}
 \;,\\
W_{R}^\mathrm{MA} 
\,\equiv & \ 
2+\cos 2\delta  -  2 \left( R_{W}^{} -1 \right)^2  \cos 2 \delta
\label{eq:W_R}
\;.
\end{align}
Here, $W_{R}^\mathrm{MA}$  represents the  contribution of 
the two-body fluctuation, 
 and  $\Theta_{\uparrow\uparrow}^{}$ 
which is defined  in Eq.\ \eqref{eq:Theta_UU}  
represents  the contribution of three-body fluctuations. 
The coefficient $C_{R}^\mathrm{MA}$ does not depend 
on the anti-parallel component of three-body correlation 
$\Theta_{\uparrow\downarrow}^{}$ 
similarly to the coefficient  $C_{T}^{}$ for quantum dots 
given  in Eq.\ \eqref{eq:dimensionless_Ct_h=0}.

In our formulation,  low-temperature expansion of  the thermopower $S$ 
can be carried out just for the leading  $T$-linear term.
It is determined by the derivative of the density of states 
at the Fermi energy  $\omega=0$, 
and can be written in the following form at zero magnetic field, 
\begin{align}
\mathcal{S} \,=& \  
\frac{\pi^2}{3} \,\frac{\rho_{d}'}{\rho_d^{}}\, \frac{T}{|e|}
 \ +O(T^3) \;.
\end{align}
Here  $\rho_{d}' $ is the derivative with respect to $\omega$, 
defined in  Eq,\ \eqref{eq:rho_d_omega_2}.

\begin{figure}[t]
\begin{minipage}{1\linewidth}
\includegraphics[width=0.6\linewidth]{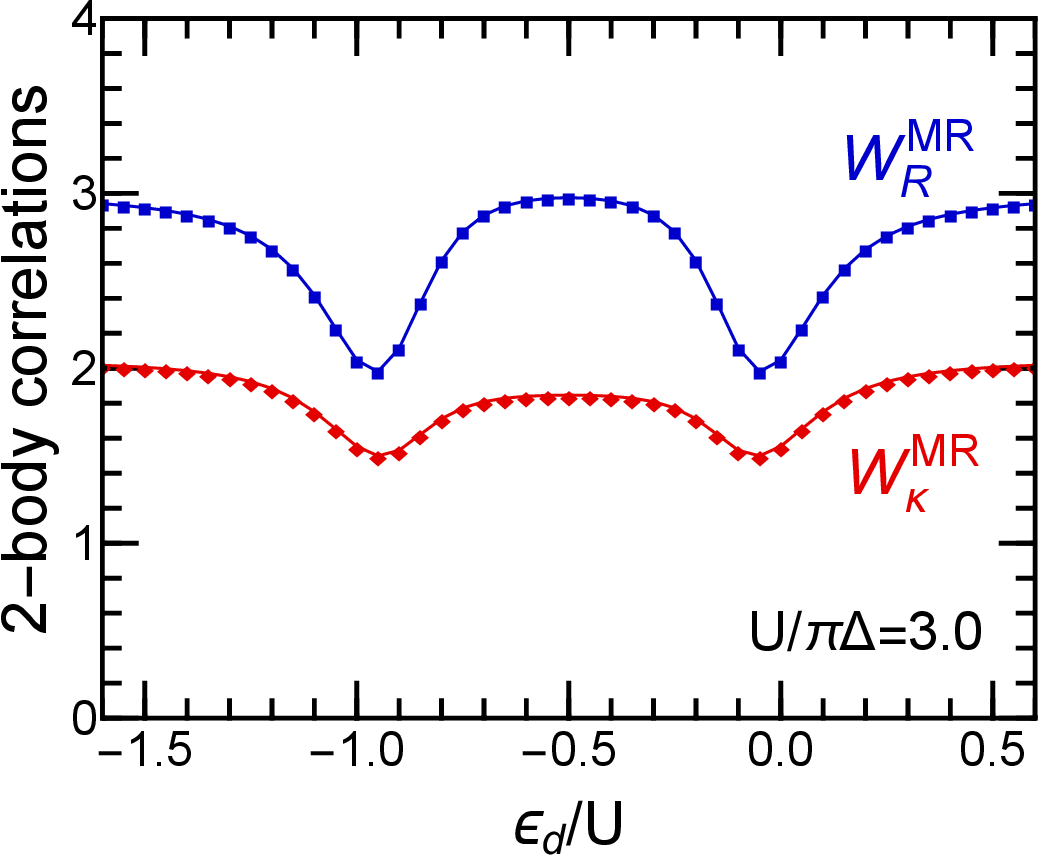}
\end{minipage}
 \caption{
(Color online) 
Contributions of the two-body fluctuation parts  
$W_{R}^\mathrm{MA}$ and $W_{\kappa}^\mathrm{MA}$, 
defined in Eqs.\ \eqref{eq:W_R} and \eqref{eq:W_kappa}, 
are plotted vs $\epsilon_{d}/U$ for $U=3.0 \pi \Delta$.
}
 \label{fig:cos_part_for_DMR}
\end{figure}

\begin{figure}[t]
\begin{minipage}{1\linewidth}
\includegraphics[width=0.6\linewidth]{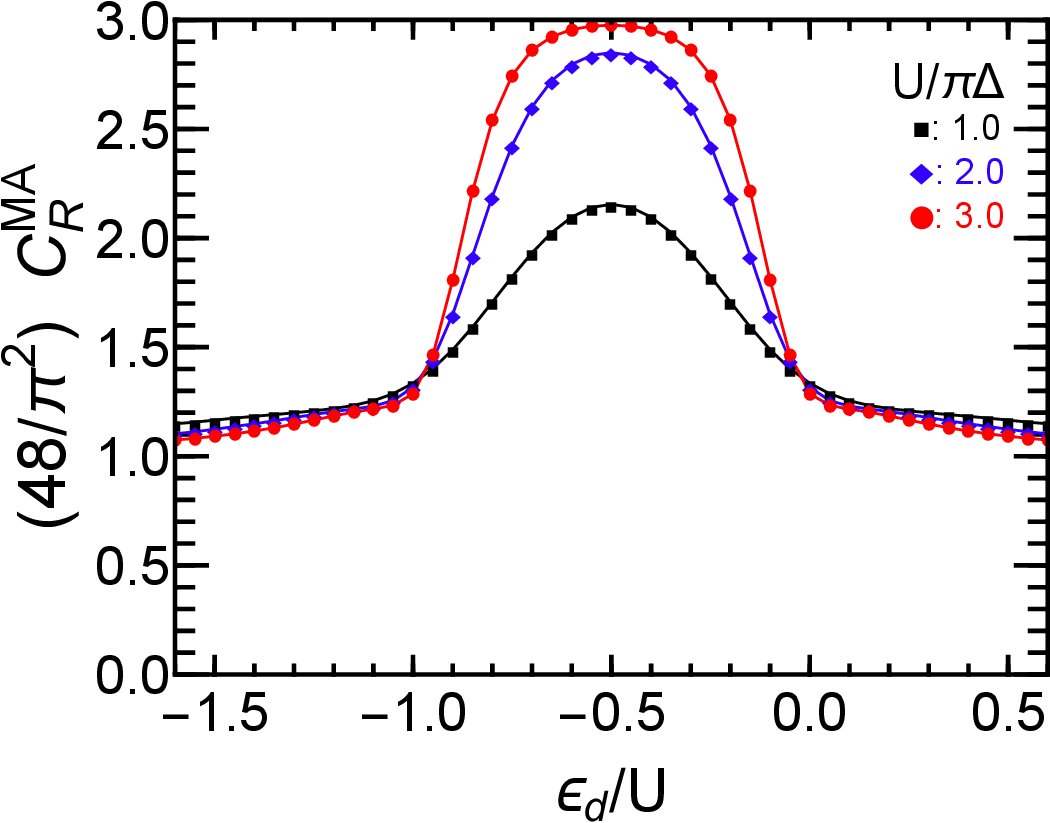}
\\
\includegraphics[width=0.6\linewidth]{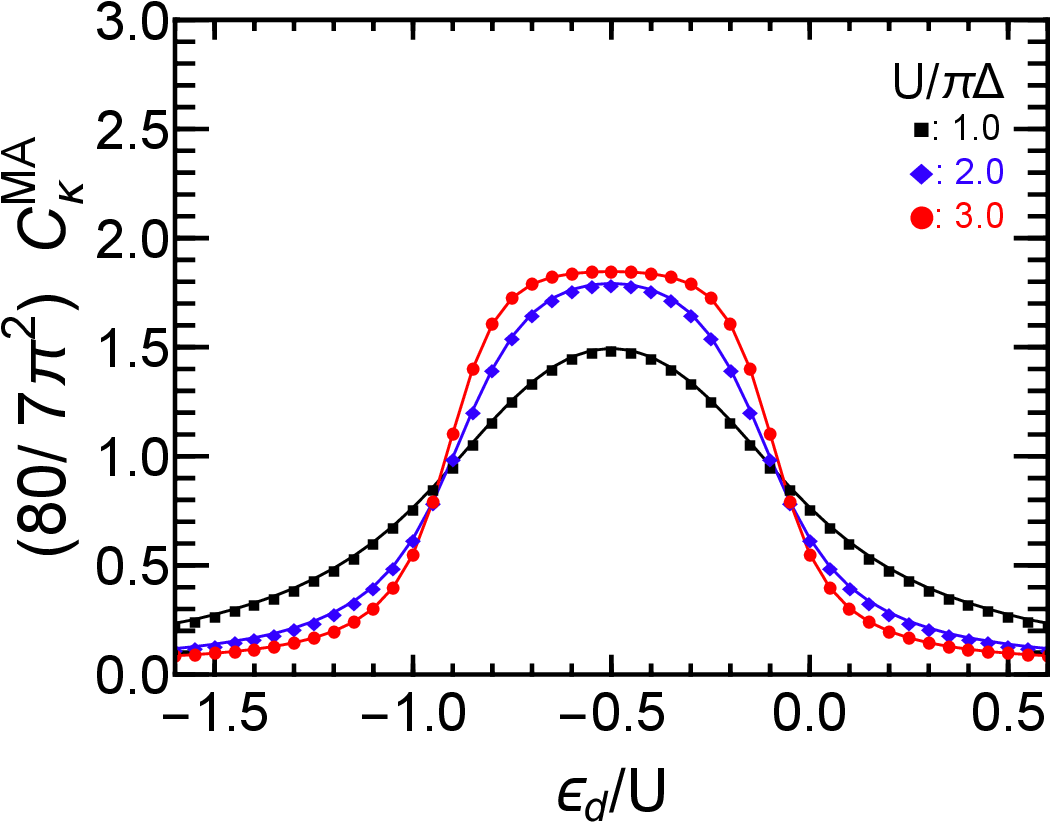}
\end{minipage}
 \caption{
(Color online) 
Coefficients  $C_{R}^\mathrm{MA}$ and $C_{\kappa}^\mathrm{MA}$, 
defined in Eqs.\ \eqref{eq:C_R_rescale} and \eqref{eq:C_kappa_rescale}, 
for the electric resistance and thermal conductivity are plotted 
vs $\epsilon_d/U$ for $U/\pi \Delta = 1.0$, $2.0$, and $3.0$.
These coefficients converge towards 
$(48/\pi^2) \,C_{R}^\mathrm{MA} \to 1$ and 
 $(80/7\pi^2)\, C_{\kappa}^\mathrm{MA} \to 1/21$ 
for $|\epsilon_{d}^{}| \to \infty$.
}

 \label{fig:Cr_Ck}
\end{figure}

The thermal conductivity $\kappa$ can be deduced up terms of order $T^3$ 
through  Eq.\ \eqref{eq:thermal_coefficients}.   
At $h=0$,  the leading $T^4$ term of the ratio 
 $(\sum_{\sigma}\mathcal{L}_{1,\sigma})^2/
\sum_{\sigma}\mathcal{L}_{0,\sigma}$  is given by 
\begin{align}
\frac{\left(\sum_{\sigma}
\mathcal{L}_{1,\sigma}\right)^2}{\sum_{\sigma} \mathcal{L}_{0,\sigma}} 
\,=\, 
\frac{8\pi^2}{9} 
\cot^2 \delta \, \chi_{\uparrow\uparrow}^2\, 
\frac{(\pi T)^4 }{\pi\Delta \rho_d^{} }
\, + O(T^6) \;. 
\end{align}
Using this ratio  and  $\mathcal{L}_{2,\sigma}$ given in Eq.\ \eqref{eq:L2_result}, 
 we can explicitly  write  the thermal conductivity at zero magnetic field,   
\begin{align}
\kappa 
\,=& \  
\frac{2\pi^2\eta_0^{}}{3}
\frac{T}{\sin^2 \delta} \,  
\left[\,
1 
+ 
\frac{c_{\kappa}^\mathrm{MA}}{\sin^2 \delta} 
%
\, \left(\pi T\right)^2  
\,\right] + O(T^5)\,, \\
c_{\kappa}^\mathrm{MA}
\,\equiv &\ 
\frac{7\pi^2}{5}
\biggl[\,
\frac{32+ 11\cos 2\delta}{21}
\, \chi_{\uparrow\uparrow}^2
- \frac{6}{7}\cos 2\delta  \,\chi_{\uparrow\downarrow}^2
\nonumber \\
& \qquad \quad   
+  
\frac{\sin 2\delta}{2\pi}
\left(
\frac{\partial \chi_{\uparrow\uparrow}}{\partial \epsilon_{d}^{}} 
-
\frac{8}{21}\,
\frac{\partial \chi_{\uparrow\downarrow}}{\partial \epsilon_{d}} 
\right)
\,\biggr]  .
\label{eq:thermal_conductivity_FL}
\end{align}
Here, the sign and normalization of  $c_{\kappa}^\mathrm{MA}$ 
 has been determined in such a way that the thermal resistivity, 
the reciprocal of $\kappa$,  is  written in the following  form, 
\begin{align}
\frac{1}{\kappa} 
\,=& \  
\frac{3}{2\pi^2\eta_0^{} T}
\left[\,
 \sin^2 \delta
\,- \,c_{\kappa}^\mathrm{MA}
\, \left(\pi T\right)^2  
\,\right] + O(T^3)\,.
\end{align}
In the particle-hole symmetric case, 
Eq.\ \eqref{eq:thermal_conductivity_FL} 
 reproduces the expression that can be deduced from the result of Yamada-Yosida, 
\begin{align}
\kappa 
\, \xrightarrow{\,\xi_d\to 0 \,}  & \  
\frac{2\pi^2\eta_0^{}}{3}
\, T \, 
\Biggl[
1
+
\frac{7\widetilde{\chi}_{\uparrow\uparrow}^2
+ 6  \,\widetilde{\chi}_{\uparrow\downarrow}^2
}{5}  \left( \frac{\pi T}{\Delta}\right)^2  
\Biggr] + O(T^5) \,.
\end{align}

We also introduce the dimensionless coefficient in the same 
way as that for the coefficient  $C_{R}^\mathrm{MA}$ of  the electric resistance 
\begin{align}
& 
\!\!\!\!\!
C_{\kappa}^\mathrm{MA} 
\equiv   c_{\kappa}^\mathrm{MA} (T^*)^2
\! = 
\frac{7\pi^2}{80}
\left(\!
W_{\kappa}^\mathrm{MA}
+
\Theta_{\uparrow\uparrow}^{} 
\! +
\frac{8}{21}
\Theta_{\uparrow\downarrow}^{} 
\! \right)  \! ,  \!\! 
\label{eq:C_kappa_rescale}
\\
& \!\!\!\!
W_{\kappa}^\mathrm{MA} 
\,\equiv \,  
\frac{32+ 11\cos 2\delta}{21}
\, - \frac{6}{7}  \left( R_{W}^{} -1 \right)^2  \cos 2\delta.
\label{eq:W_kappa}
\end{align}
Both the parallel and anti-parallel components of the three-body fluctuation, 
 $\Theta_{\uparrow\uparrow}^{}$ and   $\Theta_{\uparrow\downarrow}^{}$ 
contribute to the thermal conductivity. 
The dependence of these three-body correlation functions    
on $\epsilon_{d}^{}$ has been shown in 
Figs.\ \ref{fig:Wt_Wv_Theta_for_u3} and  \ref{fig:FL_UU_UD_several_U}.

We also show 
 the $\epsilon_{d}^{}$ dependence of the two-body-fluctuation part 
of the electric resistance and the thermal conductivity,  
$W_{R}^\mathrm{MA}$ and $W_{\kappa}^\mathrm{MA}$,  
in Fig.\  \ref{fig:cos_part_for_DMR}  for $U=3.0 \pi \Delta$.  
The contributions of the two-body fluctuation 
 reach the unitary-limit 
value  $W_{R}^\mathrm{MA} \xrightarrow{\mathrm{Kondo}} 3$ and 
$W_{\kappa}^\mathrm{MA}  \xrightarrow{\mathrm{Kondo}} 13/7$ 
in the Kondo regime where   $\delta \to \pi/2$ and $R_{W} \to 2$.
Both $W_{R}^\mathrm{MA}$ and $W_{\kappa}^\mathrm{MA}$ 
do not change  sign  
in contrast to  $W_{T}^{}$ and $W_{V}^{}$ for the 
quantum-dot conductance shown in Fig.\ \ref{fig:Wt_Wv_Theta_for_u3}  
but  have  a  minimum at the transient region between 
the Kondo regime and empty  (fully-occupied) orbital regime 
at $\epsilon_{d}^{} \simeq 0$ ($\epsilon_{d}^{} \simeq -U$). 
In the  opposite empty-orbital (EO)
  limit  $|\epsilon_{d}| \to \infty$
at which   $\cos 2\delta \to 1$ and $R_{W} \to 1$, 
the two-body contributions   
approach  $W_{R}^\mathrm{MA}  \xrightarrow{\mathrm{EO}} 3$   
and   $W_{\kappa}^\mathrm{MA} \xrightarrow{\mathrm{EO}} 43/21$.

The coefficients $C_{R}^\mathrm{MA}$ and $C_{\kappa}^\mathrm{MA}$ 
are determined by  the sum of the two-body and  three-body contributions. 
Figure \ref{fig:Cr_Ck} shows the NRG result. 
Contributions of the two-body fluctuations which enter through 
$W_{R}^\mathrm{MA}$  and $W_{\kappa}^\mathrm{MA}$ 
dominate  for  $-1.0 \lesssim \epsilon_{d}/U \lesssim 0.0$. 
In the Kondo regime, these contributions determine the 
total value such that  $(80/7\pi^2) C_{R}^\mathrm{MA} \to 3$ and
 $(80/7\pi^2) C_{\kappa}^\mathrm{MA} \to 13/7$. 
However, outside of this region  $|\epsilon_{d}/U+0.5| \gtrsim 1.0$, 
the three-body fluctuations, especially the parallel spin 
component $\Theta_{\uparrow\uparrow}^{}$,   
give negative contributions and suppress  the net value of 
 $C_{R}^\mathrm{MA}$ and $C_{\kappa}^\mathrm{MR}$.
In the  $|\epsilon_{d}^{}|\to \infty$ limit 
of  the  empty (or fully-occupied) orbital regime,  
these coefficients converge towards 
\begin{align}
\lim_{|\epsilon_{d}^{}| \to \infty} \frac{48}{\pi^2}\, C_{R}^\mathrm{MA} = 1, 
\quad \ \  
\lim_{|\epsilon_{d}^{}| \to \infty} \frac{80}{7\pi^2}\, 
C_{\kappa}^\mathrm{MA} = \frac{1}{21} .  
\end{align}

\begin{figure}[t]
\begin{minipage}{1\linewidth}
\includegraphics[width=0.6\linewidth]{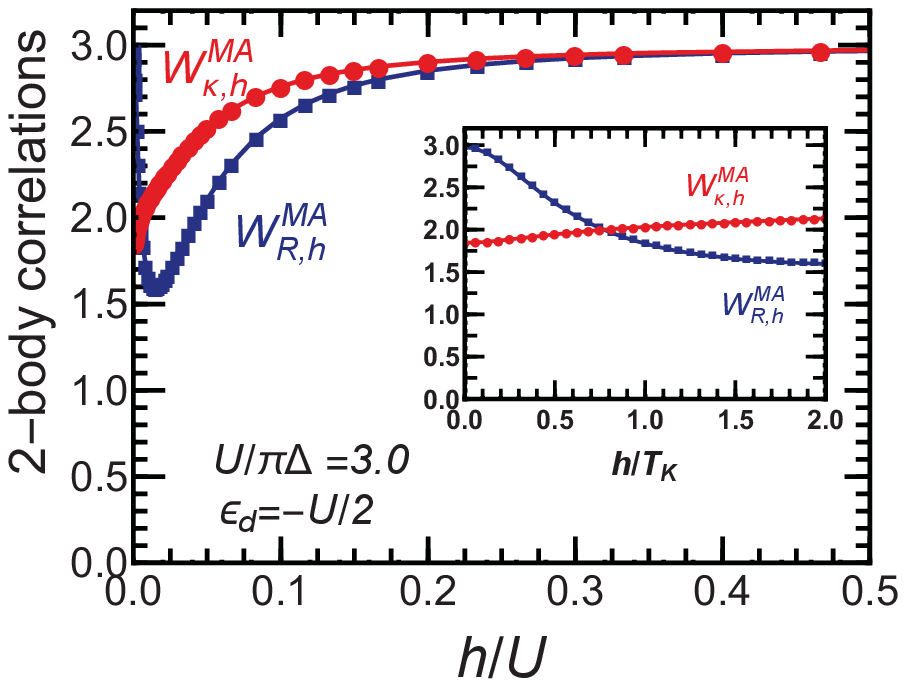}
\\
\includegraphics[width=0.6\linewidth]{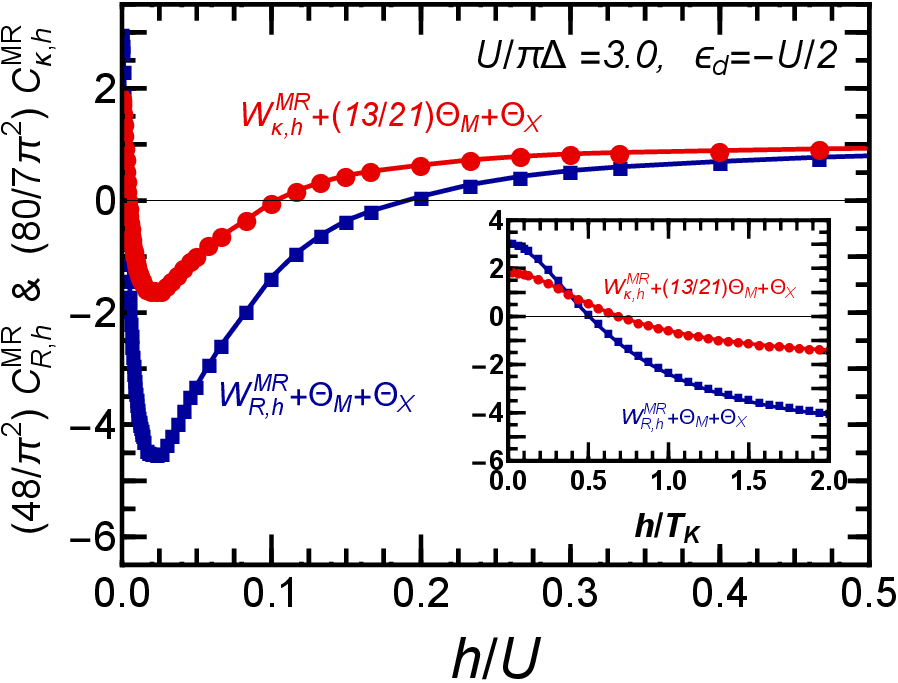}
\end{minipage}
 \caption{
(Color online) 
Thermoelectric transport coefficients are plotted vs $h/U$ 
 at half-filling $\epsilon_{d}^{}=-U/2$ for  $U/\pi \Delta = 3.0$.   
 Inset shows an enlarged view of the small $h$ region, 
for which the horizontal axis is scaled by $T_K = 0.02 \pi \Delta$ ($=0.0066U$)  
determined at  $h=0$. 
Upper panel shows the contributions of  two-body fluctuations 
$W_{R,h}^\mathrm{MR}$ and $W_{\kappa,h}^\mathrm{MR}$. 
Lower panel shows the coefficients 
$(48/\pi^2) \,C_{R,h}^\mathrm{MA} = 
W_{R,h}^\mathrm{MA}  
+ \Theta_{M}^{} + \Theta_{X}^{}$, 
and  
$(80/7\pi^2) \,C_{\kappa,h}^\mathrm{MA} = 
W_{\kappa,h}^\mathrm{MA} 
+(13/21) \,\Theta_{M}^{} + \Theta_{X}^{}$  
for the electric resistance $R_\mathrm{MA}^{}$  and thermal conductivity $\kappa$. 
For the thermal conductivity, 
the two-body contribution  
becomes smallest  $W_{\kappa,h}^\mathrm{MA}=13/7$  at $h=0$ 
and it increases with $h$.   
Magnetic-field dependence of the three-body contributions  $\Theta_{M}^{}$ 
and   $\Theta_{X}^{}$ are shown 
 in Fig.\  \ref{fig:FL_mag}. }
 \label{fig:FL_mag_thermo}
\end{figure}

\subsection{Thermoelectric effects at finite magnetic fields}

We next examine  thermoelectric effects at finite magnetic fields, 
specifically at half-filling  $\epsilon_{d}^{}=-U/2$. 
The thermopower vanishes $\mathcal{S}=0$ at half-filling also for $h \neq 0$ 
because the contributions of the two different spin states 
cancel out $\mathcal{L}_{1\uparrow}+\mathcal{L}_{1\downarrow}=0$.
It can also be explained from 
the fact  $\rho_{d\uparrow}'+\rho_{d\downarrow}' =0$. 
In this case, the density of states can be written in terms of 
the induced local moment, as 
$\rho_{d}^{}=\cos^2 \left(\frac{\pi m_d}{2}\right)/\pi \Delta$.

The magneto-resistance and thermal conductivity can be expressed 
in the following form at half-filling, 
\begin{align}
&R_\mathrm{MA}^{}
 =    
{R_\mathrm{MA}^{0}} 
\left[\,
\cos^2 \left(\frac{\pi m_d}{2}\right) \, - \, 
C_{R,h}^\mathrm{MA}
 \left(\frac{\pi T}{T^*}\right)^2  \, \right]
 +  O(T^4)\;, 
\\
&\kappa 
 =   
\frac{2\pi^2\eta_0^{}}{3}
\frac{T}{\cos^2 \left(\frac{\pi m_d}{2}\right) } \,  
\left[\,
1 
+ 
\frac{C_{\kappa,h}^\mathrm{MA}}{\cos^2 \left(\frac{\pi m_d}{2}\right) } 
\, \left(\frac{\pi T}{T^*}\right)^2  
\,\right] \nonumber \\ 
& \qquad  + O(T^5)\,.
\label{eq:thermal_conductivity_FL_h}
 \end{align}
Here,    $T^*=1/(4 \chi_{\uparrow\uparrow})$  is the field-dependent energy scale 
used in the previous section. 
 The dimensionless coefficient for the electric resistance $R_\mathrm{MA}^{}$  is given by  
 \begin{align}
&
\!\!
 C_{R,h}^\mathrm{MA} \,\equiv  \,  
\frac{\pi^2}{48} 
\left(
\, 
W_{R,h}^\mathrm{MA} 
+ \Theta_{M}^{}  +  \Theta_{X}^{} 
\right),
\label{eq:C_R_rescale_h}
\\
& 
\!\!
W_{R,h}^\mathrm{MA} 
\,=   \,   
2-\cos  \left(\pi m_d\right)  
+ 2 \left( R_{W}^{} -1 \right)^2  \cos  \left(\pi m_d\right) ,
\!\!
\label{eq:W_R_h}
\end{align}
and that for  the  thermal conductivity $\kappa$ is 
\begin{align}
& C_{\kappa,h}^\mathrm{MA} 
\equiv 
\frac{7\pi^2}{80}
\left(
W_{\kappa,h}^\mathrm{MA}
+
\frac{13}{21}\,
\Theta_{M}^{} 
 +  \Theta_{X}^{}  
\right)  , 
\label{eq:C_kappa_rescale_h}
\\
& W_{\kappa,h}^\mathrm{MA} 
=  
2 - \cos \left(\pi m_d\right)
 + \frac{6}{7}  \left( R_{W}^{} -1 \right)^2  \cos  \left(\pi m_d\right) .
\!\!
\label{eq:W_kappa_h}
\end{align}
The parameters   $W_{R,h}^\mathrm{MA}$ and  
$W_{\kappa,h}^\mathrm{MA}$  represent the contribution of 
the two-body fluctuations, 
as determined by the induced local magnetization  $m_d$ and  
the Wilson ratio $R_{W}^{}$. 
The three-body contributions 
 $\Theta_{M}^{}$  and  $\Theta_{X}^{}$  
have been defined in Eqs.\ \eqref{eq:Theta_M} and  \eqref{eq:Theta_X}, 
respectively,  and the magnetic field dependence of 
these functions have also been described in Fig.\  \ref{fig:FL_mag}. 
For large Coulomb interactions $U \gtrsim 2 \pi \Delta$ 
at zero field $h=0$, the dimensionless coefficients take the values,         
  $(48/\pi^2)C_{R,h}^\mathrm{MA}  =  3$ and 
  $(80/7\pi^2)C_{\kappa,h}^\mathrm{MA} = 13/7$,   
as $m_{d}=0$,  $R_{W}^{}\to 2$,   $\Theta_{M}^{}\to 0$, 
and  $\Theta_{X}^{}\to 0$. 
In the high-field limit $h \to \infty$,  
these two coefficients 
approach  the noninteracting values, 
  $(48/\pi^2)C_{R,h}^\mathrm{MA}  \to  1$ and 
  $(80/7\pi^2)C_{\kappa,h}^\mathrm{MA} \to 1$,   
as  $m_{d}\to 1$,  $R_{W}^{}\to 1$,  $\Theta_{M}^{}\to 0$, 
and  $\Theta_{X}^{} \to -2$.

Figure \ref{fig:FL_mag_thermo} shows the $h$ dependence of these parameters 
for $U=3.0 \pi \Delta$. 
The two-body contributions 
  $W_{R,h}^\mathrm{MA}$  and   $W_{\kappa,h}^\mathrm{MA}$ 
are positive and vary  in a  relatively small range from the high-field value  $3.0$.
For the thermal conductivity,   
it takes a minimum  $W_{\kappa,h}^\mathrm{MA}=13/7$  at $h=0$ 
and increases with $h$.   The electric resistance part 
has a minimum $W_{R,h}^\mathrm{MA}\simeq 1.6$ at a finite field $h \simeq 0.015U$. 
In contrast, the three-body contribution  $\Theta_{M}^{}$   has a much bigger  
dip  as shown in Fig.\  \ref{fig:FL_mag}.  Therefore, 
the coefficients  $C_{R,h}^\mathrm{MA}$ and $C_{\kappa,h}^\mathrm{MA}$ 
become negative in an intermediate region of  the magnetic fields, 
typically $T_K \lesssim h \lesssim 0.1U$,  
while both of these two coefficients are positive outside of this region.
The behavior in the high-filed limit is determined by the two-body contributions 
$W_{R,h}^\mathrm{MA}$  and $W_{\kappa,h}^\mathrm{MA}$, 
and the three-body contributions from  $\Theta_{X}^{}$.

\section{Summary}
\label{sec:summary_III}

In summary,  we have studied low-energy properties  
of  the steady-state Keldysh Green's function 
in the situations where  both the  bias voltage and magnetic field are finite.  
The  $(eV)^2$ real part of the self-energy 
has been deduced from the non-equilibrium Ward identities,  
using the previous result of  the $\omega^2$ real part of the self-energy.
\cite{FilipponeMocaVonDelftMora,ao2017_2_PRB} 
We have also shown that the  $(eV)^2$-correction 
and  the  $T^2$ correction of the self-energy 
are  determined by a common correlation function, 
  $\widehat{D}^2  \Sigma_{\mathrm{eq},\sigma}^{--}(\omega)
\equiv \Psi_{\sigma}^{--}(\omega)$. 
It indicates that these two corrections arise as a  linear combination, 
 $(\pi  T)^2+(3/4)(eV)^2$,  in the case where 
the bias voltages are applied such that  $\alpha=0$. 
This output has previously been pointed out by FMvDM,\cite{FilipponeMocaVonDelftMora}  
and our result provides an alternative proof.

We have applied the low-energy asymptotic form of the Green's function 
given in Eqs.\ \eqref{eq:self_imaginary}--\eqref{eq:self_real_ev_mag}   
to explore the non-linear magneto-conductance of quantum dots,   
and also the  electric resistance and thermal conductivity of dilute magnetic alloys. 
The Fermi-liquid corrections in the general case are determined 
by  two different types of contributions:  
the two-body-fluctuation contribution 
described by the susceptibilities  $\chi_{\sigma\sigma'}$ 
and the three-body-fluctuation contribution  
enters through the non-linear  susceptibilities 
$\chi_{\sigma_1\sigma_2\sigma_3}^{[3]}$.  
Using the NRG,  we have examined  the  $T^2$ and  $(eV)^2$ corrections 
of  the transport coefficients  for some  particle-hole asymmetric cases.
At zero field,   
the two-body fluctuations dominate the corrections  in the Kondo regime 
where  $n_{d\uparrow}^{}+n_{d\downarrow}^{} \simeq 1$
and the Wilson ratio is almost saturated  $R_{W}^{} \simeq 2$.
The contribution of  the three-body fluctuations 
become significant  far away from half-filling, 
especially in the valence-fluctuation  regime and empty-orbital regime.  
Furthermore, we have also reexamined  a controversial 
problem of  the zero-bias peak of $dI/dV$ at finite magnetic fields.
\cite{HewsonBauerOguri,HewsonBauerKoller,FilipponeMocaVonDelftMora}
In this case, the three-body fluctuations give a  contribution 
that is comparable to the two-body contribution even for small magnetic fields.
The three-body contribution also  plays essential role 
in  a splitting of the zero-bias peak occurring at a magnetic field, 
 $h \sim T_K$, of the order of the Kondo energy scale $T_K$. 
This observation based on the formula Eq.\ \eqref{eq:c_V_mag_half} 
 is consistent with our previous result of the 
  second-order renormalized perturbation theory.\cite{HewsonBauerOguri}

Furthermore, we have also  studied the Fermi-liquid corrections for  
 the magneto-resistance  and  thermal conductivity 
of dilute magnetic alloys away from half-filling.  
The NRG result shows that the contributions of 
the two-body fluctuations dominate in the Kondo regime, whereas  
in the valence-fluctuation regime far away from half-filling    
the  contribution of three-body fluctuations 
become  comparable to the two-body contribution. 
We have also provided the formulas for higher-order Fermi-liquid corrections 
for  the  Anderson impurity with $N$ {\it flavor} components  
in Appendix \ref{sec:D2_Psi_causal_detail}.
Further details of the multi-component case will be discussed  elsewhere.

\begin{acknowledgments}
We wish to thank J.\ Bauer and R.\ Sakano 
for valuable discussions, and C.\ Mora and J.\ von Delft for 
sending us Ref.\ \onlinecite{FilipponeMocaVonDelftMora} prior to publication.
This work was supported by JSPS KAKENHI (No.\ 26400319) and 
  a Grant-in-Aid for Scientific Research (S) (No.\ 26220711).
\end{acknowledgments}

\appendix

\section{The zero-frequency limit of $\Psi_{\sigma}^{--}(\omega)$ 
for an  $N$-component Anderson impurity}  
\label{sec:D2_Psi_causal_detail}

It  has been shown  in Eq.\ \eqref{eq:d2_T0} 
that   $\widehat{D}^2  \Sigma_{\mathrm{eq},\sigma}^{--}(\omega) 
\equiv \Psi_{\sigma}^{--}(\omega)$, 
which indicates that  the common coefficient 
 in the  $(eV)^2$ and  $T^2$ corrections to 
 $\Sigma_{\mathrm{eq},\sigma}^{--}(\omega)$ 
is determined by the  $\lim_{\omega \to 0}\Psi_{\sigma}^{--}(\omega)$.
In this appendix, we calculate this value. 
In order to give a general derivation, which can also be applied to 
an Anderson impurity with a number of components,  
we extend the impurity part of the Hamiltonian such that    
\begin{align}
\mathcal{H}_d^{(N)} =& \  
 \sum_{\sigma=1}^N
 \epsilon_{d\sigma}^{}\, n_{d\sigma}   +
\frac{1}{2} \sum_{\sigma \neq \sigma'} U_{\sigma\sigma'}^{} \, 
n_{d\sigma}^{} n_{d,\sigma'}^{} \;. 
\label{eq:Hd_multi_1}
\end{align}
The inter-electron interaction $U_{\sigma\sigma'}^{}$ 
 generally depends on $\sigma$ and $\sigma'$,    
with the requirements $U_{\sigma'\sigma}^{}=U_{\sigma\sigma'}^{}$ 
for $\sigma'\neq \sigma$. 
For  $N=2$,   it describes the single-orbital Anderson model 
for spin $1/2$ fermions which we have considered so far.
The remaining part of the Hamiltonian takes the same form as 
 Eqs.\ \eqref{eq:Ham_cond} and \eqref{eq:Ham_mix} 
but  the index runs over  $\sigma=1, 2, \ldots, N$.  
Namely,  the free conduction band  $\mathcal{H}_c$ also consists of $N$ 
{\it flavor} components,  
and  $\mathcal{H}_\mathrm{T}$ 
describes the tunnelings that  preserve the index $\sigma$.   
One of the  features of interest in the multi-component impurity 
is that for  $N>2$  the three-body correlations     
  $\chi_{\sigma_1\sigma_2\sigma_3}^{[3]}$ 
among  three different 
 components $\sigma_1\neq \sigma_2 \neq \sigma_3 \neq \sigma_1$ 
also contribute to the low-energy properties.

\begin{figure}[t]
 \leavevmode
\begin{minipage}{1\linewidth}
\includegraphics[width=0.5\linewidth]{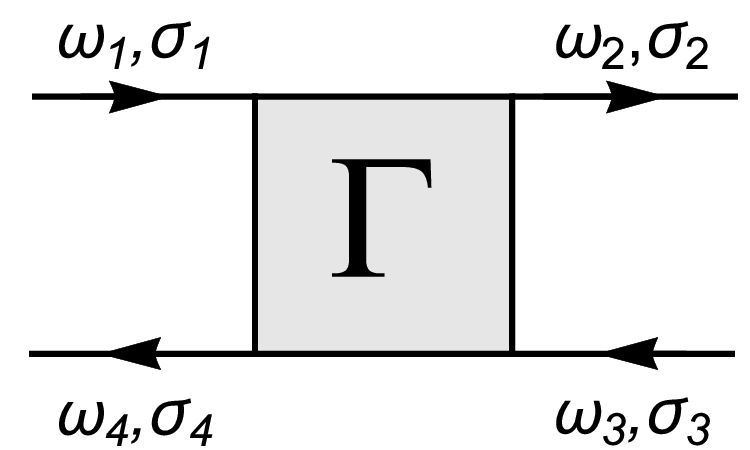}
\end{minipage}
 \caption{Vertex function 
$\Gamma_{\sigma_1\sigma_2;\sigma_3\sigma_4}^{}
(\omega_1, \omega_2; \omega_3, \omega_4)$. 
}
 \label{fig:vertex}
\end{figure}

For general $N$, the function 
 $\Psi_{\sigma}^{--}$  is defined by   
\begin{align}
\Psi_{\sigma}^{--}(\omega) 
\equiv 
 \lim_{\omega' \to 0}
\frac{\partial}{\partial \omega'} 
 \sum_{\sigma'=1}^N
\Gamma_{\sigma \sigma';\sigma' \sigma}(\omega, \omega'; \omega', \omega) 
\rho_{d\sigma'}^{}(\omega') ,
\label{eq:Psi_T0_N}
 \end{align}
in terms of the vertex function illustrated in Fig.\ \ref{fig:vertex}.
We show  in the following that the zero-frequency  limit is given by   
\begin{align}
 \lim_{\omega \to 0} 
\Psi_{\sigma}^{--}(\omega)  
 =
 \frac{1}{\rho_{d\sigma}^{}} \! 
\sum_{\sigma'(\neq \sigma)}
\frac{\partial \chi_{\sigma\sigma'}}{\partial \epsilon_{d\sigma'}} 
 - i  
\frac{3\pi}{\rho_{d\sigma}^{}} \! 
\sum_{\sigma' (\neq \sigma)}
\! 
\chi_{\sigma\sigma'}^2
\,  \mbox{sgn}(\omega)  
\label{eq:Psi_result_+_N}
\end{align}

\begin{figure}[t]
\leavevmode 
\begin{minipage}{1\linewidth}
\includegraphics[width=0.3\linewidth]{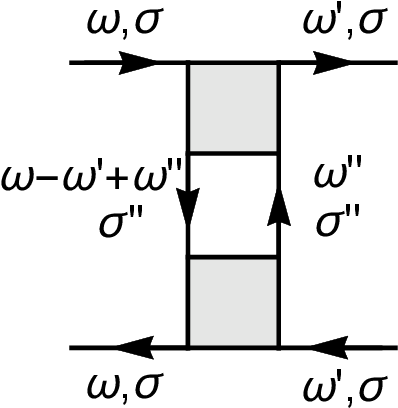}
\rule{0.02\linewidth}{0cm}
\includegraphics[width=0.3\linewidth]{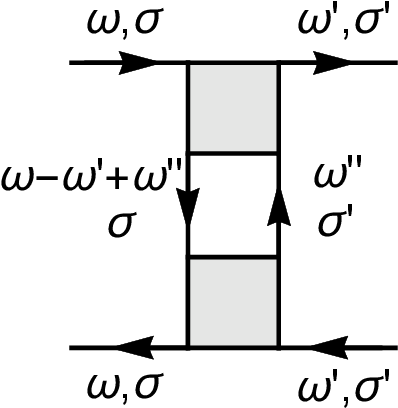}
\rule{0.02\linewidth}{0cm}
\includegraphics[width=0.3\linewidth]{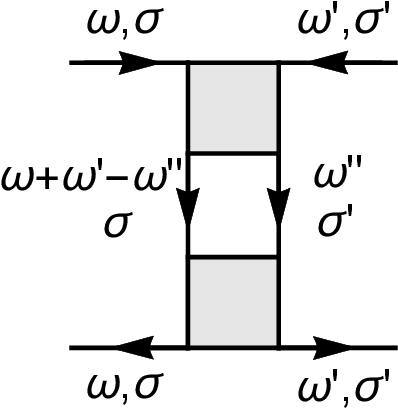}
\end{minipage}
 \caption{
Feynman diagrams which provide the imaginary part  to  the vertex function 
$\Gamma_{\sigma\sigma';\sigma'\sigma}^{} (\omega, \omega'; \omega', \omega)$ 
 for small $\omega$ and $\omega'$.
The shaded square represents the  zero-frequency  
vertex part  $\Gamma_{\sigma,\sigma'';\sigma''\sigma}^{}(0, 0; 0, 0)$, 
which for  $\sigma'' = \sigma$ identically vanishes 
 $\Gamma_{\sigma\sigma;\sigma\sigma}^{}(0, 0; 0, 0)=0$.  
The singular  $\mathrm{sgn}\, (\omega-\omega')$  
term arises from the intermediate particle-hole excitation 
with $\sigma'=\sigma$  shown in the left panel, 
and also from  the particle-hole pair  with $\sigma' \neq \sigma$ in the middle panel.
Another singular term  $\mathrm{sgn}\, (\omega+\omega')$  
arises from the  particle-particle pair excitation with $\sigma' \neq \sigma$
in the right panel. 
}
  \label{fig:vertex_singular_su2}
\end{figure}

The vertex function  
 $\Gamma_{\sigma\sigma';\sigma'\sigma}(\omega,\omega';\omega',\omega)$ 
has  lines of singularities  along  $\omega-\omega'=0$ 
and $\omega+\omega'=0$.\cite{Yoshimori,Eliashberg,EliashbergJETP15} 
 For small $\omega$ and $\omega'$, these singularities emerge  
  through the three diagrams shown in Fig.\ \ref{fig:vertex_singular_su2}, 
and  the imaginary part  of 
$\Psi_{\sigma}^{--}(\omega)$ 
can be calculated as\cite{ao2001PRB} 
\begin{align}
&  
\sum_{\sigma'}  
\rho_{d\sigma'}^{}\,
 \mathrm{Im}\, 
\frac{\partial}{\partial \omega'}
\Gamma_{\sigma\sigma';\sigma'\sigma}(\omega,\omega';\omega',\omega) 
\nonumber \\
& =    
-\,
\sum_{\sigma' (\neq \sigma)}
\left|\Gamma_{\sigma\sigma';\sigma'\sigma}(0,0;0,0)\right|^2 
\nonumber \\
& 
\ \times 
 \mathrm{Im} \Biggl[\  
\int \frac{d \omega''}{2\pi i}\,
G_{\mathrm{eq},\sigma'}^{--}(\omega'') \,
\frac{\partial}{\partial \omega'} \,
G_{\mathrm{eq},\sigma'}^{--}(\omega -\omega'+ \omega'')
\,\rho_{d\sigma}^{} 
\nonumber \\
& 
\ +\int \frac{d \omega''}{2\pi i}\,
G_{\mathrm{eq},\sigma'}^{--}(\omega'') \,
\frac{\partial}{\partial \omega'} \,
G_{\mathrm{eq},\sigma}^{--}(\omega -\omega'+ \omega'')
\,\rho_{d\sigma'}^{}
\nonumber \\
& \ 
+ \int \frac{d \omega''}{2\pi i}\,
G_{\mathrm{eq},\sigma'}^{--}(\omega'') \,
\frac{\partial}{\partial \omega'} \,G_{\mathrm{eq},\sigma}^{--}(\omega +\omega'-\omega'')\,
\rho_{d\sigma'}^{}
\,\Biggr]
 + \cdots
\nonumber
\\
& = \,   
-\,
\pi 
\sum_{\sigma' (\neq \sigma)}
\left|\Gamma_{\sigma\sigma';\sigma'\sigma}(0,0;0,0)\right|^2 
\nonumber\\
& \qquad \qquad \times 
\rho_{d\sigma}^{} \rho_{d\sigma'}^2
\Bigl[\, 
2\,\mbox{sgn}(\omega-\omega') 
+ 
\,\mbox{sgn}(\omega +\omega') \,
\Bigr]
 + \cdots
\nonumber
\\
& =    
-\pi 
\sum_{\sigma' (\neq \sigma)}
\frac{\chi_{\sigma\sigma'}^2}{\rho_{d\sigma}^{}}
\Bigl[\, 
 2\,\mbox{sgn}(\omega-\omega') \,
+\mbox{sgn}(\omega +\omega') \,
\Bigr]
 + \cdots
 \;.
\label{eq:Ward_2_Im_diagram}
\end{align}
In the second line,  
the three integrals correspond to  
 contributions of each diagram shown in Fig.\  \ref{fig:vertex_singular_su2}.   
The left and middle diagrams yield  
the non-analytic  $2\,\mbox{sgn}(\omega-\omega')$ contribution 
due to  the particle-hole pair  excitation,
and  the  right diagram yields the $\mbox{sgn}(\omega+\omega')$ contribution 
due to the  particle-particle pair excitation.
Taking first the limit $\omega'\to 0$ keeping 
the external frequency $\omega$ finite,  
we obtain the imaginary part  of Eq.\ (\ref{eq:d2_T0}), 
\begin{align}
\lim_{\omega \to 0}\,
\mathrm{Im}\, \Psi_{\sigma}^{--}(\omega) 
= 
- 3\pi 
\sum_{\sigma' (\neq \sigma)}
\frac{\chi_{\sigma\sigma'}^2}{\rho_{d\sigma}^{}}
\, \mbox{sgn}(\omega) \, . 
\label{eq:Ward_2_Im}
\end{align}

The real part of $\Psi_{\sigma}^{--}(\omega)$ 
does not have the non-analytic $\mathrm{sgn}(\omega)$ dependence, 
and   it can be deduced from  Eq.\ \eqref{eq:d2_T0} 
 by taking first the  $\omega \to 0$ limit, 
\begin{align}
\mathrm{Re}\, \Psi_{\sigma}^{--}(0)
 \, = & \ 
\sum_{\sigma'}  
\rho_{d\sigma'}^{}
\left.
\frac{\partial}{\partial \omega'}
\mathrm{Re}\,
\Gamma_{\sigma\sigma';\sigma'\sigma}(0,\omega';\omega',0)
 \,\right|_{\omega'= 0} 
\nonumber \\
& \ + \sum_{\sigma' (\neq \sigma)}
\Gamma_{\sigma\sigma';\sigma'\sigma}(0,0;0,0)
\,\rho_{d\sigma'}'.
\label{eq:D2_T0_real_I}
\end{align}
The second term of Eq.\ \eqref{eq:D2_T0_real_I} can be expressed in the form 
\begin{align}
\sum_{\sigma' (\neq \sigma)}
\Gamma_{\sigma\sigma';\sigma'\sigma}(0,0;0,0)
\,\rho_{d\sigma'}' 
= 
 - \sum_{\sigma'(\neq\sigma)}  
 \frac{\chi_{\sigma\sigma'}}
{\rho_{d\sigma}^{}\rho_{d\sigma'}^{}} 
\rho_{d\sigma'}'.  
\label{eq:D2_T0_real_Ib}
\end{align}
The first term of Eq.\ \eqref{eq:D2_T0_real_I} can be calculated as
\begin{align}
&
\!\!\!\!\!\!\!\!\!\!
\sum_{\sigma'}  
\rho_{d\sigma'}^{}
\left.
\frac{\partial}{\partial \omega'}\,  
\mathrm{Re}\,
\Gamma_{\sigma\sigma';\sigma'\sigma}(0,\omega';\omega',0)
\,
 \,\right|_{\omega'= 0} 
\nonumber \\
& = \,
\sum_{\sigma'}  
\rho_{d\sigma'}^{}
\left.
\frac{\partial}{\partial \omega'}\,  
\mathrm{Re}\,
\Gamma_{\sigma'\sigma;\sigma\sigma'}(\omega',0;0,\omega')
\,
 \,\right|_{\omega'= 0} 
\nonumber 
\\
&=  \,  
\sum_{\sigma'(\neq\sigma)}  
\rho_{d\sigma'}^{}
\left.
\frac{\partial}{\partial \omega'}\,  
\mathrm{Re}\,
\Gamma_{\sigma'\sigma;\sigma\sigma'}(\omega',0;0,\omega')
 \,\right|_{\omega' = 0} 
 \nonumber 
\\
& =\,  
- 
\sum_{\sigma'(\neq\sigma)}  
\rho_{d\sigma'}^{}
\left.
\frac{\partial}{\partial \omega}\,  
\frac{1}{\rho_{d\sigma}^{}}
\frac{\partial \,\mathrm{Re}\,\Sigma^{--}_{\text{eq},\sigma'}(\omega)}
{\partial \epsilon_{d\sigma}^{}}
 \,\right|_{\omega = 0} 
\nonumber \\
& =  \,
\sum_{\sigma'(\neq\sigma)}  
\frac{\rho_{d\sigma'}^{}}{\rho_{d\sigma}^{}}
\frac{\partial }{\partial \epsilon_{d\sigma}^{}}
\! \left.
\left(1-
\frac{\partial\, \Sigma^{--}_{\text{eq},\sigma'}(\omega)}{\partial \omega}
\right) \right|_{\omega = 0} 
\nonumber \\
& = \,   
\sum_{\sigma'(\neq\sigma)}  
\frac{\rho_{d\sigma'}^{}}{\rho_{d\sigma}^{}}
\frac{\partial \widetilde{\chi}_{\sigma'\sigma'}}{\partial \epsilon_{d\sigma}^{}}
=
\sum_{\sigma'(\neq\sigma)}  
\frac{\rho_{d\sigma'}^{}}{\rho_{d\sigma}^{}}
\frac{\partial \widetilde{\chi}_{\sigma'\sigma}}{\partial \epsilon_{d\sigma'}^{}}
\nonumber \\
 = & 
 \frac{1}{\rho_{d\sigma}^{}} 
\sum_{\sigma'(\neq\sigma)}  
\left(
\frac{\partial \chi_{\sigma\sigma'}}{\partial \epsilon_{d\sigma'}^{}} 
- \frac{\chi_{\sigma\sigma'}}{\rho_{d\sigma'}^{}}\, 
\frac{\partial \rho_{d\sigma'}^{}}{\partial \epsilon_{d\sigma'}^{}}
 \right) . 
\label{eq:D2_T0_real_Ia}
\end{align}
Note that   $\Gamma_{\sigma\sigma';\sigma'\sigma}
(0,\omega'; \omega',0) = 
\Gamma_{\sigma'\sigma;\sigma\sigma'}
(\omega',0; 0, \omega') $,  
the symmetric property of the vertex function 
has been used to obtain the second line.
To obtain the third line, we have used the property  
that  the vertex function for the parallel spins 
$\Gamma_{\sigma\sigma;\sigma\sigma}
(\omega,0;0,\omega)$ has no $\omega$-linear 
real part, which has been shown in {\it paper II}.\cite{ao2017_2_PRB}  
Therefore we obtain the following result 
from Eqs.\ \eqref{eq:D2_T0_real_I}--\eqref{eq:D2_T0_real_Ib}, 
using Eq.\ \eqref{eq:rho_d_omega_2} for the density of states:  
\begin{align}
\mathrm{Re}\, \Psi_{\sigma}^{--}(0)
\, = \, 
 \frac{1}{\rho_{d\sigma}^{}} 
\sum_{\sigma'(\neq\sigma)}  
\frac{\partial \chi_{\sigma\sigma'}}{\partial \epsilon_{d\sigma'}^{}} 
\;.
\label{eq:D2_T0_real}
\end{align}
The $(eV)^2$ and $T^2$ contributions of   
$\mathrm{Re}\,  \Sigma_{\mathrm{eq},\sigma}^{--}(0)$   
arise from  the intermediate single-particle excitation which carries 
the different {\it flavor} indexes $\sigma'$ from  the external one $\sigma$.

From these results, 
the vertex function for 
 $\Gamma_{\sigma\sigma';\sigma'\sigma}(\omega, \omega'; \omega' ,\omega)$  
can also be deduced. For $\sigma'=\sigma$, it takes the form
\begin{align}
\!\!\!\!\!
\Gamma_{\sigma\sigma;\sigma\sigma}(\omega , \omega'; \omega', \omega) 
\rho_{d\sigma}^{2}
 =
 i \pi 
\sum_{\sigma'(\neq \sigma)}
\chi_{\sigma\sigma'}^2
\bigl|\omega-\omega' \bigr| 
+ \cdots \!  , \!\!
 \label{eq:GammaUU_general_omega_dash_N}
\end{align}
and it for $\sigma'\neq \sigma$ is 
\begin{align}
& \Gamma_{\sigma\sigma';\sigma'\sigma}(\omega, \omega'; \omega' ,\omega) 
\,\rho_{d\sigma}^{}\rho_{d\sigma'}^{}
\  =  
\nonumber \\ 
& \qquad 
 -
\chi_{\sigma\sigma'}^{}
+ 
\rho_{d\sigma}^{}
\frac{\partial \widetilde{\chi}_{\sigma\sigma'}}
{\partial \epsilon_{d\sigma}^{}} \, \omega  
+ 
\rho_{d\sigma'}^{}
\frac{\partial \widetilde{\chi}_{\sigma'\sigma}}
{\partial \epsilon_{d\sigma'}^{}} \, \omega'   
\nonumber \\ 
& \qquad 
 + 
i \pi \,
\chi_{\sigma\sigma'}^2 
\Bigl(
\,\bigl|  \omega - \omega'\bigr| 
-
\,\bigl| \omega + \omega' \bigr| 
\Bigr)
+ \cdots
.
\label{eq:GammaUD_general_omega_dash_N}
\end{align}
These results and the $\omega^2$ contribution of  the self-energy 
are related each other via the Ward identity, given in Eq.\ \eqref{eq:YYY_T0_causal}.

\section{Coefficients $\alpha_{2\sigma}$ \& $\phi_{2\sigma}^{}$
of FMvDM}
\label{subsec:coefficient_FMvDM}

In this appendix, we summarize the relation between 
the parameters used in the description of FMvDM 
and the derivative of the susceptibilities.
The coefficients $\alpha_{1\sigma}$ and  $\phi_{1}^{}$  
are the parameters which  were introduced by  Nozi\`{e}res for 
his  phenomenological description,  
\begin{align}
\frac{\alpha_{1\sigma}}{\pi}\,=\, 
  \chi_{\sigma\sigma}^{} \,,
  \qquad \ \ 
\frac{\phi_{1}^{}}{\pi}\,=\, 
 - \chi_{\uparrow\downarrow}^{} \,.
\label{eq:phi_1_def_Nozi}
\end{align}
Note that 
$\chi_{\uparrow\downarrow}^{} = \chi_{\downarrow\uparrow}^{}$ 
and it is an even function of $h$  because of  $\Omega$ is an even function of $h$,  
as mentioned.

The coefficients   $\alpha_{2\sigma}$ and  $\phi_{2\sigma}^{}$ 
 defined in Eqs.\ (13a)--(13d) of  the  FMvDM's paper\cite{FilipponeMocaVonDelftMora} 
can also  be written in terms of  the susceptibilities.
Substituting the charge and spin susceptibilities, 
 $\chi_{c}  \equiv 
 \chi_{\uparrow\uparrow}^{} + \chi_{\downarrow\downarrow}^{} 
+  \chi_{\uparrow\downarrow}^{} +  \chi_{\downarrow\uparrow}^{}$ and 
$\chi_{s}  \equiv \displaystyle \frac{1}{4} 
\left(  
\chi_{\uparrow\uparrow}^{} + \chi_{\downarrow\downarrow}^{} 
-  \chi_{\uparrow\downarrow}^{} - \chi_{\downarrow\uparrow}^{}
\right)$,  into  the definitions  and rescaling the magnetic field as $B=2h$,   
\begin{align}
\frac{\alpha_{2\uparrow} + \alpha_{2\downarrow}}{2\pi} 
\,=& \  
-\frac{3}{4}\frac{\partial \chi_s}{\partial \epsilon_d}
-\frac{1}{16}\frac{\partial \chi_c}{\partial \epsilon_d}
\nonumber \\
&
\!\!\!\!\!\!\!\!\!\!\!\!\!\!\!\!\!\!\!\!\!\!\!\!\!\!\!\!\!\!\!\!
=   
\frac{1}{8} 
\left(\frac{\partial}{\partial \epsilon_{d\uparrow}}+\frac{\partial}{\partial \epsilon_{d\downarrow}}\right)
\Bigl[ 
-2
\left(  
\chi_{\uparrow\uparrow}^{} + \chi_{\downarrow\downarrow}^{} 
\right)
+\left(   \chi_{\uparrow\downarrow}^{} + \chi_{\downarrow\uparrow}^{}
\right)
\Bigr]
,
\\
\frac{\alpha_{2\uparrow}- \alpha_{2\downarrow}}{2\pi} 
\,= & \   
\frac{1}{2}\frac{\partial \chi_s}{\partial B}
+\frac{3}{8}\frac{\partial \chi_c}{\partial B} 
\nonumber \\
&
\!\!\!\!\!\!\!\!\!\!\!\!\!\!\!\!\!\!\!\!\!\!\!\!\!\!\!\!\!\!\!\!
=  
\frac{1}{8} 
\left(-\frac{\partial}{\partial \epsilon_{d\uparrow}}+\frac{\partial}{\partial \epsilon_{d\downarrow}}\right)
\Bigl[ 
2
\left(  
\chi_{\uparrow\uparrow}^{} + \chi_{\downarrow\downarrow}^{} 
\right)
+\left(   \chi_{\uparrow\downarrow}^{} + \chi_{\downarrow\uparrow}^{}
\right)
\Bigr] ,
\\
\frac{\phi_{2\uparrow}^{}+ \phi_{2\downarrow}^{}}
{2\pi} \,= & \ 
-\frac{\partial \chi_s}{\partial \epsilon_d}
+\frac{1}{4}\frac{\partial \chi_c}{\partial \epsilon_d}
\nonumber \\
\,=& \  
\frac{1}{2}
\left(\frac{\partial}{\partial \epsilon_{d\uparrow}}+\frac{\partial}{\partial \epsilon_{d\downarrow}}\right)
\left(   \chi_{\uparrow\downarrow}^{} + \chi_{\downarrow\uparrow}^{}
\right)
\; , 
\\
\frac{\phi_{2\uparrow}^{}- \phi_{2\downarrow}^{}}{2\pi} 
\,=& \  
2\frac{\partial \chi_s}{\partial B}
-\frac{1}{2}\frac{\partial \chi_c}{\partial B} 
\nonumber \\
\,=& \  
\frac{-1}{2} 
\left(-\frac{\partial}{\partial \epsilon_{d\uparrow}}+\frac{\partial}{\partial \epsilon_{d\downarrow}}\right)
\left(   \chi_{\uparrow\downarrow}^{} + \chi_{\downarrow\uparrow}^{}
\right)
.
\label{eq:phi2_def_Mora_vonDelft}
\end{align}
Thus, the coefficients $\alpha_{2\sigma}$ and $\phi_{2\sigma}^{}$ can be 
expressed  in the form 
\begin{align}
\frac{\alpha_{2\sigma}}{\pi} 
\,= & \  
-\frac{1}{2}\, \frac{\partial  \chi_{\sigma\sigma}^{} }{\partial \epsilon_{d\sigma}^{}} 
\;, 
%
\qquad \  
%
\frac{\phi_{2\sigma}^{}}{\pi}
\,=\, 
 2 \, \frac{\partial  \chi_{\uparrow\downarrow}^{}}{\partial \epsilon_{d\sigma}^{}} 
\;. 
\label{eq:phi2_def_FMvDM}
\end{align}

Using these relations, the coefficient for the  $\omega^2$ real part of the self-energy,  
provided in Eqs.\ (B2a) and (B8b) of FMvDM's paper,\cite{FilipponeMocaVonDelftMora} 
can be  confirmed to agree with Eq.\ \eqref{eq:self_w2} of ours:  
\begin{align}
& 
\!\!\!
\frac{ \widetilde{R}_{\sigma,\omega}}{z_{\sigma}}\,  =  \,  
\frac{\alpha_{2\sigma}}{\pi \rho_{d\sigma}^{}} 
- \frac{\pi \rho_{d\sigma}^{}}{z_{\sigma}^2} \cot \delta_{\sigma} 
\nonumber \\
& 
\ \ = 
-\frac{1}{2\rho_{d\sigma}^{}}\!
\left(
\frac{\partial\chi_{\sigma\sigma}^{}}{\partial \epsilon_{d\sigma}^{}}
+2 \pi 
\cot \delta_{\sigma}\chi_{\sigma\sigma}^2
\right)  
 =   -\frac{1}{2}
\frac{\partial \widetilde{\chi}_{\sigma\sigma}^{}}{\partial \epsilon_{d\sigma}^{}} .
 \end{align}
We have also used Eq.\ \eqref{eq:Dren_to_Dsus_org} to obtain the last line.

\section{Comparison with  $\mathrm{Re}\, \Sigma^r(\omega,eV)$ described 
 in Ref.\ \onlinecite{ao2001PRB}
}
\label{sec:h=0_previous}

The explicit low-energy expression of  
the real part  of the self-energy  given in Eq.\ \eqref{eq:self_real_ev_mag} 
reproduces at zero magnetic field $h=0$ the previous result, 
reported  in  Eq.\ (19) of  Ref.\ \onlinecite{ao2001PRB}.  
It  was written in such a way that  
the coefficient for the $\omega^2$ real part, $b$,  
as an additional parameter that had not  been related to 
the other renormalized parameters
\begin{align} 
b \, \equiv \, 
\left.
\mathrm{Re}\, 
\frac{\partial^2 \Sigma_{\mathrm{eq}}^{r}(\omega)}{\partial \omega^2}
\right|_{\omega=0} \;. 
\end{align}
Recent development clarifies  that this coefficient  can be written in terms 
of the derivative of the static susceptibilities 
  $b = {\partial \widetilde{\chi}_{\uparrow \uparrow}}/
{\partial \epsilon_{d\uparrow}^{}}$, as mentioned for  Eq.\ \eqref{eq:self_w2}. 
With this recent knowledge, we can explicitly confirm that 
the previous result  is completely  identical to  Eq.\ \eqref{eq:self_real_ev_zero_field}. 

The coefficients for  $\omega eV$ and  $\alpha^2 (eV)^2 $ terms 
in Eq.\ (19) of  Ref.\ \onlinecite{ao2001PRB}  can be written, respectively, as  
\begin{align}
-\left(b 
-\frac{\partial \widetilde{\chi}_{\uparrow \uparrow} }{\partial \epsilon_d^{}} \right)
 = &    
-\frac{\partial \widetilde{\chi}_{\uparrow \uparrow}}{\partial \epsilon_{d\uparrow}}
 + \frac{\partial \widetilde{\chi}_{\uparrow \uparrow} }{\partial \epsilon_d^{}} 
=
\frac{\partial \widetilde{\chi}_{\uparrow \uparrow}}{\partial \epsilon_{d\downarrow}^{}} 
\nonumber \\
= & \  
\frac{\partial \widetilde{\chi}_{\uparrow \downarrow}}{\partial \epsilon_{d\uparrow}^{}} 
,
\\ 
b 
-\frac{\partial \widetilde{\chi}_{s} }{\partial \epsilon_d^{}}
= & 
\frac{\partial \widetilde{\chi}_{\uparrow \uparrow}}{\partial \epsilon_{d\uparrow}^{}}
 - \frac{\partial }{\partial \epsilon_d} \left(\widetilde{\chi}_{\uparrow \uparrow} -\widetilde{\chi}_{\uparrow \downarrow} \right)
\nonumber \\
= & \ 
- \frac{\partial \widetilde{\chi}_{\uparrow \downarrow}}{\partial \epsilon_{d\uparrow}^{}}+ 
\frac{\partial \widetilde{\chi}_{\uparrow \downarrow}}{\partial \epsilon_{d}^{}}
=
 \frac{\partial \widetilde{\chi}_{\uparrow \downarrow}}{\partial \epsilon_{d\downarrow}^{}} .
\end{align}
These coefficients agree with the corresponding results 
given in Eq.\ \eqref{eq:self_real_ev_mag} for  $\sigma=\uparrow$ and  $h=0$. 
Furthermore, the coefficient for 
 the $(eV)^2$ term  that emerges through the $\widehat{D}^2$ operator 
can be written in the form,  
 \begin{align}
 &-
\lim_{h\to 0}
\left(   b 
-\frac{\partial \widetilde{\chi}_{\uparrow \uparrow} }{\partial \epsilon_d^{}}
+ \frac{\rho_{d}'}{\rho_{d}^{}}\, \widetilde{\chi}_{\uparrow\downarrow} 
\right) 
\,  =    \,  
\lim_{h\to 0}
\left(
\frac{\partial \widetilde{\chi}_{\uparrow \downarrow}}
{\partial \epsilon_{d\uparrow}^{}}
-
\frac{\chi_{\uparrow\downarrow} }{\rho_{d}^2}\, \rho_{d}'
\right)
\nonumber \\
 & \quad 
 =  \,  
\lim_{h\to 0}
\left[
\frac{\partial}
{\partial \epsilon_{d\uparrow}^{}}
\left( \frac{\chi_{\uparrow \downarrow}}{\rho_{d\uparrow}^{}} \right)
+
\frac{\chi_{\uparrow\downarrow}}{\rho_{d\uparrow}^2}\, 
\frac{\partial \rho_{d\uparrow}^{}}{\partial \epsilon_{d\uparrow}^{}} 
\right]
=  \,   
\lim_{h\to 0}
 \frac{1}{\rho_{d\uparrow}^{}} 
\frac{\partial \chi_{\uparrow \downarrow}}{\partial  \epsilon_{d\uparrow}^{}}
\nonumber \\
& \quad   =   
\lim_{h\to 0}
 \frac{1}{\rho_{d\uparrow}^{}} 
\frac{\partial \chi_{\uparrow \downarrow}}{\partial  \epsilon_{d\downarrow}^{}}
\;.
\label{eq:D2_T0_real_previous_rev}
\end{align}
This agrees with  the general result,  given in Eq.\ \eqref{eq:D2_T0_real}.

\section{Erratum for the original version \\ 
Phys.\ Rev.\ B {\bf  97}, 035435 (2018)}

After we published the original version of this paper, 
we found some errors which should be corrected.  
We summarized the changes in this appendix,  
which have already been  corrected  in this version. 
All the errors occurred only in the applications to some special cases.
The general Fermi-liquid relations for the self-energy, vertex functions, 
and transport coefficients, described in the original paper,  
 are not affected by these revisions.

The first point is that,  in figures  4 and 5 of our paper,  
the results for the three-body correlation function 
for the up- and down-spins  $\Theta_{\uparrow\downarrow}$ 
at zero magnetic field were plotted with the wrong sign. 
These figures  have been corrected in the version 3. 

The other errors arose for the magneto-transport coefficients, 
which have been studied at half-filling $\epsilon_d^{}=-U/2$ in Sec.\ V C, 
Sec.\ VI C, and Appendix C.
The central point is that there is an additional three-body term 
at finite magnetic fields, 
which was not  taken into account properly in these sections. 
It is simply an error occurring  in the application 
as the general Fermi-liquid relations that we described naturally yield the additional term.
This term, which we write  $\Theta_{X}^{}$,  gives 
 finite contributions together with the other three-body term $\Theta_{M}^{}$ 
that has been described in detail in the original  paper: 
 Eq.\ (5.28) has been replaced by  the following form, 
\ adding the second line for $\Theta_{X}^{}$, 
\begin{align*}
\Theta_{M}^{} 
\, \equiv & \   
-\, \frac{\sin (\pi m_d)}{2\pi} 
\frac{1}{\chi_{\uparrow\uparrow}^2}
 \, \frac{\partial \chi_{\uparrow\downarrow}}{\partial h} \,,
\qquad \qquad \qquad   \quad  \ \,   (5.28a)
\\
\Theta_{X}^{} 
\, \equiv & \     
-\, \frac{\sin (\pi m_d)}{2\pi} 
\frac{1}{\chi_{\uparrow\uparrow}^2}
\frac{\partial }{\partial \epsilon_d} 
\left(
\frac{\chi_{\uparrow\uparrow} 
- \chi_{\downarrow\downarrow}}{2}
\right)
\;. 
\qquad   (5.28b)
\end{align*}
The additional contribution of $\Theta_{X}^{}$  also appears 
in  Eqs.\ (5.23), (5.24), (5.25), (6.27), (6.29), (C.5), and (C.6):  
these  equations have  been corrected in this version 4.

Correspondingly,  the NRG results for the transport coefficients, 
presented in the lower panel of Fig.\ 8 and the upper panel of Figs.\ 9 and 12, 
 have also been replaced by the new ones.
As shown in new Fig.\ 8,  $\Theta_{X}^{}$ becomes zero at $h=0$, 
and  is very small for low fields $h \lesssim T_K$. 
It becomes comparable to $\Theta_{M}^{}$ at $h\gtrsim 0.1 U$, 
and in the high-field limit $h \to \infty$ 
it approaches $\Theta_{X}\to -2$ where the interaction $U$ becomes less important.
Therefore, major changes appear only at high fields: 
the coefficients approach  $(48/\pi^2)C_{T}^{h} \to -3$ 
and  $(64/\pi^2)C_{V}^{h} \to -3$ in Fig.\ 9,  
and   $(48/\pi^2)C_{R,h}^\mathrm{MA}  \to  1$ and 
  $(80/7\pi^2)C_{\kappa,h}^\mathrm{MA} \to 1$ in Fig.\ 12. 
For low-field behavior of these coefficients, however,  
 $\Theta_{X}^{}$  does not  cause  visible changes.
We have confirmed that the contribution of  $\Theta_{X}^{}$
becomes smaller than the symbol size for the lines shown in the lower panel of Fig.\ 9, 
in which the rescaled values of  the conductance coefficients, 
  $\overline{C}_{T}^{h}$ and  $\overline{C}_{V}^{h}$,  
are plotted for  $0<h<0.5 T_K$.


%

\end{document}